\documentclass[]{pasj01}
\usepackage{url}
\usepackage{color} 
\draft

\begin{document} 
\Received{}
\Accepted{}

\title{Spectroscopic observations of active solar-analog stars having high X-ray luminosity, as a proxy of superflare stars}

\author{Yuta \textsc{Notsu}\altaffilmark{1, 2, *}}
\author{Satoshi \textsc{Honda}\altaffilmark{3}}
\author{Hiroyuki \textsc{Maehara}\altaffilmark{4}}
\author{Shota \textsc{Notsu}\altaffilmark{1, 2}}
\author{Kosuke \textsc{Namekata}\altaffilmark{1}}
\author{Daisaku \textsc{Nogami}\altaffilmark{1}}
\author{Kazunari \textsc{Shibata}\altaffilmark{5}}

\altaffiltext{1}{Department of Astronomy, Kyoto University, Kitashirakawa-Oiwake-cho, Sakyo-ku, Kyoto 606-8502}
\email{ynotsu@kwasan.kyoto-u.ac.jp}
\altaffiltext{2}{JSPS Research Fellow (DC1)}
\altaffiltext{3}{Center for Astronomy, University of Hyogo, 407-2, Nishigaichi, Sayo-cho, Sayo, Hyogo 679-5313}
\altaffiltext{4}{Okayama Astrophysical Observatory, National Astronomical Observatory of Japan, 3037-5 Honjo, Kamogata, Asakuchi, Okayama 719-0232}
\altaffiltext{5}{Kwasan and Hida Observatories, Kyoto University, Yamashina-ku, Kyoto 607-8471}



\maketitle

\begin{abstract}
Recent studies of solar-type superflare stars have suggested that 
even old slowly rotating stars similar to the Sun can have large starspots and superflares.
We conducted high dispersion spectroscopy of 49 nearby solar-analog stars (G-type main sequence stars with $T_{\rm{eff}}\approx5,600\sim6,000$ K)
identified as ROSAT soft X-ray sources, 
which are not binary stars from the previous studies. 
We expected that these stars can be used as a proxy of bright solar-analog superflare stars, 
since superflare stars are expected to show strong X-ray luminosity.
More than half (37) of the 49 target stars show no evidence of binarity, 
and their atmospheric parameters (temperature, surface gravity, and metallicity)
are within the range of ordinary solar-analog stars. 
We measured the intensity of Ca II 8542 and H$\alpha$ lines, which are good indicators of the stellar chromospheric activity. 
The intensity of these lines indicates that all the target stars have large starspots. 
We also measured $v\sin i$ (projected rotational velocity) and Lithium abundance for the target stars. 
Li abundance is a key to understanding the evolution of the stellar convection zone, 
which reflects the stellar age, mass and rotational history. 
We confirmed that many of the target stars rapidly rotate and have high Li abundance, 
compared with the Sun, as suggested by many previous studies. 
There are, however, also some target stars that rotate slowly 
($v\sin i=2\sim 3$ km s$^{-1}$) and have low Li abundance like the Sun. 
These results support that old and slowly rotating stars 
similar to the Sun could have high activity level and large starspots.
This is consistent with the results of our previous studies of solar-type superflare stars.
In the future, it is important to conduct long-term monitoring observations of these active solar-analog stars
in order to investigate detailed properties of large starspots from the viewpoint of stellar dynamo theory.
\end{abstract}

\section{Introduction}\label{sec:intro}
\ \ \ \ \ \ \
Flares are energetic explosions on the surface of the stars 
and are thought to occur by impulsive releases of magnetic energy stored around starspots, 
like solar flares (e.g., \cite{Priest1981}; \cite{Shibata2011}; \cite{Shibata2016}).
Many stars are known to show stellar magnetic activity including flares.
The relations among stellar rotation, activity, and Li abundances of solar-type stars 
have long been under investigation in many studies 
(e.g., \cite{Skumanich1972}; \cite{Noyes1984}; \cite{Pizzolato2003}; \cite{Takeda2010}; \cite{Mishenina2012}; \cite{Xing2012}).
These previous studies show that young rapidlly rotating stars 
($P_{\rm{rot}}\sim$a few days and $v \sin i\gtrsim$10 km s$^{-1}$) 
tend to show high magnetic activity (for example, strong X-ray luminosity), 
and magnetic fields of a few kG are considered to be distributed in large regions on the stellar surface (\cite{Gershberg2005}; \cite{ReidHawley2005}; \cite{Benz2010}). 
In contrast, the Sun rotates slowly ($P_{\rm{rot}}\sim$25 days and $v \sin i\sim$2 km s$^{-1}$), 
and the mean magnetic field is weak (a few G). 
Then it has been thought that slowly rotating stars like the Sun basically does not have high magnetic activity.
\\ \\
\ \ \ \ \ \
Recently, however, many superflares on solar-type stars (G-type main sequence stars) have been discovered 
(\cite{Maehara2012} \& \yearcite{Maehara2015}; \cite{Shibayama2013}; \cite{Candelaresi2014}; \cite{Wu2015}; \cite{Balona2015}; \cite{Davenport2016})
by using the high-precision photometric data of Kepler space telescope (\cite{Koch2010}).
Superflares are flares that have a total energy of $10^{33}\sim10^{38}$ erg (\cite{Schaefer2000}), 
$10\sim10^6$ times larger than that of the largest solar flares on the Sun ($\sim10^{32}$erg; \cite{Emslie2012}).
Here solar-type stars are defined as stars that have a surface temperature 
of $5100\leq T_{\rm{eff}}\leq6000$ K and a surface gravity of $\log g\geq4.0$.
The analyses of Kepler data enabled us to discuss statistical properties of superflares since a large number of flare events were discovered.
Many superflare stars show quasi-periodic brightness variations 
with a typical period of from one day to a few tens of days and the amplitude of 0.1\% $\sim$10\%,
and they can be explained well by the rotation of a single star with fairly large starspots \citep{YNotsu2013}.
We clarified that the superflare energy is related to the total coverage of the starspots, 
and that the superflare energy can be explained by the magnetic energy stored around these large starspots. 
We then found that energetic superflares (with energy comparable to $10^{34}\sim10^{35}$ erg) can occur 
on stars rotating as slowly as the Sun ($P_{\rm{rot}}=20\sim30$ days), 
even though the frequency is low (once in a few thousand years), 
compared with rapidly rotating stars \citep{YNotsu2013}.
In addition, we suggested, on the basis of theoretical estimates,
that the Sun can generate a large magnetic flux 
that is sufficient for causing superflares with an energy of $10^{34}$ erg within one solar cycle ($\sim$11 yr) \citep{Shibata2013}.
\\ \\
\ \ \ \ \ \
The results of Kepler described here are now supported by the spectroscopic studies.
We observed 50 solar-type superflare stars using Subaru/HDS 
(\cite{SNotsu2013}; \cite{Nogami2014}; \cite{YNotsu2015a} \& \yearcite{YNotsu2015b}; \cite{Honda2015}). 
We found that more than half of the targets have no evidence of a binary system, 
and stellar atmospheric parameters (temperature, surface gravity, and metallicity) of these stars 
are in the range of ordinary solar-type stars \citep{YNotsu2015a}. 
\citet{Nogami2014} found that spectroscopic properties ($T_{\rm{eff}}$, $\log g$, [Fe/H], and rotational velocity) of
the two superflare stars KIC9766237 and KIC9944137 are very close to those of the Sun.
More important, \citet{YNotsu2015b} supported the above interpretation that 
the quasi-periodic brightness variations of superflare stars are explained by the rotation of a star with large starspots,
by measuring $v\sin i$ (projected rotational velocity) and the intensity of Ca II 8542 line.
Existence of large starspots on superflare stars were also supported by \citet{Karoff2016} using Ca II H\& K observations with LAMOST telescope. 
We also conducted the Lithium (Li) abundance analysis of these superflare stars \citep{Honda2015}.
As mentioned above, Li abundance is known to be a clue for investigating the stellar age 
(e.g., \cite{Skumanich1972}; \cite{Takeda2010}; \cite{Mishenina2012}).
Many of the superflare stars tend to show high Li abundance, 
but there are some objects that have low Li abundance and  rotate slowly. 
These results indicate that superflare stars are not only connected to young stars but also old stars like our Sun.
\\ \\
\ \ \ \ \ \ \
These results of superflare studies support that even old slowly rotating stars similar to the Sun 
can have large starspots and superflares.
We also found that there is a positive correlation between starspot size and energy of superflares \citep{YNotsu2013}. 
We can say that the existence of large starspots is a key to understand superflares.
As a next step of a series of superflare research, 
we need to observe the detailed properties of such stars with large starspots in order to understand their properties  
(e.g., lifetime of starspots, period of the activity cycle like the solar 11-year cycle, differential rotation)
from the viewpoint of stellar dynamo theory. 
For this, more detailed (e.g., long-term monitoring) observations are necessary, 
but many of the stars investigated by Kepler prime mission are relatively faint (typically, $V=12\sim14$ mag) 
and all of the stars are distributed in one limited field in the constellations Cygnus and Lyra (\cite{Koch2010}; \cite{Brown2011}). 
It is difficult to conduct spectroscopic monitoring observations of these relatively faint stars with 2$\sim$4m-class telescope,
and it is better to have more bright target stars in various area of the sky 
for conducting long-term monitoring observations efficiently in the near future.
\\ \\
\ \ \ \ \ \ \
In this context, we have now performed spectroscopic observations of the 49 bright active solar-analog stars 
(G-type main sequence stars with $T_{\rm{eff}}\approx5,600\sim6,000$ K) 
identified as ROSAT soft X-ray All-Sky Survey sources (\cite{Voges1999} \& \yearcite{Voges2000}), 
by using the High Dispersion Echelle Spectrograph (HIDES) attached at the 1.88-m reflector of Okayama Astronomical Observatory (OAO).
Solar-type stars in an active phase having superflares are expected to show strong X-ray luminosity, 
and thus these stars could be used as a proxy of bright solar-analog superflare stars.
In this study, we measure the detailed properties 
(e.g., atmospheric parameters, rotation velocity, chromospheric activity, Li abundance) of these 49 stars, 
and investigate whether there are slowly rotating stars 
among the target active solar-analog stars as suggested by superflare studies.
This is the first step of detailed observations of bright superflare stars mentioned above.
\\ \\
\ \ \ \ \ \ \
We describe the selection of the target stars and the details of our observations in Section \ref{sec:target-and-obs}.
In Section \ref{sec:stpara}, we checked the binarity of the target list, and after that, 
estimated stellar various parameters on the basis of spectroscopic data (The details of analyses and results are described in Appendix \ref{sec-apen:Detail-ana}). 
After these analyses, in Section \ref{sec:discussion}, 
we comment on the estimated stellar parameters, 
and then discuss rotational velocity, chromospheric activities, and Li abundances of 
the target active solar-analog stars having high X-ray luminosity.
Finally, we summarize the results of this paper and mention implications for future studies in Section \ref{subsec:dis-fututre}.

\section{Targets and Observation}\label{sec:target-and-obs}
\subsection{Target stars}\label{subsec:target-stars}
\begin{figure}[htbp]
 \begin{center}
  \FigureFile(82mm,82mm){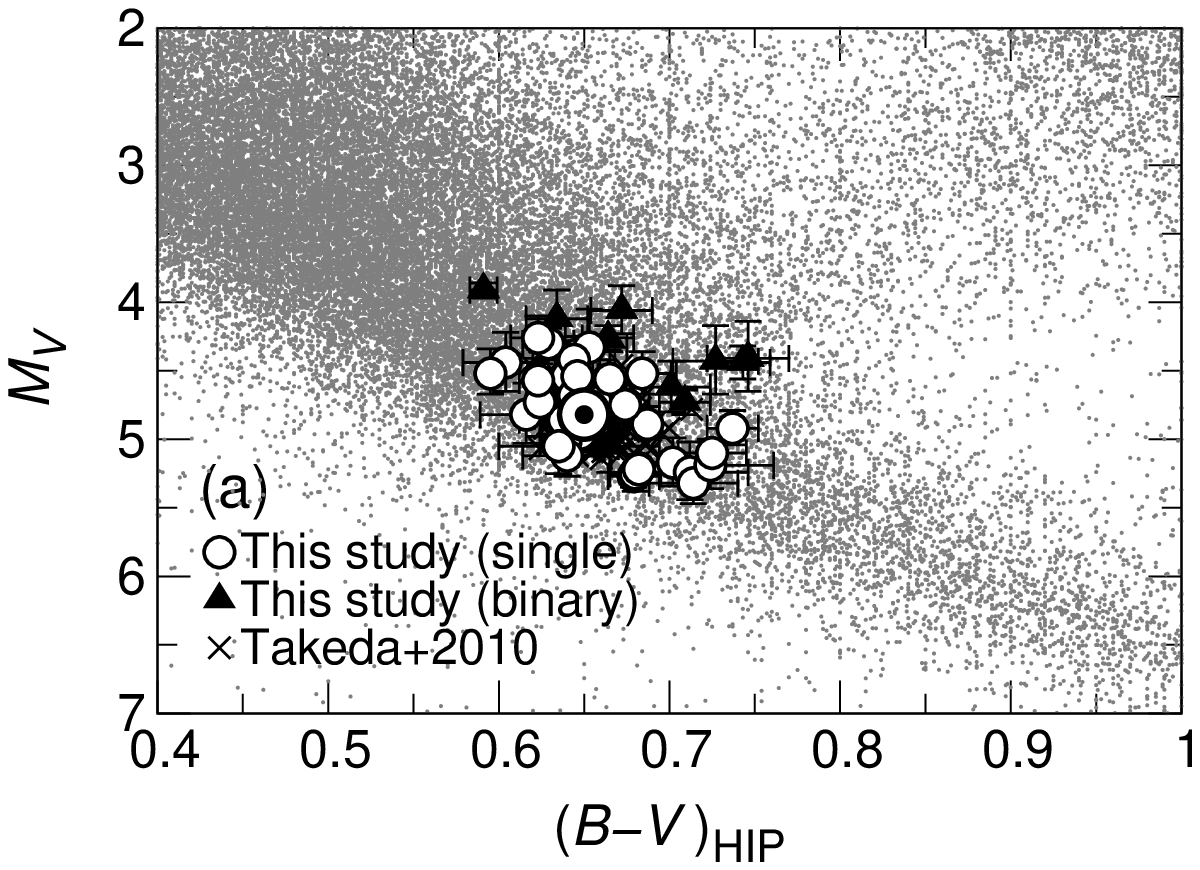}
  \FigureFile(82mm,82mm){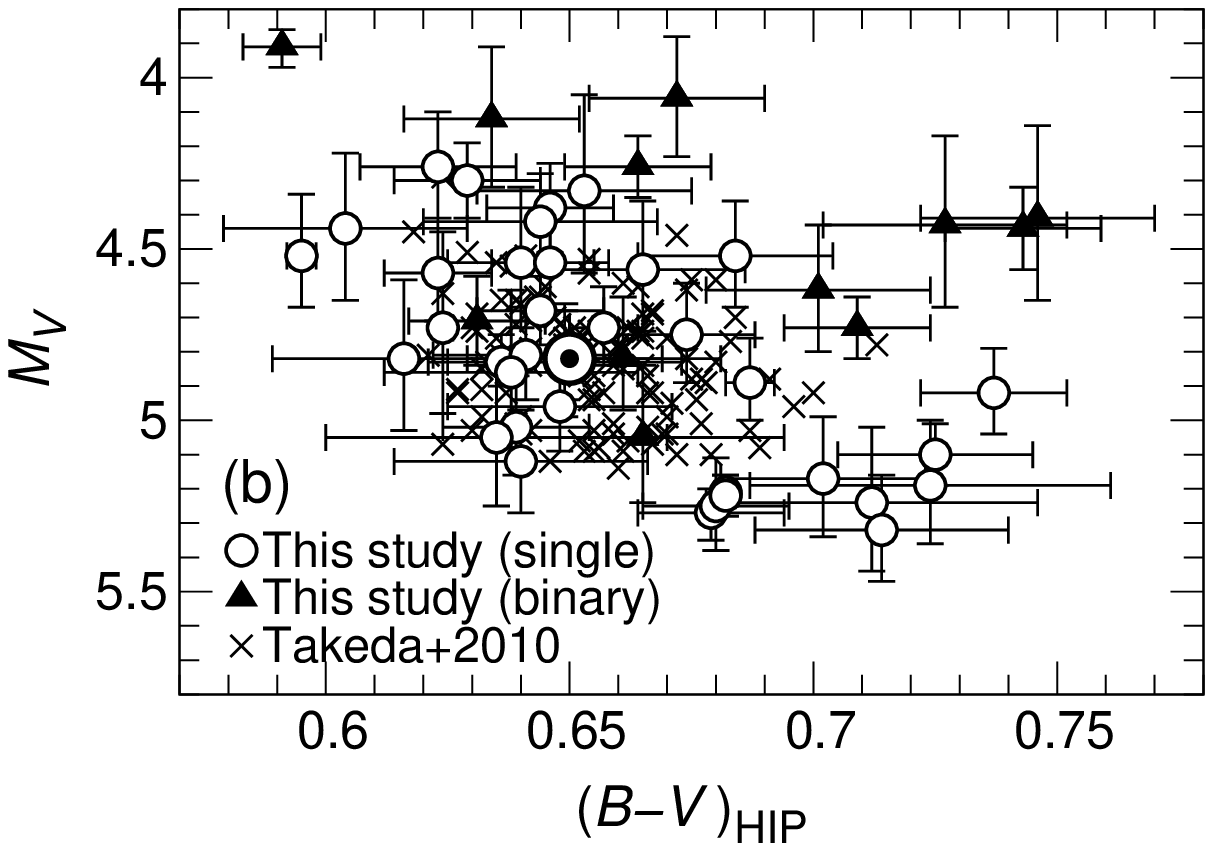}
  \FigureFile(82mm,82mm){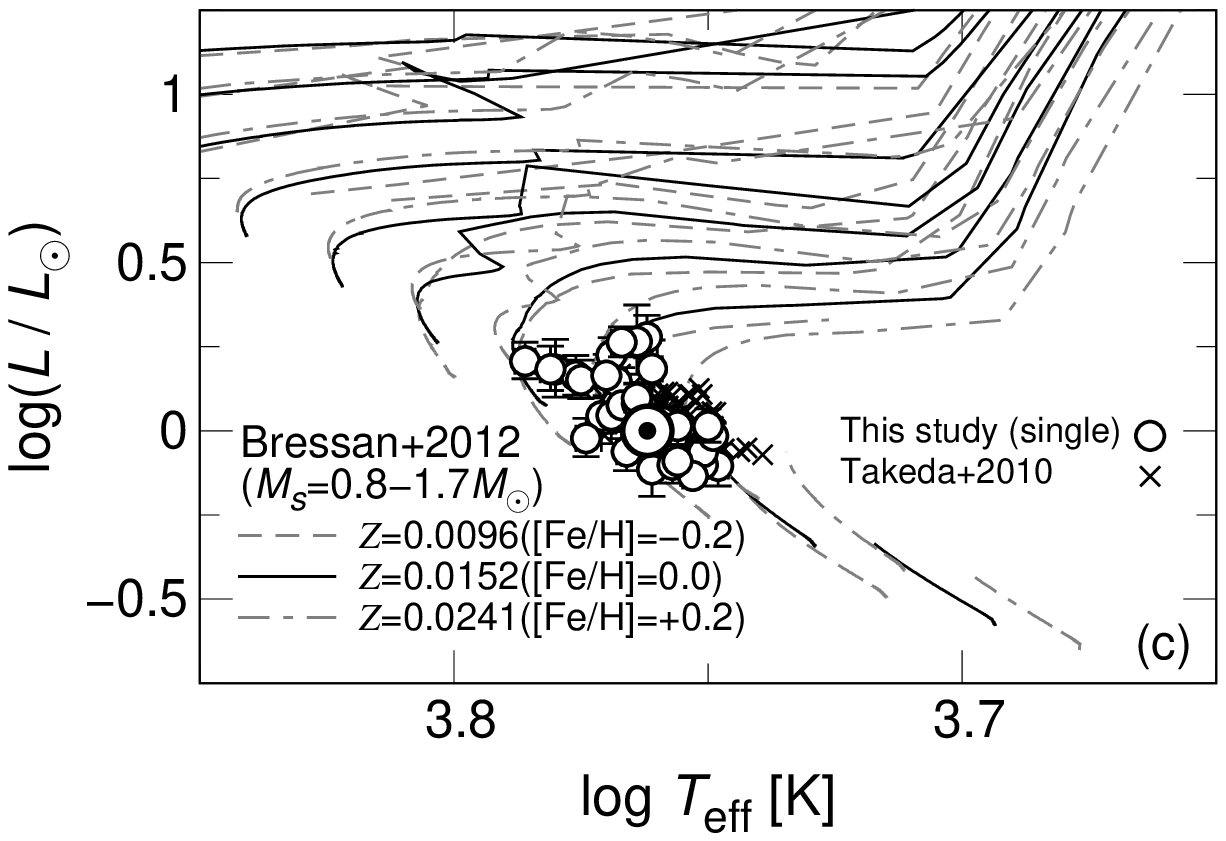}
  \FigureFile(82mm,82mm){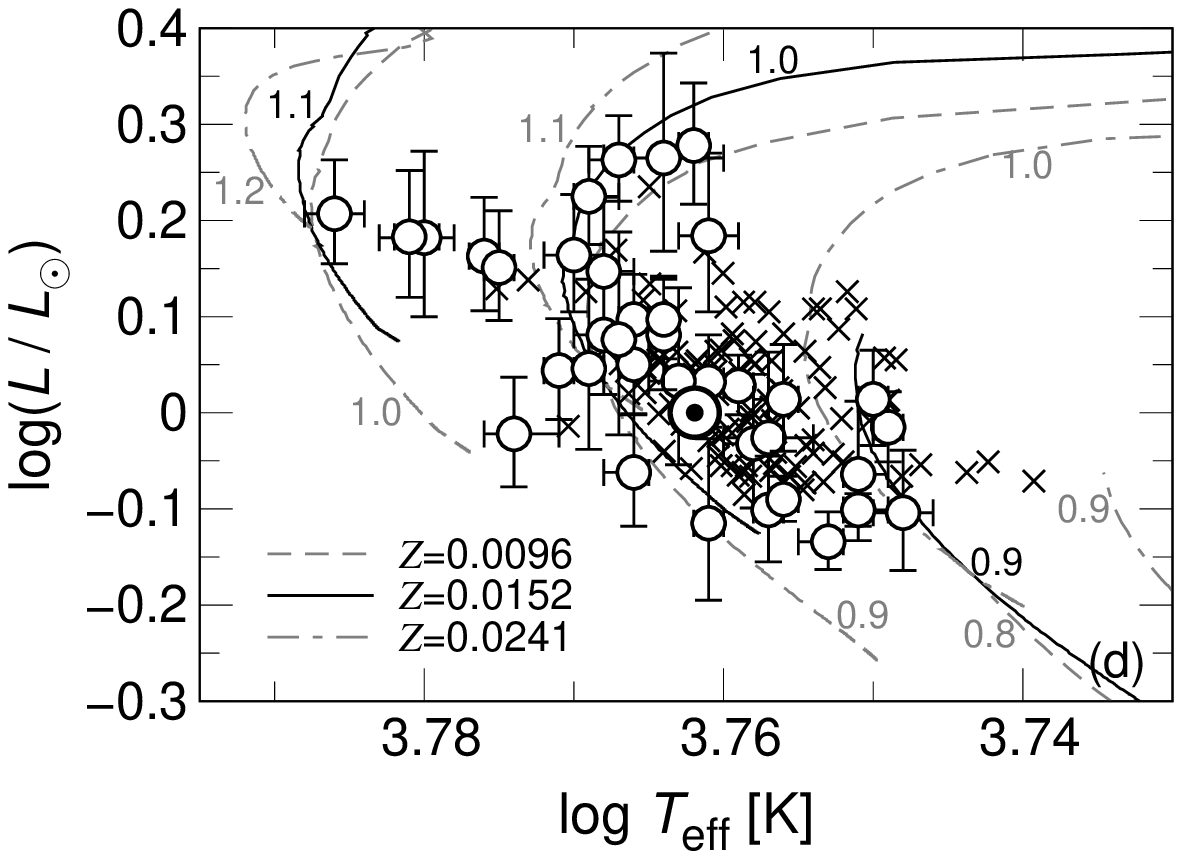}
\end{center}
\caption{(a) Scatter plot of absolute $V$ magnitude ($M_{V}$) vs. $B-V$ color ($(B-V)_{\rm{HIP}}$) 
of the target stars in this study (open circles and filled triangles). 
Open circles correspond to the stars identified as single stars in the following of this study
while filled triangles are those identified as binary stars (For the details, see Appendix \ref{subsec-apen:ana-binarity}).
All stars listed in the Hipparcos Catalogue \citep{ESA1997} are plotted with gray small circles in order to depict the HR diagram.
In addition, ordinary solar-analog stars investigated by \citet{Takeda2010} 
(cross marks) and the Sun (the circled dot point) are also plotted for reference.
\\
(b) An Extended figure of (a) without the gray small circles. 
The plot range is limited to 0.57$\leq (B-V)_{\rm{HIP}}\leq$0.78 and 3.8$\leq M_{V}\leq$5.8.
\\ 
(c) The 37 target single stars (open circles) plotted on the theoretical HR diagram 
(absolute stellar luminosity $\log (L/L_{\odot})$ vs. temperature $T_{\rm{eff}}$) 
with the theoretical evolutionary tracks (lines) deduced from PARSEC isochrones of \citet{Bressan2012}.
The theoretical tracks are drawn for 10 mass values ($M_{\rm{s}}$) ranging from 0.8 to 1.7 $M_{\odot}$ 
with a step of 0.1 $M_{\odot}$ for three different metallicities: $Z=0.0096$ ([Fe/H] $\approx -0.2$, dashed line),
$Z=0.0152$ ([Fe/H] $\approx 0.0$, solid line), $Z=0.0241$ ([Fe/H] $\approx +0.2$, dash dotted line).
The values of ordinary solar-analog stars investigated by \citet{Takeda2010} (cross marks) and 
that of the Sun (the circled dot point) are also plotted for reference.
For the details of this figure, see Appendix \ref{subsec-apen:ana-luminosity-radius} and \ref{subsec-apen:ana-age-mass}.
\\
(d) Enlarged figure of (c).
}\label{fig:BV-Mv}
\end{figure}

\noindent
\ \ \ \ \ \ \
As our target stars, we observed 49 solar-analog stars identified 
as the ROSAT X-ray All-Sky Survey sources (\cite{Voges1999}, \yearcite{Voges2000}).
The names of these 49 stars and their basic stellar parameters are listed in Table \ref{table:basic-data} \footnote{
Tables \ref{table:basic-data} $\sim$ \ref{table:period-BYDra} are in Appendix \ref{sec-apen:tables}}.
The way that we selected these 49 stars as targets is summarized in the following.
\\ \\
\ \ \ \ \ \ \
We listed the stars included both in the ROSAT X-ray All-Sky Survey Source Catalogue (\cite{Voges1999} \& \yearcite{Voges2000})
and in the Hipparcos Catalogue \citep{ESA1997}.
We here judged that a star identified as a ROSAT source is the same object as that in the Hipparcos Catalogue  
when the distance between the position (coordinate) in the Hipparcos Catalogue 
and that in the ROSAT Catalogue ($\theta_{\rm{ROSAT-HIP}}$ in Table \ref{table:basic-data}) is within 20 arcsec.
We used this criterion, considering the typical angular separation between a ROSAT source 
and its optical counterpart shown in Figure 8 of \citet{Voges1999}.
\\ \\
\ \ \ \ \ \ \
From these stars, we then listed only the stars whose X-ray photon count rate on the ROSAT detector (energy band: 0.1$\sim$2.4 keV)
is $\geq 0.1$ counts s$^{-1}$ if it is placed at a distance of 10 pc from the Earth. 
This value corresponds to $L_{\rm{X}}\gtrsim 10^{28}$erg s$^{-1}$ ($L_{\rm{X}}$: total X-ray Luminosity), 
which is much larger than the solar value in its maximum ($L_{\rm{X}}\approx 4.7\times 10^{27}$erg s$^{-1}$; \cite{Peres2000}).
After that, we finally selected the 49 stars with the following four criteria.
(1) $V\lesssim 9.5$ mag. (2) $0.58\lesssim B-V\lesssim 0.72$ mag. (3) $4.0 \lesssim M_{V}\lesssim 5.4$ mag.\footnote{
$M_{V}$(absolute magnitude) was derived from $V$ band magnitude and parallax ($\pi$) taken from the Hipparcos Catalogue \citep{ESA1997}. 
In this target selection process, we did not consider the effect of interstellar reddening ($A_{V}$) on $M_{V}$ 
since this effect is small ($\lesssim$0.1 mag; cf. $A_{V}$ in Table \ref{table:atmos}) for the stars treated here ($\lesssim$100 pc).
} 
(4) The target star is not a binary star (e.g., RS CVn-type binary, eclipsing binary, and visual binary)
on the basis of some previous studies (e.g., GCVS (General Catalogue of Variable Stars) database \citep{Samus2015}).\footnote{
We note here that our prior investigation of binarity on the basis of previous catalogues is not necessarily exhaustive one.
For example, HIP58071, one of the 49 target stars, were already reported as a spectroscopic binary star by \citet{Pourbaix2004}.
}
We used the criterion (1) since the target stars should be bright enough to conduct high dispersion spectroscopic observations
with reasonable exposure time.
We used the criteria (2), (3), and (4) 
since in this study, we aim to investigate stars whose stellar parameters are similar to the Sun 
$((B-V)_{\odot}=0.65$ mag and $(M_{V})_{\odot}=4.82$ mag; \cite{Cox1999}). 
For establishing these criteria, we referred to Section 2 of \citet{Takeda2007}. 
The positions of these finally selected 49 target stars on the HR diagram are shown in Figure \ref{fig:BV-Mv}
\footnote{
In Figure \ref{fig:BV-Mv} (c) \& (d), the values of stellar luminosity $L$ and temperature $T_{\rm{eff}}$ are plotted 
with the theoretical evolutionary tracks. 
These values are deduced in the following analyses of this paper, but we plotted these values ($L$ vs. $T_{\rm{eff}}$) here
to clearly show the place of the target stars in the HR diagram. 
For the details of this $L$ vs. $T_{\rm{eff}}$ diagram, see Appendix \ref{subsec-apen:ana-luminosity-radius} and \ref{subsec-apen:ana-age-mass} 
with Figure \ref{fig:HR}.}.
\\

\newpage
For these 49 target stars, four are also listed in the Swift X-ray Telescope Point Source Catalog \citep{Evans2014},
and they are HIP25002, HIP112364, HIP40562, and HIP77528. 
These four stars are listed in Table \ref{table:Swift-catalog} for reference.
We also checked the XMM-Newton Serendipitous Source Catalogue 3XMM-DR5 \citep{Rosen2016} 
and The Chandra Source Catalog Release 1.1 \citep{Evans2010}, 
but none of the 49 target stars are listed in these two catalogues.

\subsection{Details of observations and data reduction}\label{subsec:obs-data}
\noindent
\ \ \ \ \ \ \
Our spectroscopic observations were carried out by using 
the High Dispersion Echelle Spectrograph (HIDES: \cite{Izumiura1999}) 
attached at the 1.88-m reflector of Okayama Astrophysical Observatory (OAO).
The observations were done during the semester 2014A (2014 March 22, 23, April 22, 23, 24, and May 16, 17, 18, 19), 
2014B (2014 August 12, September 9, 10, November 5, 6, and December 29, 30), and 2015A (2015 June 3, 4).
We used a red cross disperser and a mosaic of three 2K$\times$4K CCDs.
The spectral coverage on the observation date except for 2014 December 29 and 30 was about 5600$\sim$9100\AA.
This wavelength range includes the lines sensitive to chromospheric activity 
such as H$\alpha$ 6563\AA~and Ca II infrared triplet 8498, 8542, 8662\AA.
The spectral coverage on 2014 December 29 and 30 was about 4300$\sim$7700\AA, and does not include Ca II infrared triplet 8498, 8542, 8662\AA.
We used a high-efficiency fiber-link (hereafter HIDES-Fiber) system with an image slicer (\cite{Kambe2013}) 
in order to attain spectroscopic resolution ($R=\lambda/\Delta\lambda$) of $R\sim$59,000 efficiently.
Data reduction (bias subtraction, flat fielding, aperture determination, scattered light subtraction, spectral extraction, wavelength calibration, normalization by the continuum, 
and heliocentric radial-velocity correction) was conducted using the ECHELLE package of the 
IRAF\footnote{IRAF is distributed by the National Optical Astronomy Observatories,
which is operated by the Association of Universities for Research in Astronomy, Inc., under cooperate agreement with the National Science Foundation.} software.
\\ \\
\ \ \ \ \ \ \
The observation date of each target star, the exposure time of each observation,
and the obtained signal-to-noise ratio (S/N) are shown in 
Supplementary Table 1\footnote{
Supplementary tables are available only in the online edition as ``Supporting Information"
}.
We observed the 49 solar-analog target stars identified 
as ROSAT X-ray sources as mentioned above, and 42 stars among them were observed multiple times.
\\ 
\\
\ \ \ \ \ \ \
In addition to these 49 target stars, 
we also observed six bright ($V<8$ mag) late F-type $\sim$ early G-type stars and the Moon. 
Their basic stellar parameters are in Table \ref{table:comp-stpara}.
These stars are observed as comparison stars in the discussion of stellar atmospheric parameters and chromospheric activity, as we have also done in \citet{YNotsu2015b}.
Among them, 18Sco (HIP79672), HIP100963, and HIP71813 are ``solar-twin" stars reported by the previous studies (\cite{King2005}; \cite{TakedaTajitsu2009}). 
59 Vir rotates fast and has the strong average magnetic field ($\sim$500 G), 
while 61 Vir rotates slowly and no magnetic field is detected \citep{Anderson2010}.
KIC7940546 (HIP92615) is a late F-type star  
identified as a Swift X-ray source (see Table \ref{table:Swift-catalog}). 
This star was also observed by Kepler, and atmospheric parameters of this star were investigated 
by previous researches (e.g., \cite{Molenda-Zakowic2013}). 

\section{Analyses and Results}\label{sec:stpara}
\noindent
\ \ \ \ \ \ \
For the first step of our analyses, we checked the binarity of each target star.
We checked the shape of the line profiles and time variations such as radial velocity (RV) shifts of the target stars, as we have also done in \citet{YNotsu2015a}.
The details of the analyses of binarity are described in Appendix \ref{subsec-apen:ana-binarity}.
As a result, we regard 12 target stars as binary stars. 
These 12 binary stars are shown in the 12th column of Table \ref{table:basic-data}. 
We treat the remaining 37 target stars as ``single stars" in this paper, 
since they do not show any evidence of binarity within the limits of our analyses.
In the following, we conduct the detailed analyses only for these 37 ``single" stars.
\\
\\
\ \ \ \ \ \ \
We then estimated various stellar parameters of the 37 ``single" target stars on the basis of our spectroscopic data \footnote{
Spectroscopic data used here for estimating stellar parameters are available in Supplementary Data 1 
(For the details, see Appendix \ref{subsec-apen:ana-binarity}).
Supplementary data are available only on the online edition as ``Supporting Information".}.
Our analysis methods used here are basically on the basis of the previous studies such as 
\authorcite{YNotsu2015a} (\yearcite{SNotsu2013}, \yearcite{YNotsu2015a} \& \yearcite{YNotsu2015b}), 
\citet{Honda2015}, \citet{Nogami2014}, and \authorcite{Takeda2007} (\yearcite{Takeda2007} \& \yearcite{Takeda2010}).
The details of the analyses and results are described in Appendix \ref{subsec-apen:ana-atmos-para} $\sim$ \ref{subsec-apen:para-comparison} of this paper.
We estimate stellar atmospheric parameters ($T_{\rm{eff}}$, $\log g$ and [Fe/H]) using Fe I/II lines in Appendix \ref{subsec-apen:ana-atmos-para}, 
and projected rotational velocity ($v\sin i$) in Appendix \ref{subsec-apen:ana-vsini}.
We show measurement results of the intensity of Ca II 8542 and H$\alpha$ lines in Appendix \ref{subsec-apen:ana-CaHa} and \ref{subsec-apen:ExcessFlux},
and describe the analysis of Li abundance of the target stars in Appendix \ref{subsec-apen:ana-Li}.
On the basis of stellar parameters in Section \ref{subsec-apen:ana-atmos-para}, 
we also estimate the stellar luminosity and radius with Hipparcos parallax in Appendix \ref{subsec-apen:ana-luminosity-radius}, 
the mass and age from the stellar evolutionary tracks in Appendix \ref{subsec-apen:ana-age-mass}, 
and X-ray luminosity from ROSAT data in Appendix \ref{subsec-apen:ana-Xray}. 
In Appendix \ref{subsec-apen:para-comparison}, we performed some analyses in order to check 
whether these spectroscopically derived values are good sources with which to discuss the actual properties of stars.
\\
\\
\ \ \ \ \ \ \
These estimated parameters are listed in Tables \ref{table:atmos} $\sim$ \ref{table:comp-stpara}. 
Examples of spectra around photospheric lines, including Fe I 6212, 6215, 6216, 6219 are shown 
in Figure \ref{fig:specsg-Fe6212} in Appendix \ref{subsec-apen:ana-binarity}.
The observed spectra of these 37 stars around Ca II 8542, H$\alpha$ 6563, and Li I 6708 are shown in 
Figures \ref{fig:specsg-Ca8542}, \ref{fig:specsg-Ha}, and \ref{fig:specsg-Li6708}, respectively.

\section{Discussion}\label{sec:discussion}
\subsection{Estimated stellar parameters}\label{subsec:dis-atmos}
\begin{figure}[htbp]
 \begin{center}
  \FigureFile(82mm,82mm){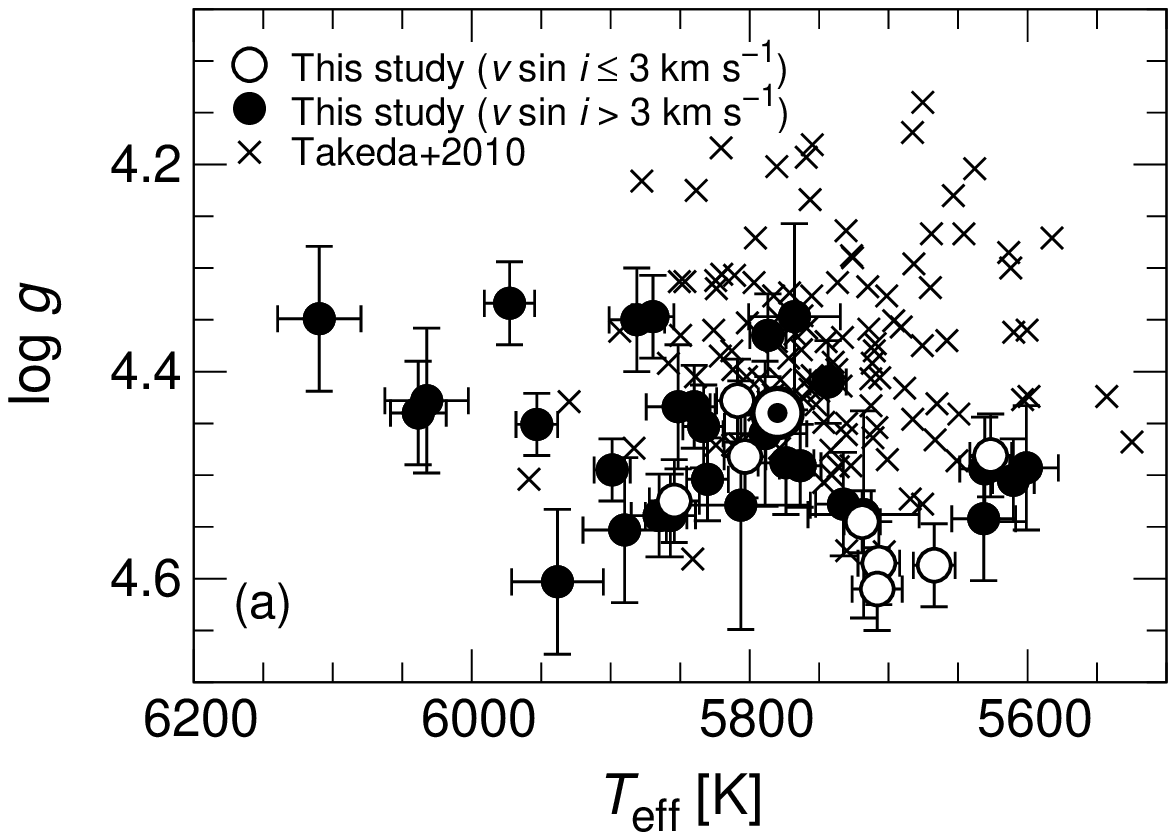}
  \FigureFile(82mm,82mm){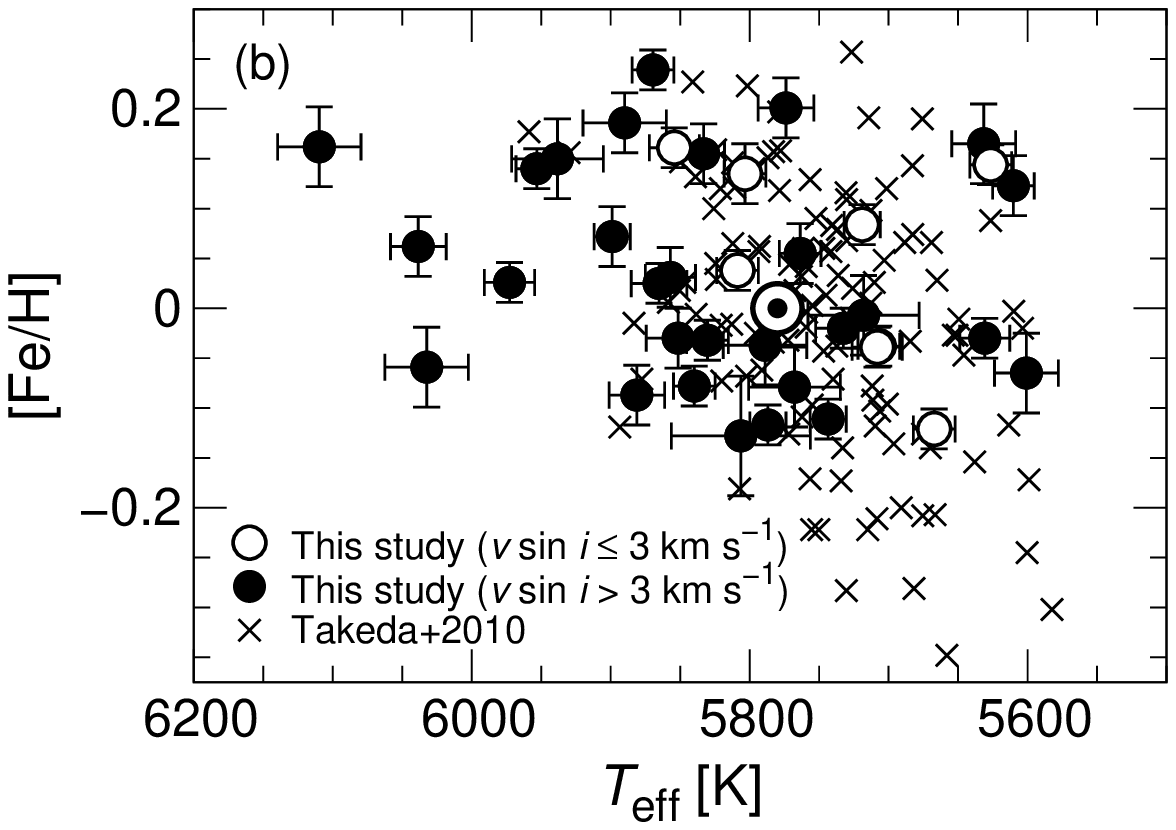}
\end{center}
\caption{
Temperature ($T_{\rm{eff}}$), surface gravity ($\log g$), and metallicity ([Fe/H]) 
of the 37 single target stars in this study (open and filled circles). 
Target stars with $v\sin i\leq$3 km s$^{-1}$ and with $v\sin i>$3 km s$^{-1}$are 
shown by using open circles and filled circles, respectively.
The values of ordinary solar-analog stars investigated by \citet{Takeda2010} 
(cross marks) and the Sun (the circled dot point) are also plotted for reference.
}\label{fig:T-logg-FeH}
\end{figure}

\ \ \ \ \ \ \
In Section \ref{sec:stpara} and Appendix \ref{sec-apen:Detail-ana}, 
we estimated the stellar parameters such as $T_{\rm{eff}}$, $\log g$, and [Fe/H] of the 37 target stars,
which are considered as single stars in Appendix \ref{subsec-apen:ana-binarity}.
The measured values of $T_{\rm{eff}}$, $\log g$, and [Fe/H]
are in the range of $5,600\sim6,200$ K, $4.2\sim4.7$, and $-0.2\sim0.3$, respectively (Figure \ref{fig:T-logg-FeH}).
This means that the stellar parameters of these target stars are in the range of ordinary solar-analog stars \footnote{
HIP3342 is identified as (1) pre-main sequence Star Candidate in SIMBAD Astronomical Database (http://simbad.u-strasbg.fr/simbad/) 
(See also Table \ref{table:basic-data} of this paper), but our spectroscopic data show no such features. 
}, and no clear ``metal-rich" or ``metal-poor" stars are included in our target stars.
Ages of the target stars are also estimated by comparing the position 
on the HR diagram (cf. Figure \ref{fig:HR} (a) \& (b) in Appendix \ref{subsec-apen:ana-luminosity-radius}).
Figure \ref{fig:HR} (c) shows ages of them are in the range of $1\sim10$ Gyr,
and we can say that the target stars are not necessarily young, though the error values of ages are large.

\noindent
\ \ \ \ \ \ \
Close binary stars such as RS CVn-type stars have been widely known as magnetically active stars
because of the high rotation rate thanks to the tidal interaction between the primary and companion stars (e.g., \cite{Walter1981}).
It is very important whether our target stars are such close binary stars or not,
especially for considering whether single stars like the Sun can really show strong magnetic activities.
The 37 target stars show no double-lined profile or radial velocity shifts, as explained in Appendix \ref{subsec-apen:ana-binarity}.
It is then highly possible that almost all the target stars are not close binary stars, though more detailed observations are needed.

\subsection{Rotation velocity}\label{subsec:dis-rot}
\begin{figure}[htbp]
 \begin{center}
   \FigureFile(82mm,82mm){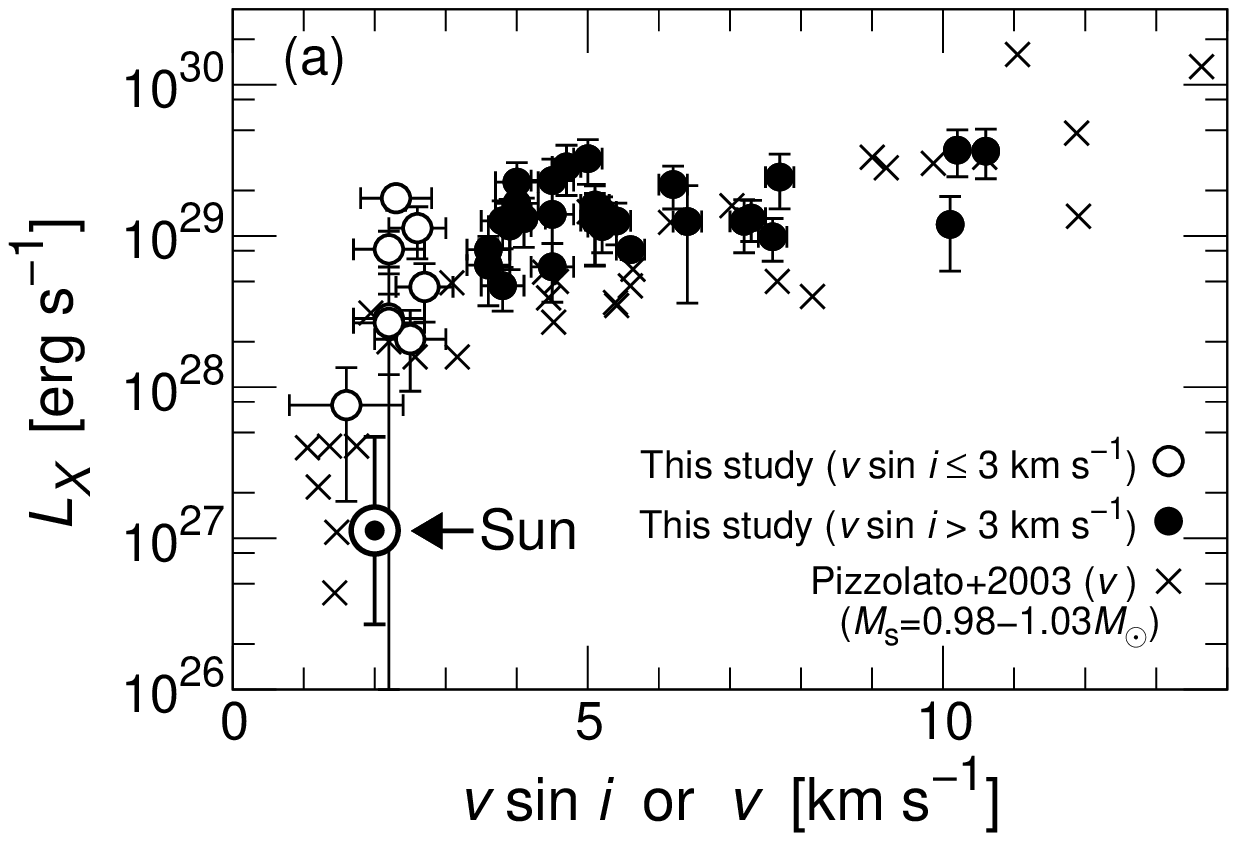}
   \FigureFile(82mm,82mm){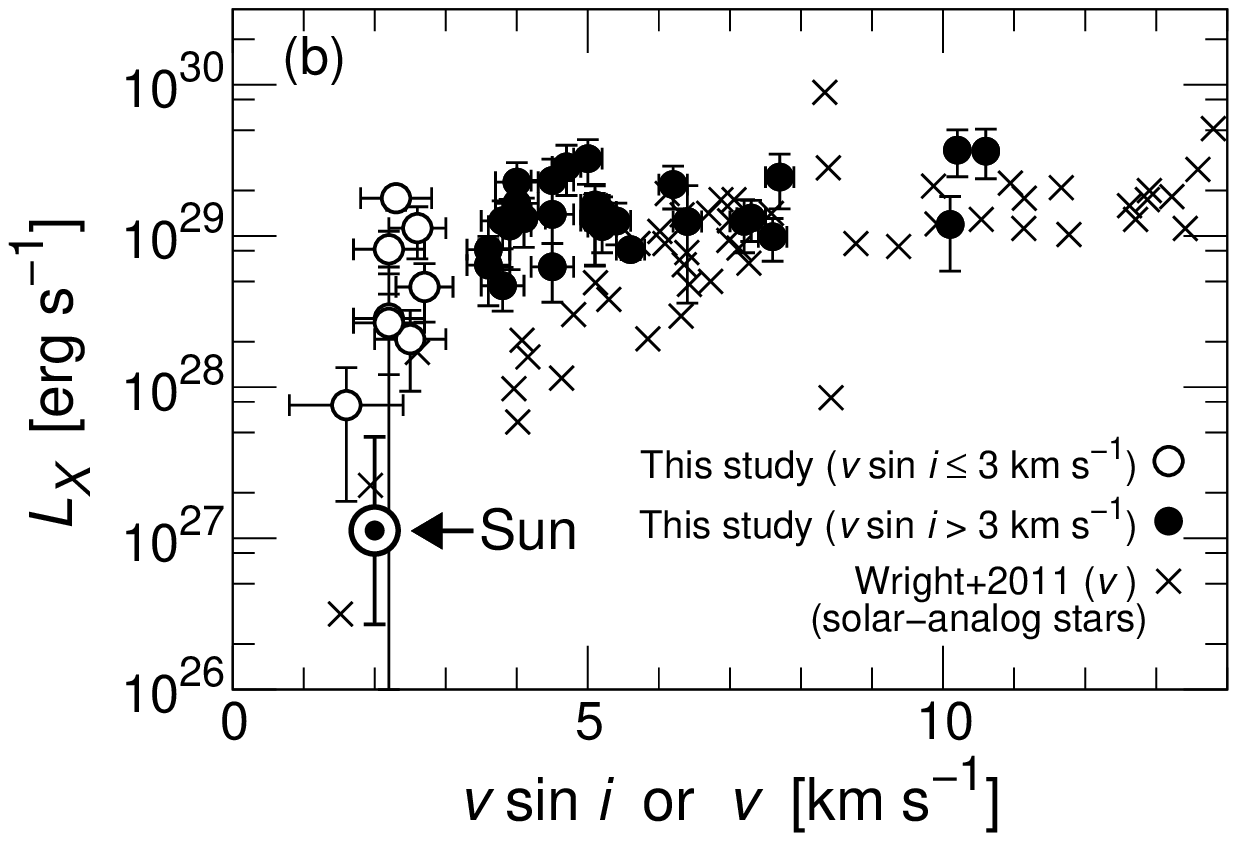}   
 \end{center}
\caption{
(a)
X-ray luminosity ($L_{X}$) on the basis of ROSAT data as a function of projected rotation velocity ($v\sin i$) of the 37 target stars (open and filled circles). 
Target stars with $v\sin i\leq$3 km s$^{-1}$ and with $v\sin i>$3 km s$^{-1}$are shown by using open circles and filled circles, respectively.
Cross marks correspond to the $L_{X}$ vs. $v$ of the ordinary solar-type stars 
whose masses are $0.98\sim1.03~M_{\odot}$, taken from \citet{Pizzolato2003}.
In \citet{Pizzolato2003}, the values of $L_{X}$ and rotation period ($P$) are listed, so we here assume $v\approx 2\pi R_{\odot}/P$ 
in order to plot the data points in this figure.
The Sun's X-ray luminosity in the ROSAT band estimated by \citet{Peres2000} 
($2.7 \times 10^{26}$ erg s$^{-1}$ and $4.7 \times 10^{27}$ erg s$^{-1}$ at solar 11-year minimum and maximum, respectively) is also plotted for reference (the circled dot point).
\\
(b)
Circles and the circled dot point are the same as (a).
Cross marks correspond to the $L_{X}$ vs. $v$ of the solar-analog stars 
with 5600$\leq T_{\rm{eff}}\leq$6000 K and $\log g\geq$4.0, taken from \citet{Wright2011}. 
Here the values of $v$ are estimated from the rotation period ($P$) 
and stellar radius ($R_{\rm{s}}$) listed in \citet{Wright2011}: $v=2\pi R_{\rm{s}}/P$ .
}\label{fig:vsini-Lx}
\end{figure}

\ \ \ \ \ \ \
We now have soft X-ray luminosity ($L_{X}$) on the basis of ROSAT data, and stellar projected rotation velocity ($v\sin i$) of the 37 target stars.
In Figure \ref{fig:vsini-Lx}, we plot $L_{X}$ as a function of $v\sin i$, 
and  we can confirm that all the target stars have higher X-ray luminosity compared with that of the Sun.
Previous studies (e.g., \cite{Pallavicini1981}; \cite{Pizzolato2003}; \cite{Wright2011}) reported a positive correlation between $L_{X}$ and stellar rotational velocity,
and our target stars show a similar correlation, though the value of our target stars in the horizontal axis 
of Figure \ref{fig:vsini-Lx} is not exact velocity ($v$) but ``projected" rotational velocity ($v\sin i$).
In this figure, most of our target stars tend to rapidly rotate compared with the Sun, 
and this suggests that such stars can be a bit younger than the Sun, 
assuming a well-known correlation between stellar age and rotational velocity 
(cf. \cite{Skumanich1972}; \cite{Soderblom2010}; \cite{Meibom2015}).
This result is consistent with well-known results 
that young rapidly rotating stars have high magnetic activity (e.g., \cite{Skumanich1972}).
Some target stars, however, have projected rotational velocity as slow as the Sun ($v\sin i\sim$2 km s$^{-1}$) 
(See open circle data points in Figure \ref{fig:vsini-Lx}).
This suggests that such stars, having higher X-ray luminosity compared with that of the Sun, can be as old as the Sun, 
though the velocity in Figure \ref{fig:vsini-Lx} is only ``projected" rotational velocity ($v\sin i$).

\begin{figure}[htbp]
 \begin{center}
   \FigureFile(82mm,82mm){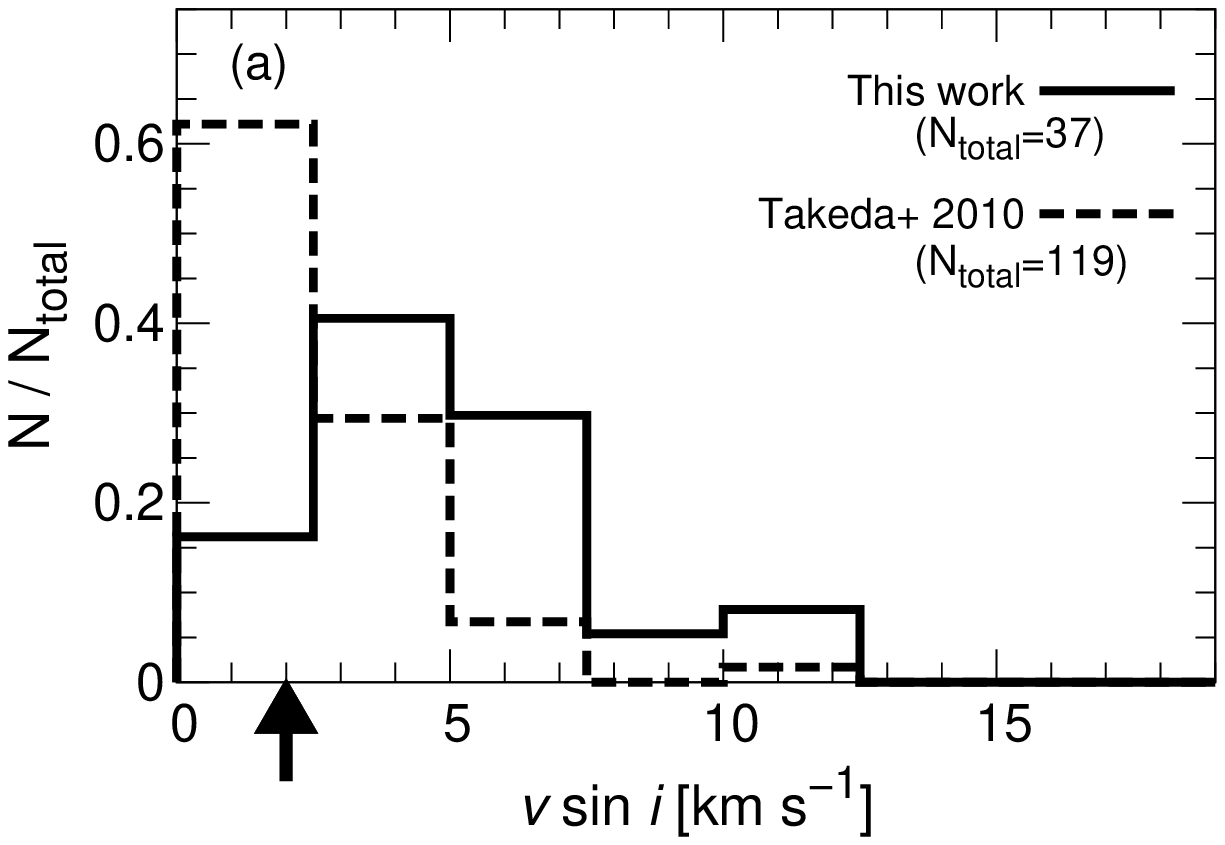}
  \FigureFile(82mm,82mm){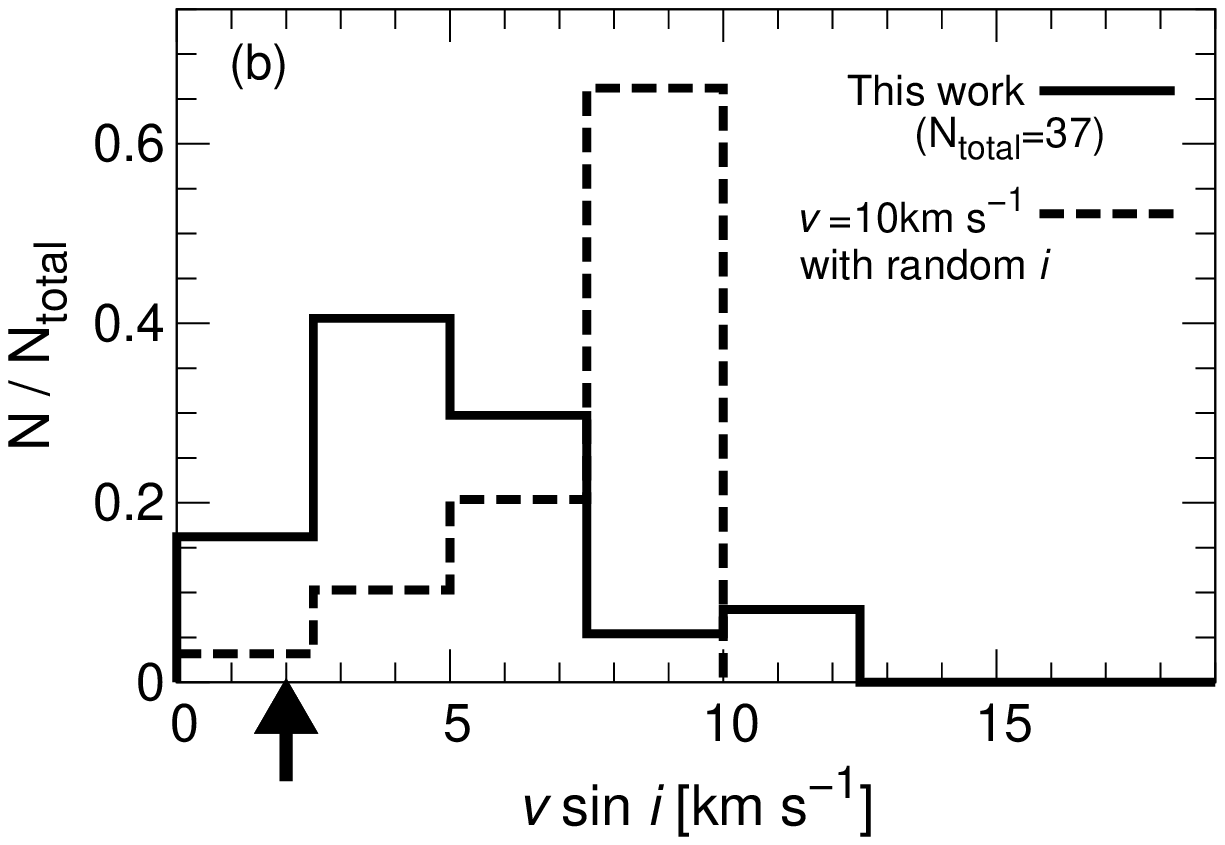}
  \FigureFile(82mm,82mm){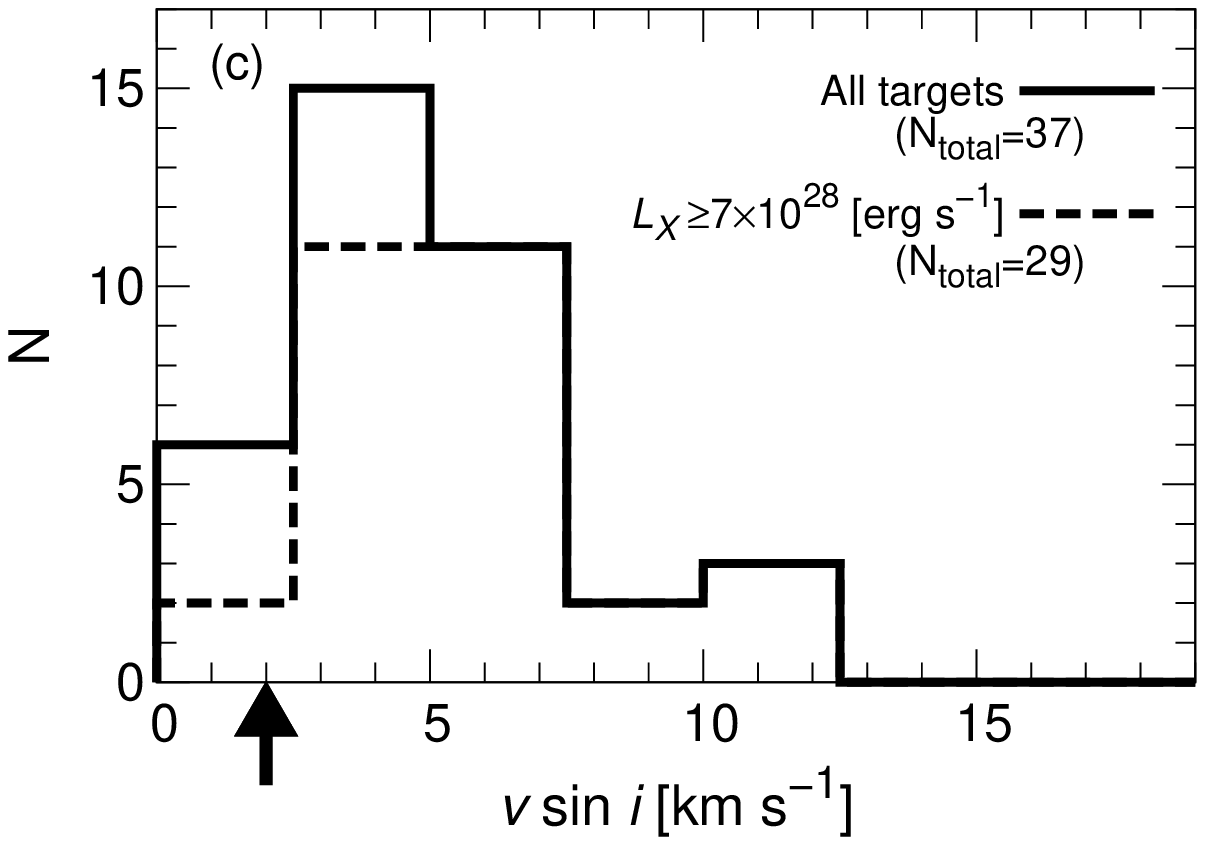}
 \end{center}
\caption{
Histograms showing the distribution of $v \sin i$. 
The solar $v\sin i$ value ($\sim$2 km s$^{-1}$) is roughly indicated by a black upward arrow.\\
(a) The filled line shows the data of all the single target stars in this study, and 
the dashed line shows that of ordinary solar-analog stars reported in \citet{Takeda2010}.
\\
(b) The filled line shows the data of all the single target stars in this study, which is the same as (a). 
The dashed line corresponds to the estimated distribution 
if we assume that all the stars have $v\sin i=10$~km s$^{-1}$ and random distribution of inclination angle $i$.
\\
(c) The filled line shows  the data of all the single target stars in this study, which is the same as (a). 
The dashed line corresponds to the single target stars with $L_{X}\geq 7\times 10^{28}$erg (cf. Figure \ref{fig:vsinir0} (a)).
}\label{fig:vsini-bunpu}
\end{figure}

\noindent
\ \ \ \ \ \ \
The distribution histogram of $v\sin i$ for the target stars are shown in Figure \ref{fig:vsini-bunpu} (a) (solid line).
The data of 119 ordinary solar-analog stars reported in \citet{Takeda2010} are also plotted in this figure for reference (dashed line).
We can regard the data of \citet{Takeda2010} as a random sample of ordinary solar-analog stars, 
considering their target selection method as explained in \citet{Takeda2007}. 
Comparing these two data in Figure \ref{fig:vsini-bunpu} (a), 
we can say $v\sin i$ of the observed target stars tends to be higher than the sample of ordinary solar-analog stars.
In particular, three of the 37 target stars have extremely high $v\sin i$ values ($v\sin i>10$ km s$^{-1}$), and 13 stars have $5< v\sin i\leq10$ km s$^{-1}$.
In contrast, 21 stars of the 37 target stars have low $v\sin i$ values ($v\sin i<5$ km s$^{-1}$), 
and they include the stars as slow as the Sun ($v\sin i\sim2$ km s$^{-1}$).
The velocity discussed here, however, are only the ``projected" rotational velocity ($v\sin i$), 
and the effect of inclination angle ($i$) should be taken into consideration
since rapidly rotating stars can have small $v\sin i$ if they have small inclination angle $i$.
Then in Figure \ref{fig:vsini-bunpu} (b), we plot the estimated distribution 
if we assume that all the stars have $v\sin i=10$~km s$^{-1}$ and random distribution of inclination angle $i$ (dashed line).
The value of $v\sin i=10$~km s$^{-1}$ is used here, 
since young stars whose age is a few hundred Myr have $v\sin i\sim10$~km s$^{-1}$ 
on the basis of the well-known age-rotation relationship (\cite{Skumanich1972}; \cite{Meibom2015}).
Comparing the filled and dashed lines in this figure, 
we can say that the number of stars with $v\sin i<$ 5 km s$^{-1}$ cannot be explained 
if we assume that all the target stars rapidly rotate ($v\sin i\gtrsim10$~km s$^{-1}$) and they have random distribution of $i$.
This suggests that the existence of stars having low $v\sin i$ values ($v\sin i<5$ km s$^{-1}$)
seems to be real, and slowly rotating stars are included in our target stars with strong soft X-ray emissions.
\\ \\
\ \ \ \ \ \ \
In addition, Figure \ref{fig:vsini-bunpu} (c) is the distribution histogram of $v\sin i$ 
for the target stars having $L_{X}\geq 7\times 10^{28}$ erg.
This threshold of $L_{X}$ is used to select extremely high activity level compared with the Sun (See the dotted line in Figure \ref{fig:vsinir0} (a)). 
In Figure \ref{fig:vsini-bunpu} (c), the number of stars with solar-level $v\sin i$ values ($v\sin i\sim$2 km s$^{-1}$) decrease, 
but large number of the stars with relatively low $v\sin i$ values ($v\sin i<5$ km s$^{-1}$) still exist 
and the main suggestions mentioned for (a) \& (b) in the above do not change.

\subsection{Chromospheric activities}\label{subsec:dis-CaHa}
\begin{figure}[htbp]
 \begin{center}
   \FigureFile(80mm,80mm){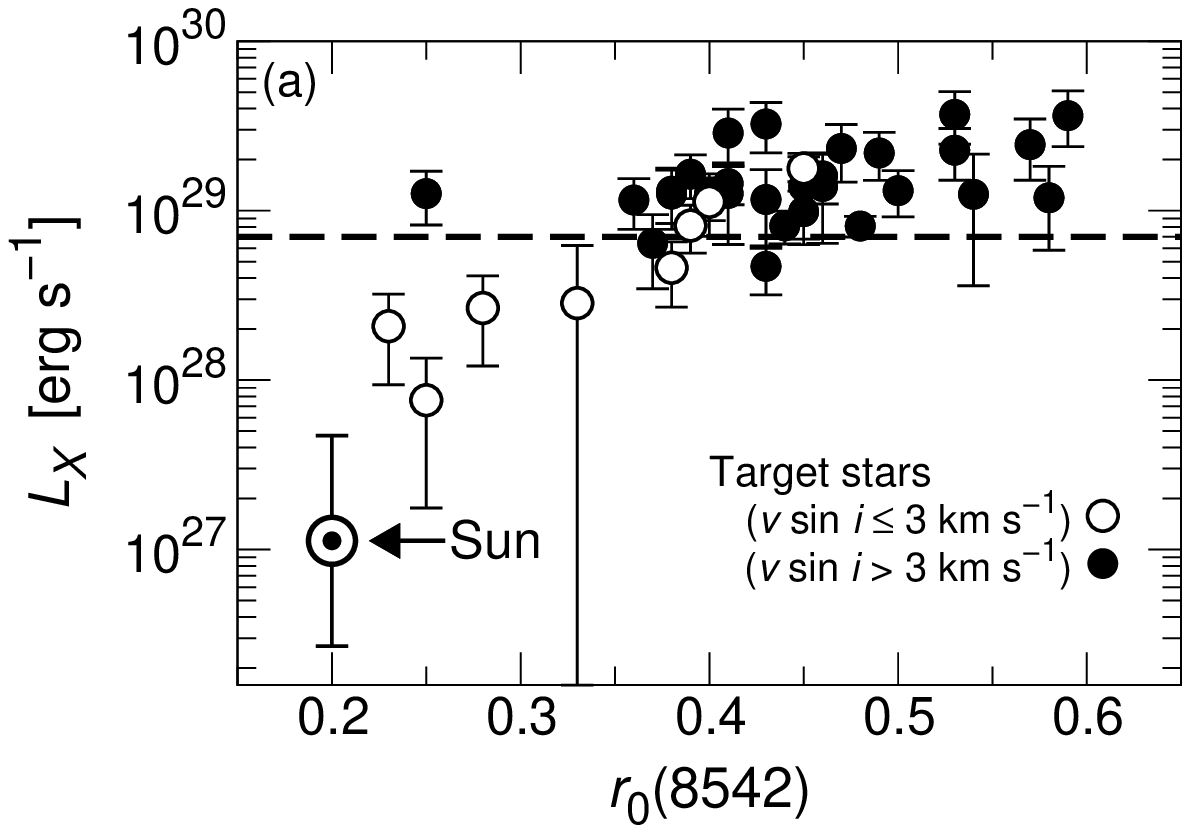}
   \FigureFile(80mm,80mm){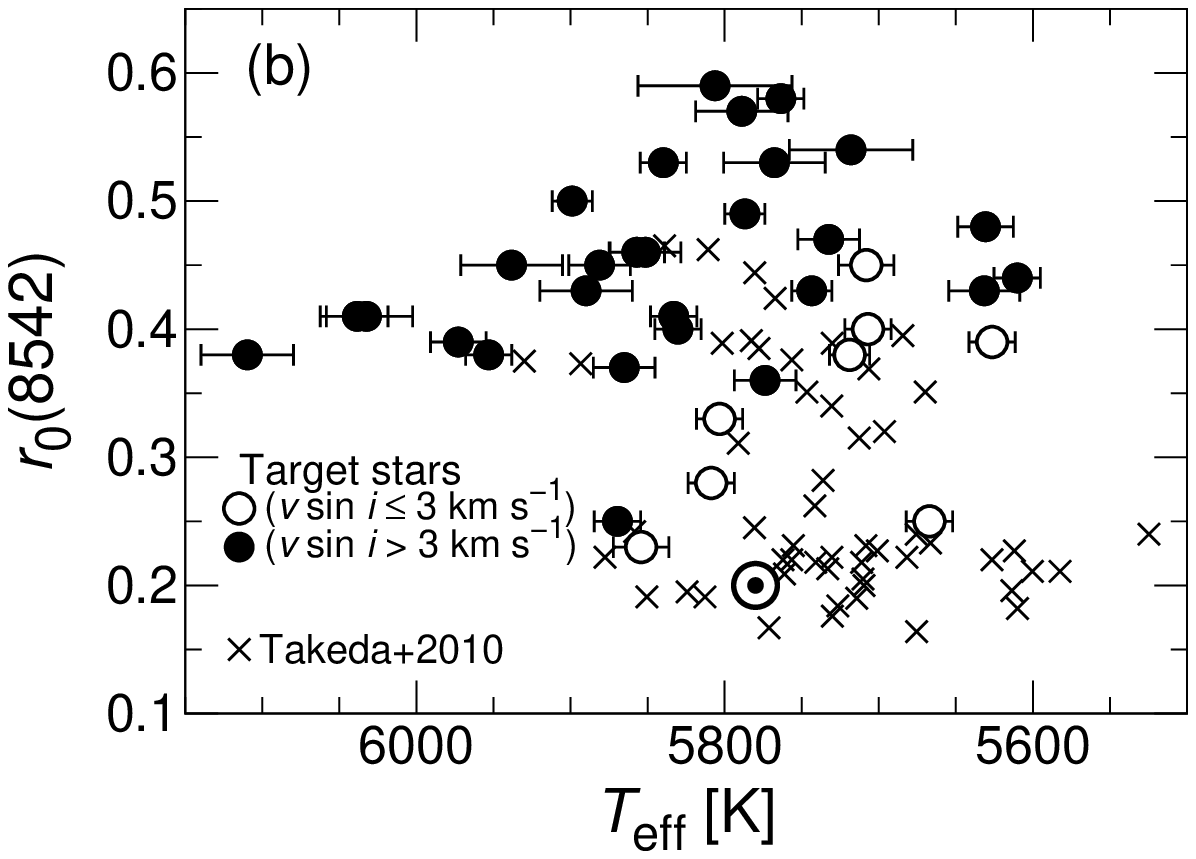}     
   \FigureFile(80mm,80mm){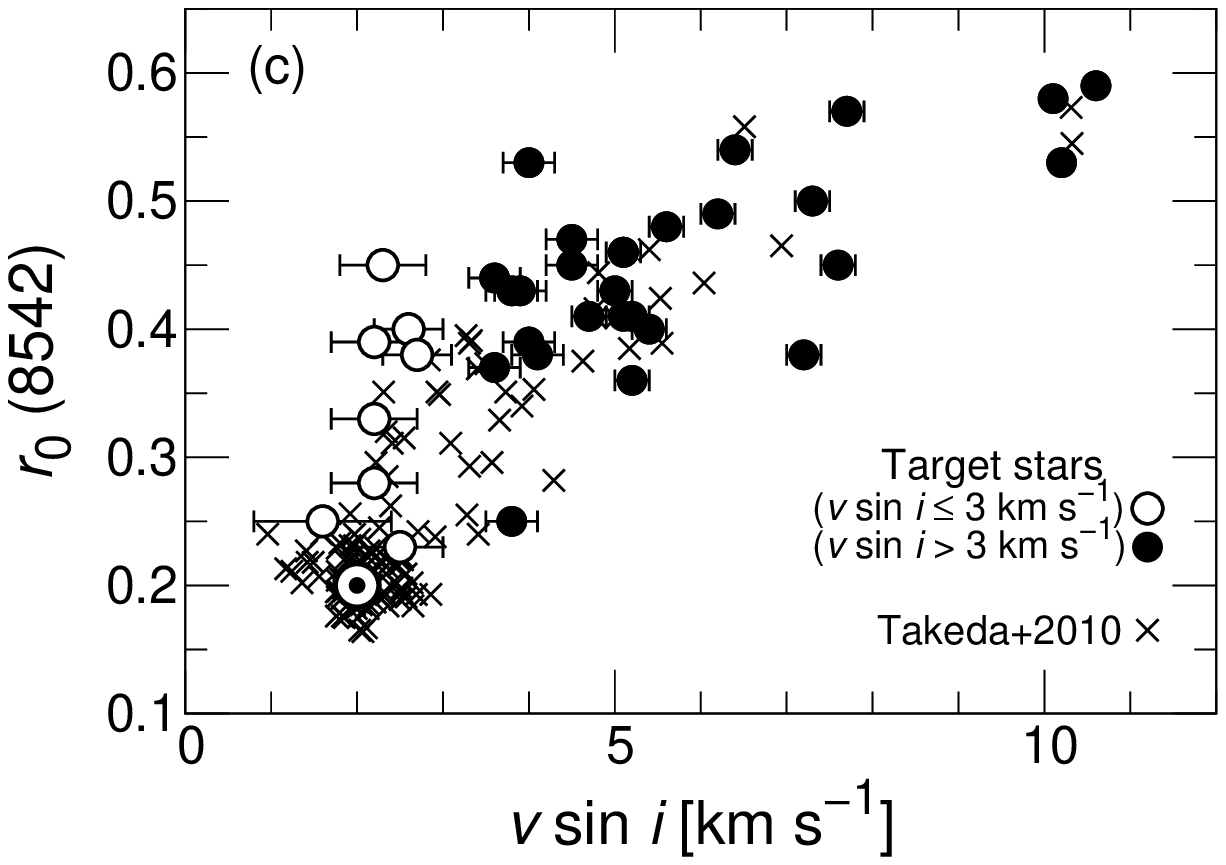}
 \end{center}
\caption{
(a) The X-ray luminosity ($L_{X}$) as a function of $r_{0}$(8542) of the 37 target stars (open and filled circles). 
Target stars with $v\sin i\leq$3 km s$^{-1}$ and with $v\sin i>$3 km s$^{-1}$are shown by using open circles and filled circles, respectively. 
The Sun's $r_{0}$(8542) ($r_{0}$(8542)$\sim$0.2) and $L_{X}$ values are also depicted using a circled dot point for reference.
Here we use the Sun's $L_{X}$ in the ROSAT band estimated by \citet{Peres2000} 
($2.7 \times 10^{26}$ erg s$^{-1}$ and $4.7 \times 10^{27}$ erg s$^{-1}$ at solar 11-year minimum and maximum, respectively).
The threshold of $L_{X}$ used in Figure \ref{fig:vsini-bunpu} (b) ($L_{X}\geq 7\times 10^{28}$erg) is shown with the dashed line for reference.
\\
(b) $r_{0}$(8542) as a function of $T_{\rm{eff}}$ of the stars. 
In addition to the results of the 37 target stars (open and filled circles), 
the data of ordinary solar-analog stars in \citet{Takeda2010} (cross marks) are plotted.
The solar value is also plotted using a circled dot point for reference.
\\
(c) $r_{0}$(8542) as a function of $v\sin i$.
The classification of the symbols is the same as in (b).

}\label{fig:vsinir0}
\end{figure}

\ \ \ \ \ \ \
In Figure \ref{fig:vsinir0} (a), we plot $L_{X}$ as a function of $r_{0}$(8542), 
and in \ref{fig:vsinir0} (b), we plot $r_{0}$(8542) as a function of $T_{\rm{eff}}$ of the target stars.
In Figure \ref{fig:vsinir0} (b), the data of ordinary solar-analog stars reported in \citet{Takeda2010} are also plotted for reference.
The variability range of $r_{0}$(8542) between the solar maximum and solar minimum 
is only 0.19$\sim$0.20 (\cite{Livingston2007}).
These two figures show that all the target stars of this study are more active compared with the Sun and ordinary solar-analog stars 
from the viewpoint of high $r_{0}$(8542) value.
This indicates that all the target stars, which are selected as ROSAT X-ray sources, have large starspots 
since $r_{0}$(8542) has a good correlation with starspot coverage (cf. \cite{YNotsu2015b}).
The ROSAT All sky survey was conducted 
more than twenty years ago (1990$\sim$1991: \cite{Voges1999} \& \yearcite{Voges2000}), 
but the target stars selected here still have high activity levels  
and large starspots even in the period of this observation over 2014$\sim$2015.
In  Figure \ref{fig:vsinir0} (a), we can also see a rough positive correlation between $L_{X}$ and $r_{0}$(8542),
which is consistent with the well-known correlation between soft X-ray luminosity and stellar magnetic flux (cf. \cite{Pevtsov2003})
since $r_{0}$(8542) can be a good indicator of stellar magnetic flux (cf. See Figure 4 of \cite{YNotsu2015b}).
\\

\noindent
\ \ \ \ \ \ \
In Figure \ref{fig:vsinir0} (c), we plot $r_{0}$(8542) as a function of $v\sin i$ of the target stars.
The data of ordinary solar-analog stars reported in \citet{Takeda2010} are plotted for reference also in this figure \footnote{
As mentioned in the footnote 1 of \citet{YNotsu2015b}, we consider that the variation of $r_{0}$(8542) 
due to the blurring effect caused by an increase of $v\sin i$ does not have any essential effects on the interpretation of this figure.
}. There is a rough positive correlation between $r_{0}$(8542) ($\approx$ starspot coverage) and $v\sin i$, 
and more rapidly rotating stars tend to have higher activity level.
This is consistent with many previous observational results of ordinary main-sequence stars (e.g., \cite{Skumanich1972}; \cite{Noyes1984}; \cite{Takeda2010}).
In contrast, some target stars with relatively small $v\sin i$ values ($v\sin i\lesssim 3$ km s$^{-1}$) in this figure
have relatively high chromospheric activity compared with the data of ordinary solar-analog stars.
Considering the discussions in Section \ref{subsec:dis-rot} with Figure \ref{fig:vsini-bunpu} (b), 
we may say that the effect of inclination angle ($i$) does not change the overall results.
This could suggest that even slowly rotating target stars have high activity level and large starspots.
\\ \\
\ \ \ \ \ \ \
In Appendix \ref{subsec-apen:ExcessFlux}, we also investigated the emission flux of Ca II 8542 and H$\alpha$ 6563 lines for reference, 
and confirmed that we can get basically the same conclusions if we use the emission flux values in stead of $r_{0}$(8542) index in this section.

\subsection{Li abundances}\label{subsec:dis-Li}
\subsubsection{Temperature and Li abundance}\label{subsubsec:dis-Teff-Li}
\begin{figure}[htbp]
 \begin{center}
\FigureFile(82mm,82mm){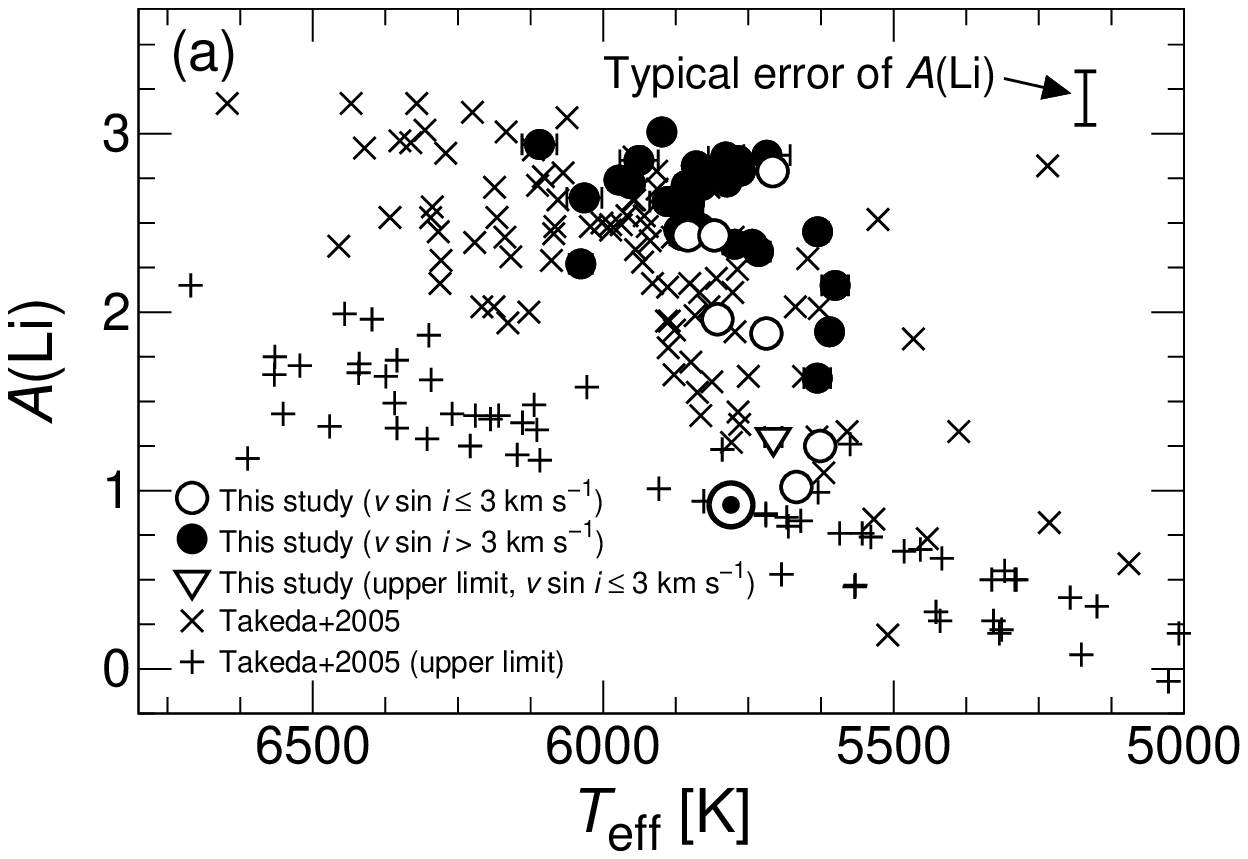}
\FigureFile(82mm,82mm){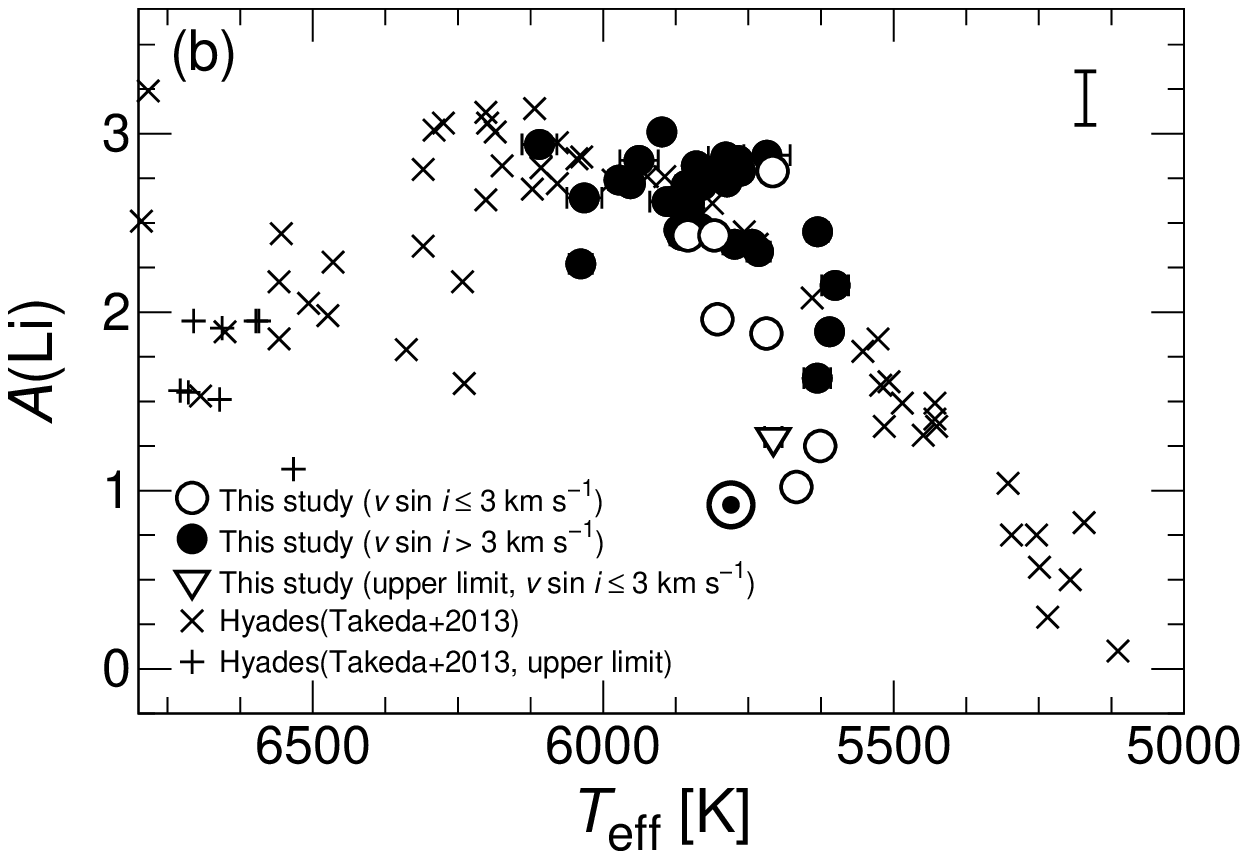}  
 \end{center}
\caption{
(a)
$A$(Li) vs. $T_{\rm{eff}}$ with ordinary F-, G-, and K-type dwarfs/subgiants.
Circles and triangles indicate the Lithium abundance [$A$(Li)] of the target stars in this study. 
Target stars with $v\sin i\leq$3 km s$^{-1}$ and with $v\sin i>$3 km s$^{-1}$are shown by using open circles/triangles and filled circles, respectively,
and open triangles correspond to the upper limit values of $A$(Li) for the unmeasurable cases. 
We plotted the data of the F-, G-, and K-type main-sequence stars reported by \citet{Takeda2005} with cross marks (x marks and plus marks), 
and the plus marks correspond to the upper limit values of $A$(Li) for the unmeasurable cases. 
The solar value is also plotted with a circled dot point for reference.
Typical error value of $A$(Li) ($\sim$0.15 dex) mentioned in Appendix \ref{subsec-apen:ana-Li} is shown 
with the error bar in the upper right of this figure, and this error bar is also shown in the following figures.
\\
(b)
$A$(Li) vs. $T_{\rm{eff}}$ with the stars in the Hyades cluster. The age of Hyades is $6.25\times 10^{8}$ yr (e.g., \cite{Perryman1998}).
The data points of circles, triangles, and the circled dot point are the same as (a).
In this figure, we plotted the data of the stars in Hyades cluster reported by \citet{Takeda2013} with cross marks (x marks and plus marks),  
and the plus marks correspond to the upper limit values of $A$(Li) for the unmeasurable cases.
}\label{fig:T-ALi-Takeda20052013}
\end{figure}

\ \ \ \ \ \ \
Figure \ref{fig:T-ALi-Takeda20052013} (a) shows the Li abundances $A$(Li) as a function of $T_{\rm{eff}}$ of the target stars 
with ordinary F-, G-, and K-type stars (\cite{TakedaKawanomoto2005}).
The Li depletion is seen in the stars whose temperature is lower than that of the Sun ($T_{\rm{eff}}\lesssim 5500$ K), 
while the stars with $T_{\rm{eff}}\gtrsim 6000$ K show no Li depletion. 
This is because as the star becomes cooler, the convection zone in the stellar atmosphere evolves and the Li is transported to a deeper hotter zone, 
where Li is easily destroyed (p, $\alpha$ reactions: $^{7}$Li, $T\geq2.5\times10^{6}$ K; $^{6}$Li, $T\geq2.0\times10^{6}$ K). 
The depletion of Li in the stellar surface caused by convective mixing increases with a lapse of time, 
and we can say stars with high $A$(Li) are young stars.
A large diversity (by more than 2 dex) of Li abundance is seen in stars similar to the Sun ($T_{\rm{eff}}\approx5500\sim6000$ K),
and among them, the solar Li abundance is quite low (cf. \cite{TakedaKawanomoto2005}).  
\\

\noindent
\ \ \ \ \ \ \
In Figure \ref{fig:T-ALi-Takeda20052013} (a), many of the target stars tend to show higher Li abundances compared with the ordinary solar-type stars.
Some of them show high Li compared with the stars in the Hyades cluster (Figure \ref{fig:T-ALi-Takeda20052013} (b)), 
and such stars are suggested to be younger than the Hyades cluster
(The age of the Hyades cluster is estimated to be $6.25\times 10^{8}$ yr (e.g., \cite{Perryman1998})). 
It is reasonable that such young stars have high activity levels. 
However, more than ten target stars do not show higher $A$(Li) values compared with the Hyades cluster,
and some of them have quite low $A$(Li) values as low as the Sun ($A$(Li)$<1.5$).

\subsubsection{Rotation and Li abundance}\label{subsubsec:dis-vsini-Li}
\begin{figure}[htbp]
 \begin{center}
\FigureFile(82mm,82mm){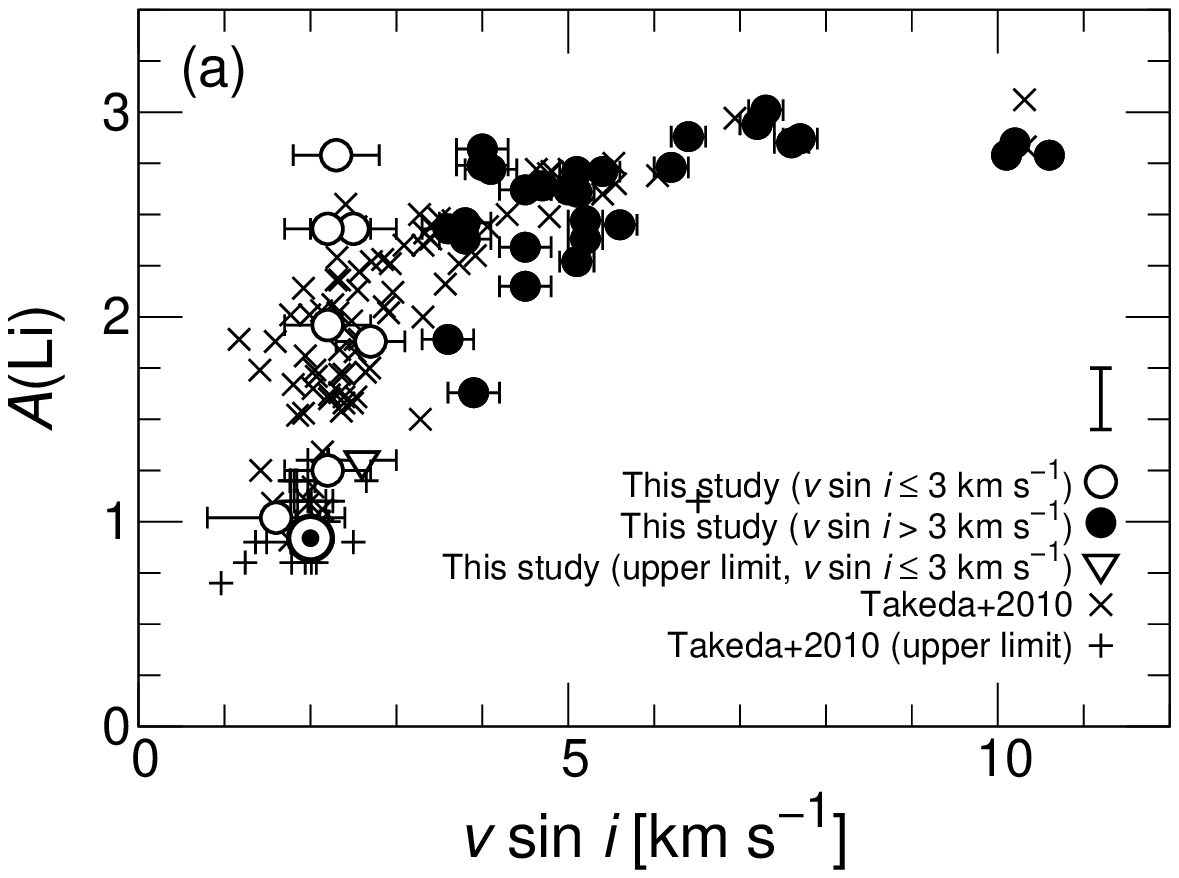}
\FigureFile(82mm,82mm){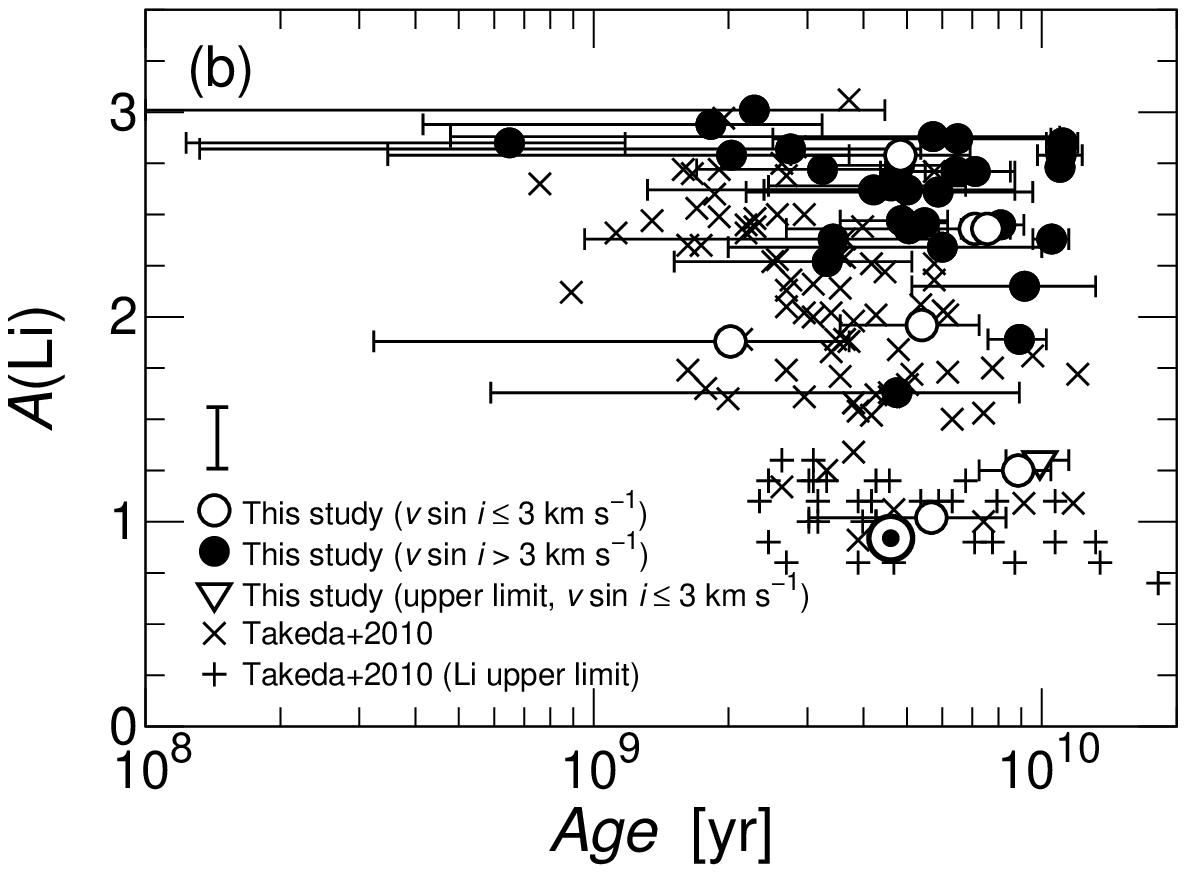} 
 \end{center}
\caption{ 
(a) $A$(Li) vs. $v\sin i$. 
(b) $A$(Li) vs. $Age$. 
\\
Circles and triangles indicate the Li abundance of the target stars in this study. 
Target stars with $v\sin i\leq$3 km s$^{-1}$ and with $v\sin i>$3 km s$^{-1}$are shown by using open circles/triangles and filled circles, respectively,
and open triangles correspond to the upper limit values of $A$(Li) for the unmeasurable cases. 
We plotted the data of the ordinary solar-analog stars observed by \citet{Takeda2010} with cross marks (x marks and plus marks),  
and the plus marks correspond to the upper limit values of $A$(Li) for the unmeasurable cases.
The solar value is also plotted with a circled dot point for reference. 
}\label{fig:ALi-vsini}
\end{figure}

\ \ \ \ \ \ \
In general, stellar rotation is considered to have a good correlation with stellar age and activity.
Solar-type stars have a decrease in rotation, activity, and Li abundance with age \citep{Skumanich1972}.
Li abundance can give important clues about the age of the star \citep{Soderblom2010}. 
Figure \ref{fig:ALi-vsini} (a) shows $A$(Li) as a function of $v\sin i$ of the target stars and ordinary solar-analog stars reported in \citet{Takeda2010}.
We can see many of the target stars tend to have high $v\sin i$ and high $A$(Li) compared with ordinary solar-analog stars.
Stellar rotation is also an index of age, 
and it seems reasonable to assume that those stars are young.
However, we also find that there are some target stars with low Li abundance [$A$(Li)$<1.5$] and small projected rotation velocity ($v\sin i=2\sim3$ km s$^{-1}$).
This result may indicate that those stars are not young.
The target stars, which are selected as active solar-analog stars on the basis of the ROSAT data, 
are not necessarily young on the basis of our spectroscopic observations.
\\

\noindent
\ \ \ \ \ \ \
In Appendix \ref{subsec-apen:ana-age-mass}, ages of the target stars are also roughly estimated 
by comparing the position on the HR diagram (cf. Figure \ref{fig:HR} (a) \& (b)).
In Figure \ref{fig:ALi-vsini} (b), we compare such age values with $A$(Li).
Previous studies (e.g., \cite{Takeda2007}) suggest a very weak dependence of $A$(Li) upon age, 
but we cannot see a clear correlation between the two in Figure \ref{fig:ALi-vsini} (b).
This large scatter is caused by the large errors of our stellar age estimation in this age range ($\gtrsim$1 Gyr).
We estimated stellar age from the stellar evolutionary tracks in Appendix \ref{subsec-apen:ana-age-mass},
but this method can severely be affected by many error sources 
(e.g., errors of $T_{\rm{eff}}$, metallicity [Fe/H], stellar distance $d_{\rm{HIP}}$ on the basis of Hipparcos parallax, 
apparent stellar magnitude $m_{V}$, and models of stellar evolutionary tracks themselves).
It is then highly possible that the error bars of age in Figure \ref{fig:ALi-vsini} (b) are underestimated. 
We should be strictly careful when considering this figure.

\subsubsection{Magnetic activity and Li abundance}\label{subsubsec:dis-r0-Li}
\begin{figure}[htbp]
 \begin{center}
   \FigureFile(82mm,82mm){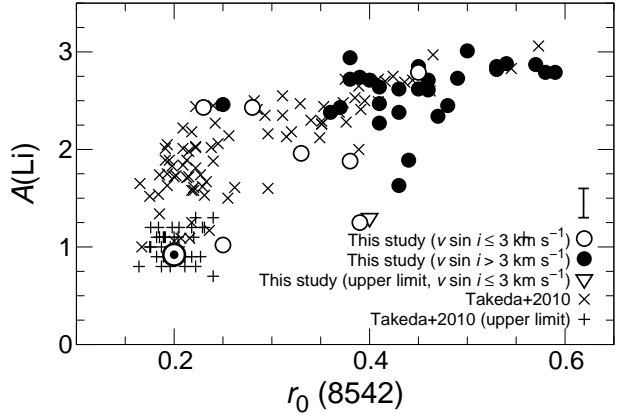}   
 \end{center}
\caption{
$A$(Li) vs. $r_{0}$(8542). 
Circles and triangles correspond to the target stars in this study.
Target stars with $v\sin i\leq$3 km s$^{-1}$ and with $v\sin i>$3 km s$^{-1}$are shown by using open circles/triangles and filled circles, respectively,
and open triangles correspond to the upper limit values of $A$(Li) for the unmeasurable cases. 
The solar value is plotted with a circled dot point for reference. 
We also plotted the data of the ordinary solar-analog stars observed by \citet{Takeda2010} with cross marks (x marks and plus marks),  
as also done in Figure \ref{fig:ALi-vsini}.
}\label{fig:ALi-Lx-8542}
\end{figure}

\ \ \ \ \ \ \
Figure \ref{fig:ALi-Lx-8542} shows the $A$(Li) as a function of $r_{0}$(8542) index for the target stars.
Many of the target stars have high $A$(Li) and high $r_{0}$(8542) values, 
while there are some stars with low $A$(Li) and high $r_{0}$(8542) values.
As mentioned above, $A$(Li) can be an indicator of stellar age, and $r_{0}$(8542) index is an indicator of starspot coverage.
This figure suggests some target stars are not young but have large starspot coverage,
which is not consistent with Skumanich's law \citep{Skumanich1972}. 
This tendency was also seen for the solar-type superflare stars investigated by \citet{Honda2015}.

\subsection{Summary and implication for superflare studies}\label{subsec:dis-fututre}
\ \ \ \ \ \ \
In this study, we conducted high dispersion spectroscopy of 49 nearby solar-analog stars (G-type main sequence stars with $T_{\rm{eff}}\approx5,600\sim6,000$ K) 
identified as ROSAT soft X-ray sources (\cite{Voges1999} \& \yearcite{Voges2000}), 
which are not binary stars (e.g., RS CVn-type binary, eclipsing binary, and visual binary) on the basis of the previous studies.
We confirmed that more than half (37) of the 49 target stars show no evidence of binarity, 
and atmospheric parameters ($T_{\rm{eff}}$, $\log g$, and [Fe/H]) of them are within the range of ordinary solar-analog stars. 
We measured the intensity of Ca II 8542 and H$\alpha$ lines, which are good indicators of stellar chromospheric activity. 
The intensity of these lines indicates that all the target stars, which are selected as ROSAT X-ray sources, have large starspots (Figure \ref{fig:vsinir0}). 
We also measured $v\sin i$ (projected rotational velocity) 
and Li abundance [A(Li)] of these target stars. 
Li abundance is a key to understand the evolution of the stellar convection zone, 
which reflects the age, mass and the rotational history of solar-type stars 
(e.g., \cite{Skumanich1972}; \cite{Soderblom2010}; \cite{Honda2015}). 
We confirmed that the target active solar-analog stars tend to rapidly rotate and 
have high Li abundance, compared with the Sun (Figures \ref{fig:vsini-bunpu} \& \ref{fig:ALi-vsini}). 
This is consistent with many previous studies (e.g., \cite{Skumanich1972}; \cite{Takeda2010}).
There are, however, also some target stars that rotate slowly ($v\sin i=2\sim 3$ km s$^{-1}$) (Figure \ref{fig:vsini-bunpu}) 
and have low Li abundance like the Sun (Figure \ref{fig:ALi-vsini}). 
These results support that old and slowly rotating stars similar to the Sun 
could have high activity level and large starspots.
\\ \\
\ \ \ \ \ \ \
Our previous studies of superflare stars on the basis of Kepler data and Subaru/HDS observations mentioned in Section \ref{sec:intro}
suggest that even old slowly rotating stars similar to the Sun can have superflares and large starspots.
This is roughly consistent with the results in this study,
especially in the point that there are stars having slow rotation velocity.
Then as a future research, 
it is important to conduct long-term monitoring observations of the bright active solar-analog stars, like the target stars in this study,  
in order to investigate the detailed properties of large starspots
(e.g., lifetime of starspots, period of the activity cycle like the solar 11-year cycle, differential rotation)
from the viewpoint of stellar dynamo theory.
In relation to this context, many of the bright target stars in this study can be observed with the future transit satellite 
such as TESS (Transiting Exoplanet Survey Satellite; \cite{Ricker2015}) and PLATO (PLAnetary Transits and Oscillations of stars; \cite{Rauer2014}).
The results of this study (e.g., atmospheric parameters, $v\sin i$, chromospheric activity level, Li abundance) 
could be compared with the data of the rotation period and starspot coverage taken by these new missions in the near future.
In particular, we can know the value of inclination angle ($i$) with spectroscopic $v\sin i$ and photometric rotation period (e.g., from TESS data),
and the discussion of the rotation velocity in Section \ref{subsec:dis-rot} can become more accurate.

\begin{ack}
\ \ \ \ \ \ \
The authors thank the anonymous referee for his/her helpful comments. 
This study is based on observational data collected with Okayama Astrophysical Observatory (OAO), 
which is operated by the National Astronomical Observatory of Japan (NAOJ). 
We are grateful to Mr. Takuya Shibayama, Dr. Eiji Kambe, and the staffs of Okayama Astrophysical Observatory
for making large contributions in carrying out our observations. 
We would also like to thank Dr. Yoichi Takeda for his opening the TGVIT and SPTOOL programs into public. 
This work was supported by JSPS KAKENHI Grant Numbers 
JP25287039, JP26400231, JP26800096, JP16H03955, JP16J00320, and JP16J06887.
\end{ack}

\appendix

\section{Details of analyses and results}\label{sec-apen:Detail-ana}

\subsection{Binarity}\label{subsec-apen:ana-binarity}
\begin{figure}[htbp]
 \begin{center}
  \FigureFile(78mm,78mm){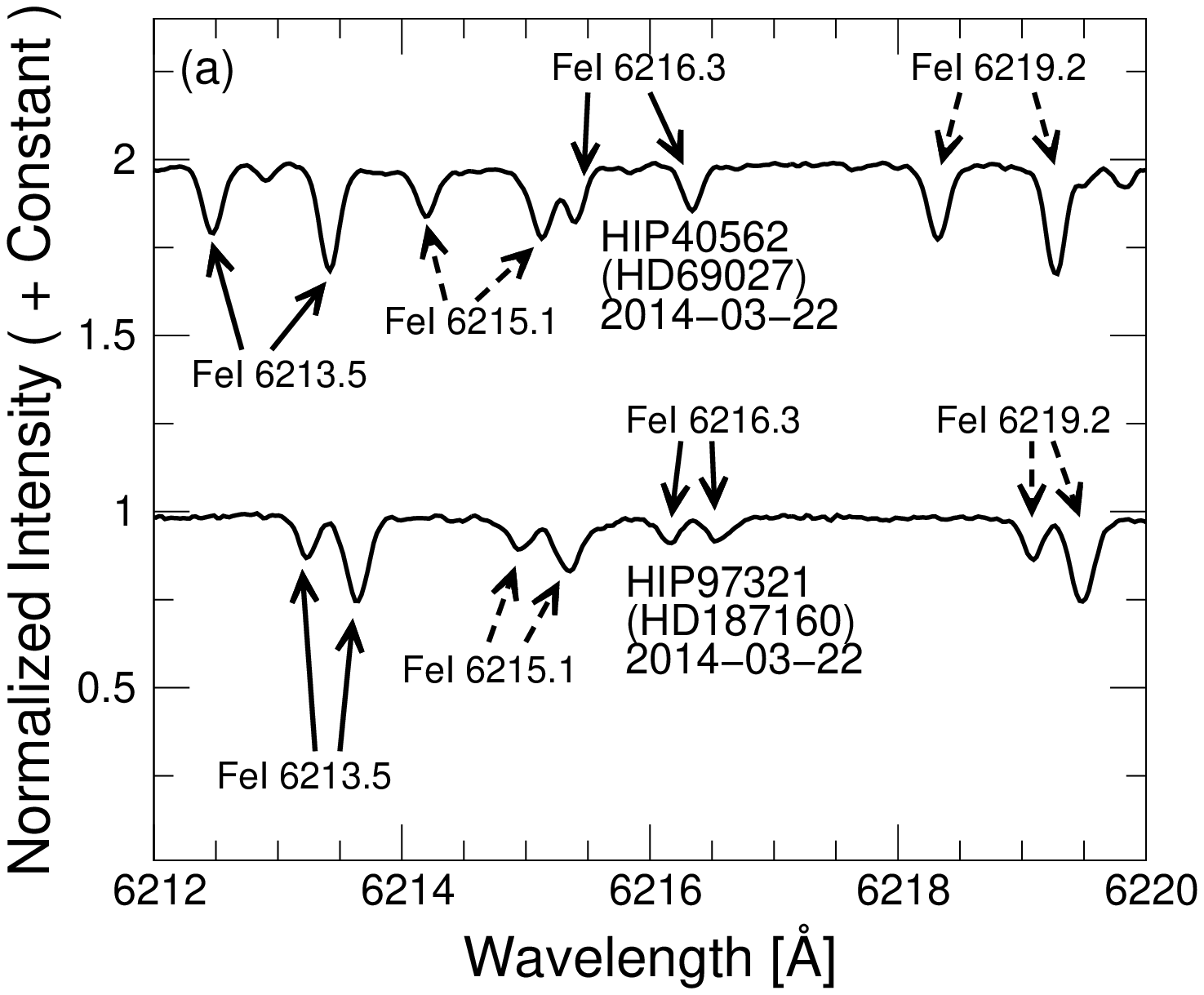}
  \FigureFile(90mm,90mm){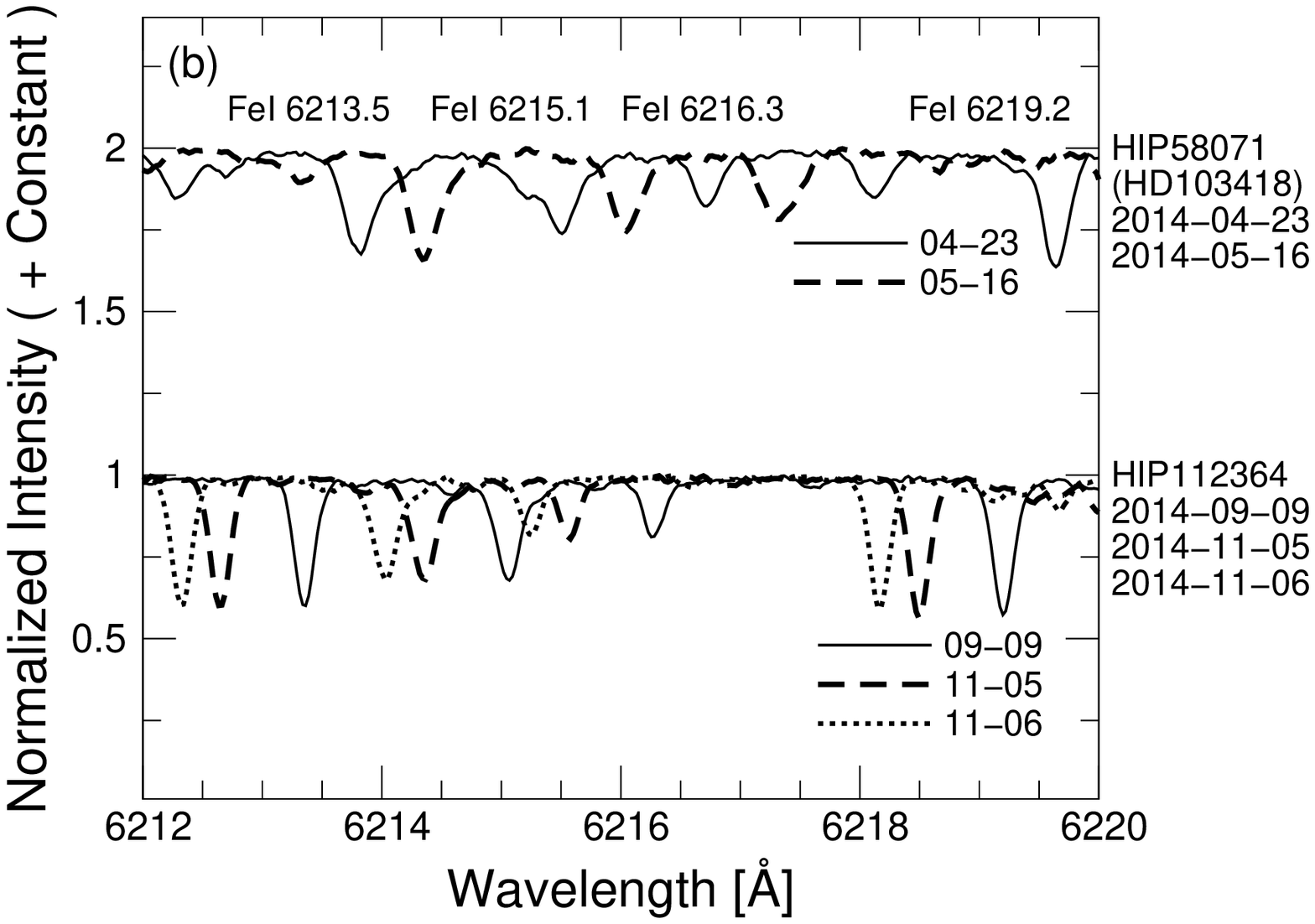}
 \end{center}
\caption{Examples of spectra of stars that we consider as spectroscopic binary stars. The wavelength scale is adjusted to the heliocentric frame. 
Numbers below each star's name (e.g., ``2014-03-22") show observation dates.
Figures of all spectroscopic binary stars are shown in Supplementary Figure 1. \\
(a) Stars that show double-lined profiles. \\
(b) Stars whose spectral lines show radial velocity shifts between the multiple observations, 
which are expected to be caused by the orbital motion in the binary system.}\label{fig:specSB2RV-Fe}
\end{figure}

\noindent
\ \ \ \ \ \ \
For the first step of our analyses, we checked the binarity of each target star, as we did in \citet{YNotsu2015a}.
We investigated the line profiles, and found that seven stars 
(HIP40562, HIP43897, HIP53980, HIP67458, HIP77528, HIP79491, and HIP97321)
have double-lined profiles. 
In this process, we visually checked the profile of the main spectral lines 
(4 Fe I (photospheric) lines in the range of 6212$\sim$6220\AA, Ca II 8542, H$\alpha$ 6563, and Li I 6708) 
that are shown in Supplementary Figure 1 \footnote{
Supplementary figures are available only in the online edition as ``Supporting Information"
}. 
We also visually checked the profile of other many ($\gtrsim$100) Fe I and II lines of the stars 
that are not classified as binary stars 
in this paper, simultaneously with measuring equivalent width values of these lines in Appendix \ref{subsec-apen:ana-atmos-para}.
The double-lined spectra of HIP40562 and HIP97321 are shown in Figure \ref{fig:specSB2RV-Fe} (a) as examples.
Spectral figures (4 Fe I (photospheric) lines in the range of 6212$\sim$6220\AA, Ca II 8542, H$\alpha$ 6563, and Li I 6708) 
of all these seven stars available in Supplementary Figure 1.
Since these double-lined profiles are caused by overlap of the radiation of multiple stars, 
we regard these seven stars as double-lined spectroscopic binary stars. 
These seven stars have ``yes (SB2)" in the 12th column of Table \ref{table:basic-data}. 
\\

\begin{figure}[htbp]
 \begin{center}
  \FigureFile(70mm,70mm){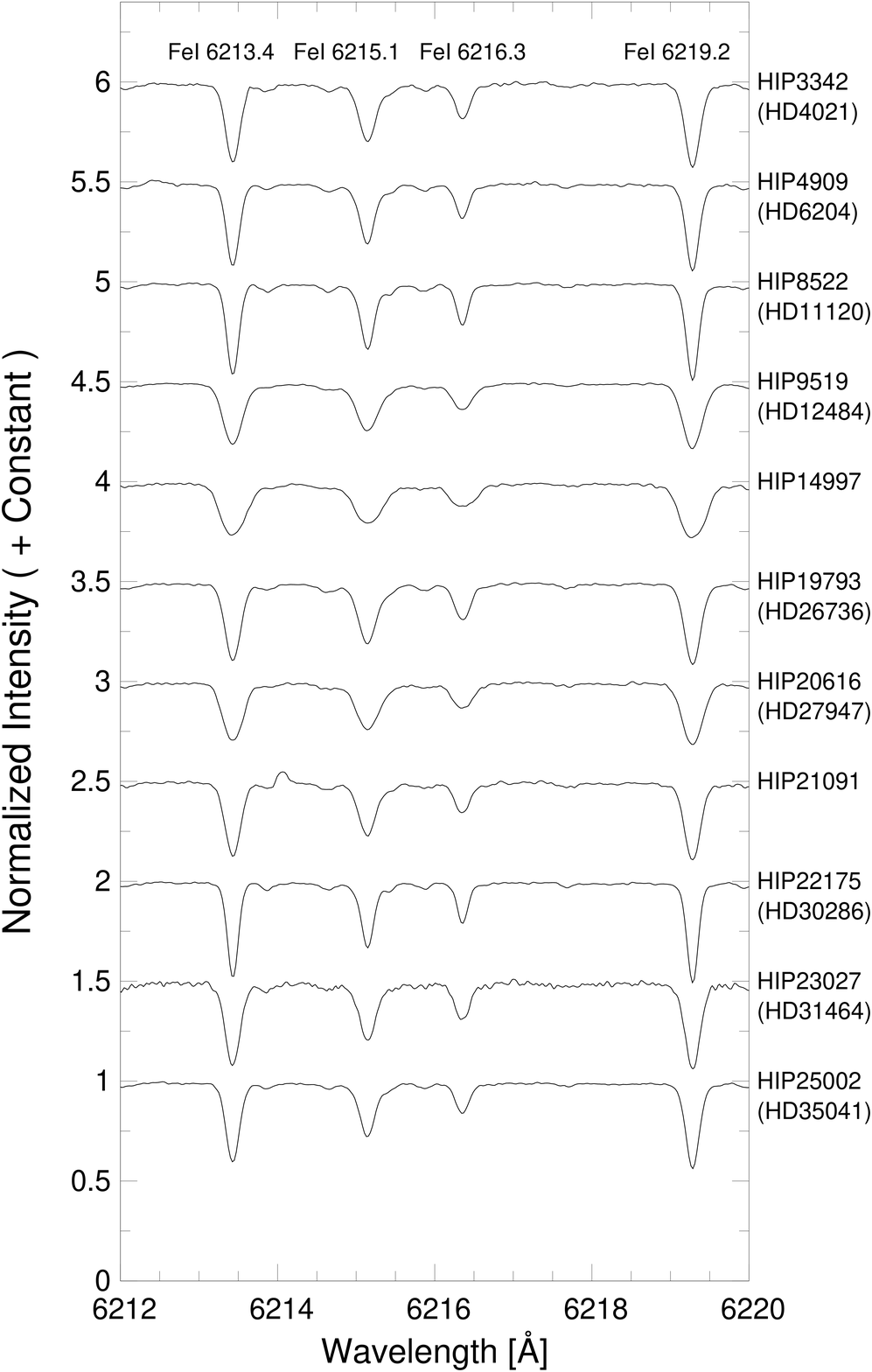}
  \FigureFile(70mm,70mm){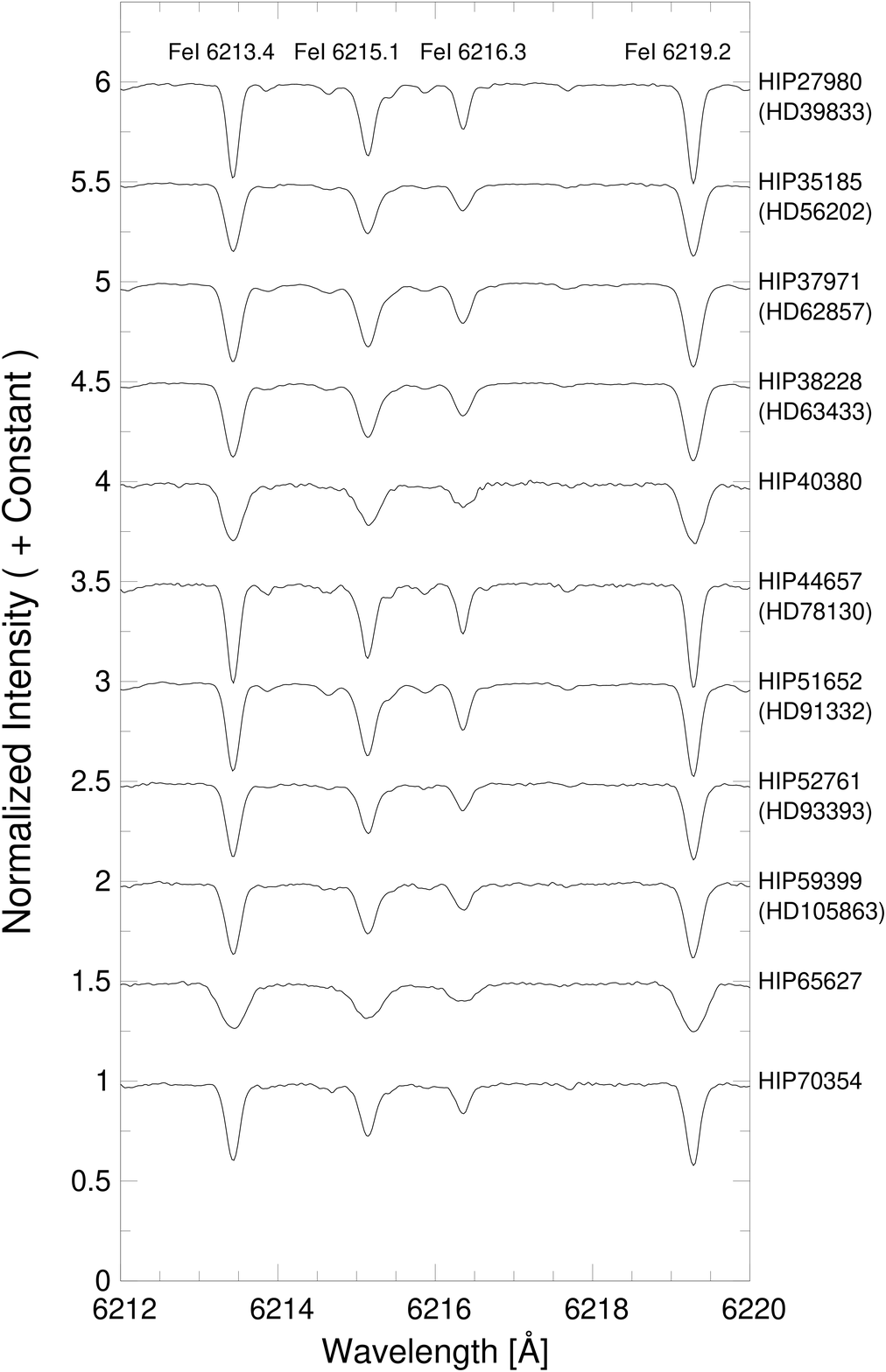}
  \FigureFile(70mm,70mm){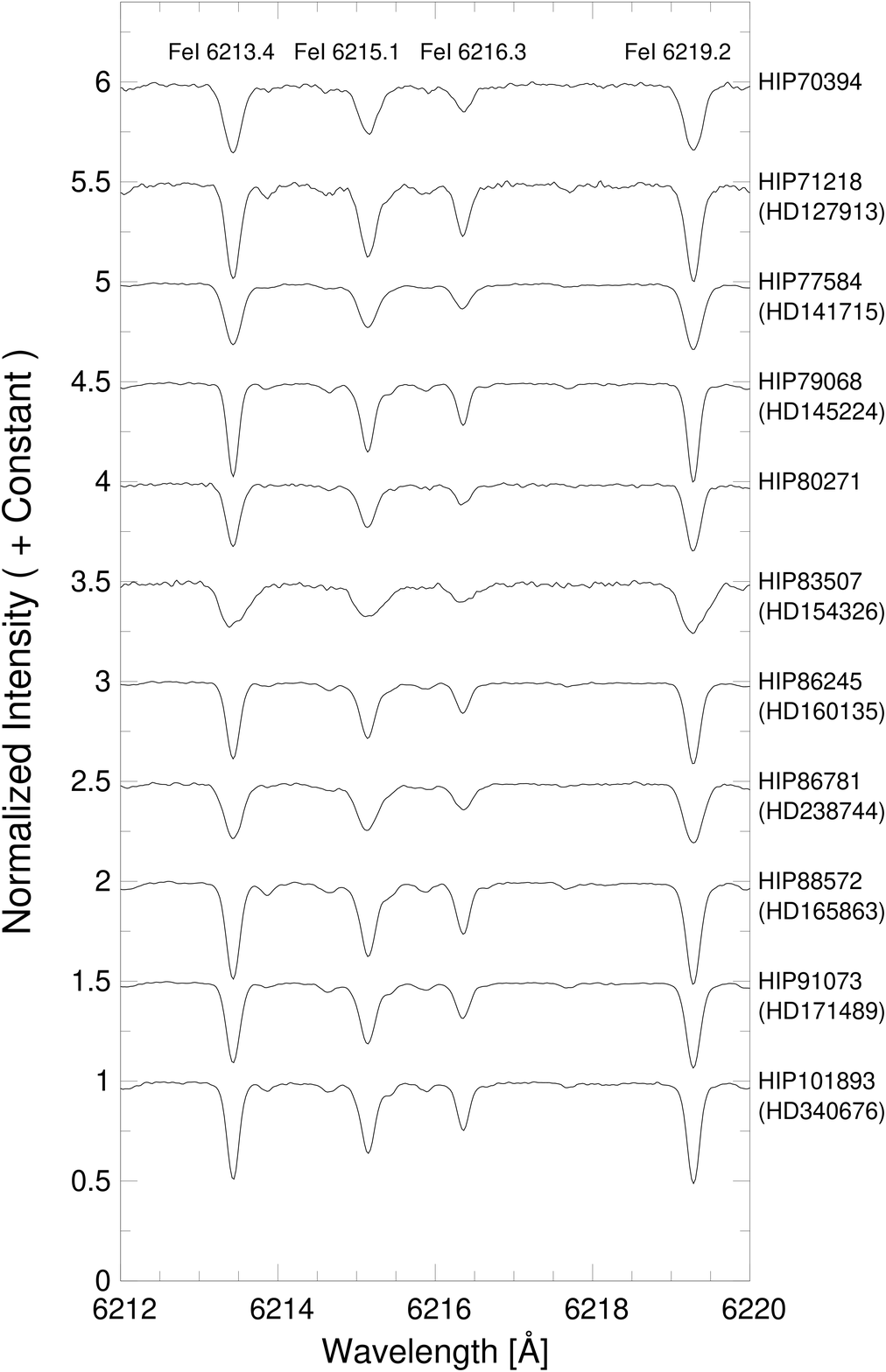}
  \FigureFile(70mm,70mm){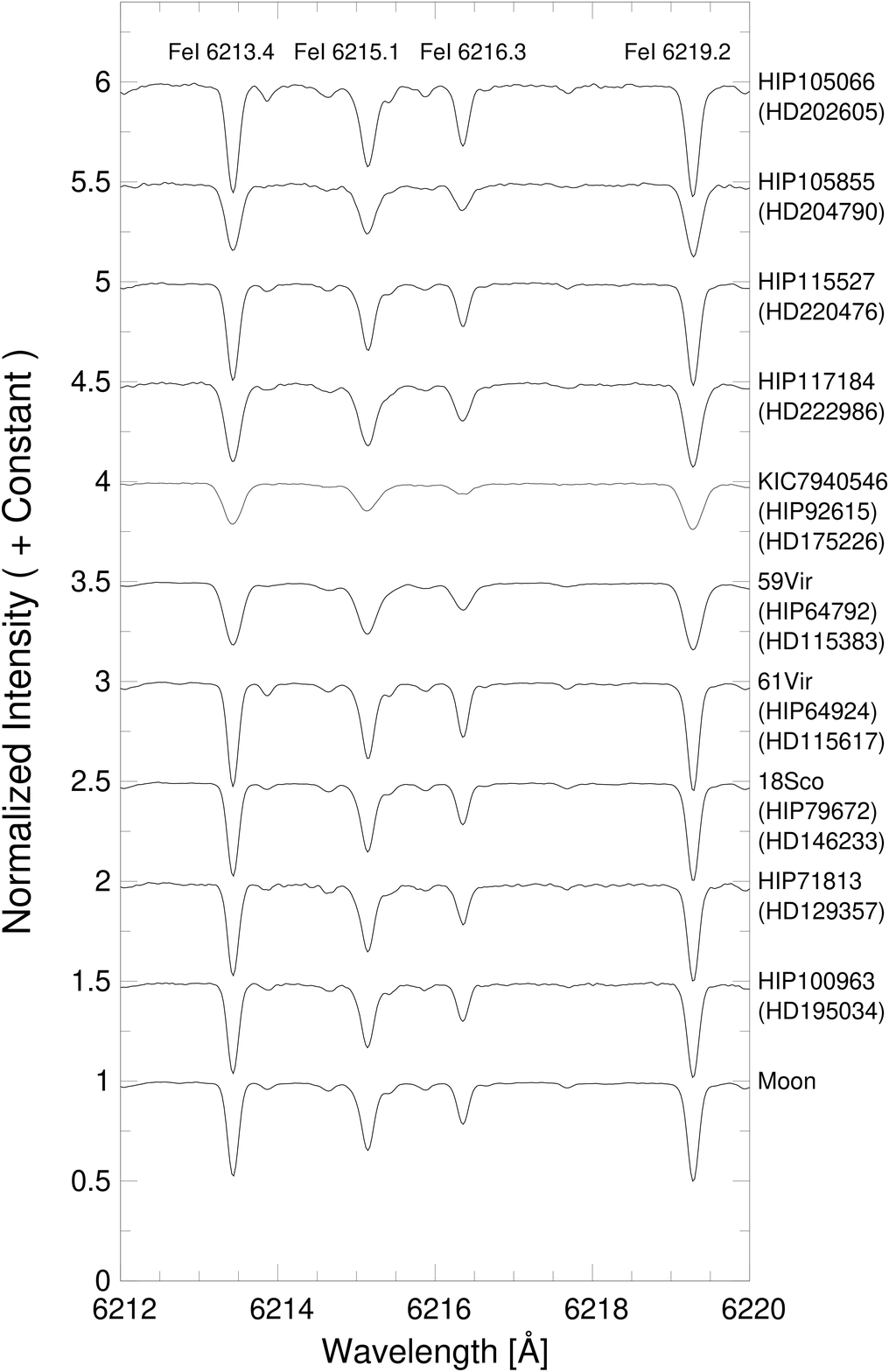}
 \end{center}
\caption{Example of photospheric absorption lines, including Fe I 6213, 6215, 6216, and 6219, 
of the 37 target stars that show no evidence of binarity, six comparison stars, and Moon. 
The wavelength scale is adjusted to the laboratory frame. 
Co-added spectra are used here in case the star was observed multiple times.}\label{fig:specsg-Fe6212}
\end{figure}

\noindent
\ \ \ \ \ \ \
Next, we investigated time variations of the line profiles between multiple observations that are expected to be 
caused by the orbital motion in the binary system. 
This investigation was for the target stars that we observed multiple times (42 stars).
We measured the radial velocity (RV) of all the target stars that were not classified 
as double-lined spectroscopic binary stars \footnote{
We do not measure the radial velocities of each component of the double-lined profiles of these stars in this paper, 
since it is not necessary for the main discussions in this paper.
}. 
We used Fe I \& II lines for measuring RV values.
The estimated RVs and the numbers of lines we used here are listed in Supplementary Table 1 \footnote{
The errors of RVs listed in Supplementary Table 1 are the standard deviations of the RV values that are calculated for individual lines. 
The typical error value of them is $\lesssim1.0$ km s$^{-1}$.
}. As a result, four stars (HIP10972, HIP58071, HIP83099, and HIP112364) show large RV changes ($\gtrsim$10 km s$^{-1}$).
Two of them (HIP58071 and HIP112364) are shown in Figure \ref{fig:specSB2RV-Fe} (b) as examples. 
Spectral figures (4 Fe I (photospheric) lines in the range of 6212$\sim$6220\AA, Ca II 8542, H$\alpha$ 6563, and Li I 6708) 
of all these four stars are available in Supplementary Figure 1.
We consider that these stars are in binary systems, and these stars have ``yes (RV)" in the 12th column of Table \ref{table:basic-data}. 
\\
\\
\ \ \ \ \ \ \
In addition to the above measurements of RV values, 
we also checked RV values of some target stars reported by the previous studies.
\citet{Holmberg2007} reported RV values of the 17 stars (
HIP19793, HIP20616, HIP22175, HIP25002, HIP27980, HIP35185, HIP38228, HIP44657, 
HIP51652, HIP79068, HIP88572, HIP91073, HIP115527, HIP67458, HIP77528, HIP79491, and HIP97321)
among our 49 target stars and six comparison stars.
We listed these values in Supplementary Data 2. 
As a result, except for the four stars 
(HIP67458, HIP77528, HIP79491, and HIP97321) that were already reported as binary in the above analyses, 
there are no stars newly identified as binary stars having large RV changes ($>$ 2-3km s$^{-1}$) compared with our observations above.
We also checked the RV values reported in SOPHIE archive (\cite{Perruchot2008}; \cite{Bouchy2013}) \footnote{http://atlas.obs-hp.fr/sophie/}, 
which reported RV values of the 16 stars (
HIP8522, HIP9519, HIP19793, HIP20616, HIP23027, HIP25002, HIP37971, HIP38228, 
HIP79068, HIP86245, HIP88572, HIP91073, HIP100259, HIP101893, HIP79491, and HIP97321) 
among our 49 target stars and four comparison stars (59Vir, 18Sco, HIP71813, and HIP100963).
We also listed the values in Supplementary Data 2. 
As a result, HIP100259 shows a long-term ($\sim$ several years time scale) RV change as listed in Supplementary Table 2. 
We consider this star (HIP100259) is in a long-period binary system, though not in a close binary system, 
and then this star has ``yes (RVarch)" in the 12th column of Table \ref{table:basic-data}. 
For the remaining stars, except for the two stars (HIP79491 and HIP97321) that were already reported as binary in the above analyses, 
there are no stars newly identified as binary stars having large RV changes ($>$ 2-3km s$^{-1}$) compared with our observations above.
\\
\\
\ \ \ \ \ \ \
In total, we regard 12 target stars as binary stars.
We treat the remaining 37 target stars as ``single stars" in this paper, 
since they do not show any evidence of binarity within the limits of the above analyses.
These stars except for HIP83507 have ``no" in the 4th column of Table \ref{table:basic-data}. 
HIP83507 has a bit asymmetric line profiles (See Figure \ref{fig:specsg-Fe6212}), and this star has ``asymmetry?" in the 12th column of Table \ref{table:basic-data}. 
We, however, treat HIP83507 as a single star in the following analyses, 
since these profiles does not show clear changes (e.g., radial velocity shifts) between the multiple observations.
In this paper, we conduct the detailed analyses only for these 37 ``single" stars including HIP83507.
\\ \\
\ \ \ \ \ \ \
Spectra of photospheric lines, including Fe I 6212, 6215, 6216, and 6219 of the 37 ``single" target stars, 
six comparison stars, and the Moon
are shown in Figure \ref{fig:specsg-Fe6212}.
We observed the 34 stars multiple times among these 37 ``single" target stars. 
Five of the six comparison stars and the Moon were also observed multiple times.
We made co-added spectra of these 40 stars (34 single target stars, 5 comparison stars, and the Moon) by conducting the following two steps.
First, we shifted the wavelength value of each spectrum to the laboratory frame on the basis of the radial velocity (RV) value of each observation. 
These RV values are listed in Supplementary Table 1. 
Second, we added up these shifted spectra to one co-added spectrum.
These co-added spectra are mentioned as ``2014-comb." in Supplementary Table 1. 
In this process of creating co-added spectra, the data whose S/N is very low (marked in the 11th column of Supplementary Table 1) were not used. 
The data taken during the semester 2015A (2015 June 3, 4) also were not used in this process 
since high S/N ratios were already attained only with the data from several observations in 2014. 
This is the case only for three stars (HIP65627, HIP70354, and HIP70394), since only these stars were observed in 2015.
In Figure \ref{fig:specsg-Fe6212}, co-added spectra of these 40 stars are used. 
Only the co-added spectra are used in this paper when we estimate stellar parameters 
from the spectral data of the 40 stars that we observed multiple times.
In addition, the spectroscopic data (1D fits file) of the 37 ``single" target stars, six comparison stars, and the Moon, which are used for estimating stellar parameters in the following analyses, are available in Supplementary Data 1. 
As for the above 40 stars that we observed multiple times, the co-added data are used in Supplementary Data 1.
\\

\noindent
\ \ \ \ \ \ \
In the above, we described that 37 stars of the 49 target stars have no evidence of binary system. 
We need to remember here that we cannot completely exclude the possibility 
that some of these 37 ``single" target stars have companions, since our analyses above are limited (cf. See also Section 4.1 of \cite{YNotsu2015a}).
For example, for many of the target stars, we performed radial velocity measurements only twice or three times,
and it is necessary to observe these target stars more repeatedly in order to exclude as much as possible the possibility of their having companion stars.
Moreover, we also cannot completely exclude the possibility that the above 37 ``single" stars have other faint neighboring stars, 
such as M-dwarfs, even if no double-lined profiles and no radial velocity shifts are confirmed in this study.
In the process of investigating whether the target stars have double-lined profiles in the above, 
we only visually checked the profile of the main spectral lines \footnote{
In Appendix \ref{subsec-apen:ana-atmos-para}, we also visually checked  
the profile of many other ($\sim$100) Fe I and II lines of the stars 
that are not classified as binary stars in the process of measuring atmospheric parameters.
} (H$\alpha$ 6563, Ca II 8542, and four Fe I lines in the range of 6212$\sim$6220\AA) that are shown in Supplementary Figure 1. 
This suggests that the classification of double-lined spectroscopic binary stars is not complete.
More detailed analyses such as the cross-correlation of the target and the template spectrum are needed 
to detect the signature of binarity caused by the existence of faint companion stars. 
We must note this point, but we consider that such detailed analyses of binarity are not necessary 
for the overall discussion of stellar properties of the stars showing high X-ray luminosity in this paper.

\subsection{Temperature, surface gravity, and metallicity}\label{subsec-apen:ana-atmos-para}
\noindent
\ \ \ \ \ \ \
We estimate the effective temperature $T_{\rm{eff}}$, surface gravity $\log g$, microturbulence $v_{\rm{t}}$, 
and metallicity [Fe/H] of the target stars and the comparison stars, 
by measuring the equivalent widths of Fe I and Fe II lines. 
The method is basically the same as the one we used in \citet{YNotsu2015a}, 
which is originally based on \citet{Takeda2002a} and \citet{Takeda2005}. 
We summarize the method in the following.
\\
\\
\ \ \ \ \ \ \
We used Fe I and Fe II lines in the range of  $5630-8370$\AA, 
selected from the line list presented in Online Data of \citet{Takeda2005}.
In the process of measuring equivalent widths, we used the code SPSHOW contained in 
SPTOOL software package\footnote{http://optik2.mtk.nao.ac.jp/$\sim$takeda/sptool/} developed by Y. Takeda, 
which was originally based on Kurucz's ATLAS9/WIDTH9 model atmospheric programs (\cite{Kurucz1993}).
For deriving $T_{\rm{eff}}$, $\log g$, $v_{\rm{t}}$, and [Fe/H] from the measured equivalent widths, 
we used TGVIT program\footnote{http://optik2.mtk.nao.ac.jp/$\sim$takeda/tgv/} developed by Y. Takeda.
The procedures adopted in this program are minutely described in \citet{Takeda2002a} and \citet{Takeda2005}.
\\ \\ 
\ \ \ \ \ \ \
The resultant atmospheric parameters ($T_{\rm{eff}}$, $\log g$, $v_{\rm{t}}$, and [Fe/H]) of the target stars and the comparison stars 
are listed in Tables \ref{table:atmos} and \ref{table:comp-stpara}, respectively.
Equivalent width values of all lines of all the stars that we measured in the above process are also listed in 
Supplementary Data 3. 

\subsection{Projected rotation velocity}\label{subsec-apen:ana-vsini}
\begin{figure}[htbp]
 \begin{center}
  \FigureFile(75mm,75mm){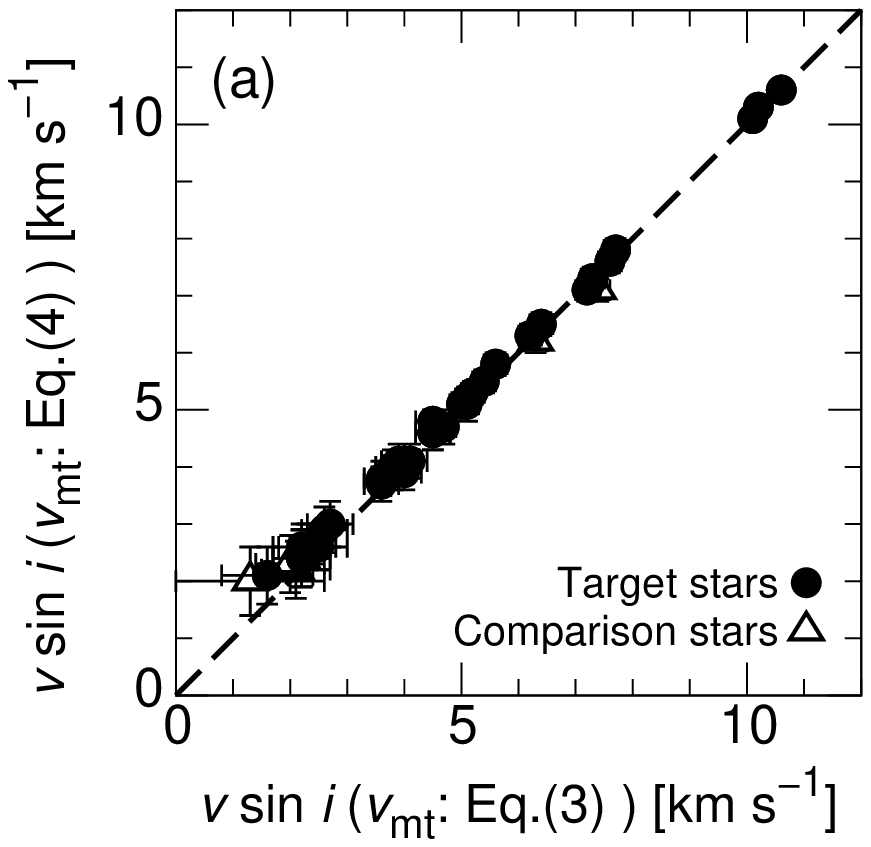}
  \FigureFile(75mm,75mm){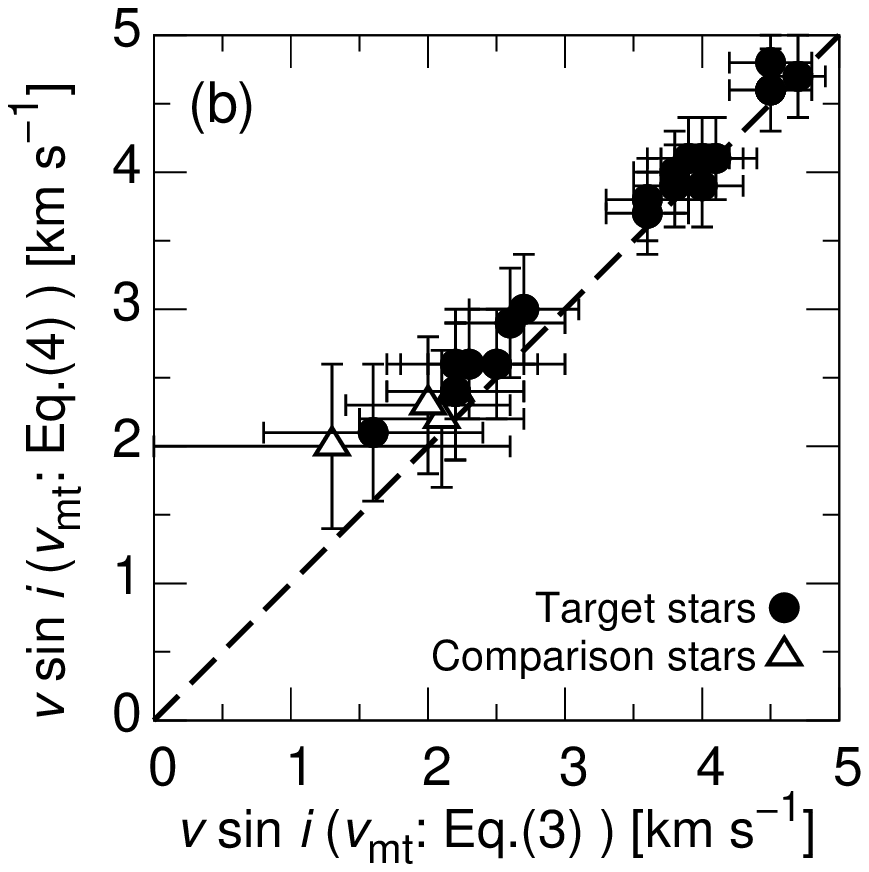}
 \end{center}
\caption{(a) Comparison of two types of $v \sin i$ value using the other equations (Equations (\ref{eq:macroT}) and (\ref{eq:Doyle})) 
for the estimation of macroturbulence velocity ($\zeta_{\rm{RT}}$) in the process of $v \sin i$ measurement. 
The black filled circles correspond to the data of the 37 target stars, while the open triangles are the data of comparison stars.\\
(b) Extended figure of (a). The plot range is limited to 0$\leq v\sin i\leq$5 km s$^{-1}$. \\
}\label{fig:vsini-vmac}
\end{figure}

\ \ \ \ \ \ \
We measured  $v \sin i$ (stellar projected rotational velocity) of the target stars and the comparison stars by using the method 
that is basically the same as in our previous studies (e.g., \cite{SNotsu2013} \& \yearcite{YNotsu2015a}). 
The method is originally based on the one described in \citet{Takeda2008}.
We summarize the method in the following.
\\ \\
\ \ \ \ \ \ \ 
We took into account the effects of macroturbulence and instrumental broadening on the basis of \citet{Takeda2008}.
According to \citet{Takeda2008}, there is a simple relationship among the line-broadening parameters, which can be expressed as 
\begin{equation}
 v_{\mathrm{M}}^{2} = v_{\mathrm{ip}}^{2}+v_{\mathrm{rt}}^{2}+v_{\mathrm{mt}}^{2} \ .
\end{equation}
Here, $v_{\rm{M}}$ is $\mathit{e}$-folding width of the Gaussian macrobroadening function 
{$f(v)\propto \exp[-(v/v_{\rm{M}})^{2}]$}, including instrumental broadening ($v_{\rm{ip}}$), 
rotation ($v_{\rm{rt}}$), and macroturbulence ($v_{\rm{mt}}$). 
We derived $v_{\rm{M}}$ by applying an automatic spectrum-fitting technique \citep{Takeda1995a}, 
assuming the model atmosphere corresponding to the atmospheric parameters estimated in Appendix \ref{subsec-apen:ana-atmos-para}.
In this process, we used the MPFIT program contained in the SPTOOL software package. 
We applied this fitting technique to 6212$\sim$6220\AA \ region to derive $v\sin i$ values.
This region has also been used in our previous studies (e.g., \cite{SNotsu2013} \& \yearcite{YNotsu2015a}). 
\\ \\
\ \ \ \ \ \ \
\ \ \ The instrumental broadening velocity $v_{\rm{ip}}$ was calculated using the following relation 
(cf. \cite{Takeda2008}),
\begin{equation}\label{eq:vip}
 v_{\mathrm{ip}} = \frac{3\times 10^{5}}{2R\sqrt{\rm{ln} 2}} \ ,
\end{equation}
where $R$(=$\lambda/\Delta\lambda$) is the resolving power of the observation.
For estimating $R$ value adopted in this process, we used Th-Ar spectrum data \footnote{
Th-Ar spectrum data were originally taken for wavelength calibration.
} around the 6112-6220\AA~region, 
where we applied the above fitting technique.
We conducted Gaussian-fitting of $\sim$50 lines in the 6180$\sim$6240\AA~region of Th-Ar spectrum, 
and estimated $R(=\lambda/\Delta\lambda)\sim 59000\pm 4000$.
In the following, we used this $R$ value to estimate $v_{\mathrm{ip}}$ with Equation (\ref{eq:vip}).
The macroturbulence velocity $v_{\rm{mt}}$ was estimated by using the relation $v_{\rm{mt}} \sim 0.42\zeta_{\rm{RT}}$~\citep{Takeda2008}.
The term $\zeta_{\rm{RT}}$ is the radial-tangential macroturbulence, and we estimate $\zeta_{\rm{RT}}$ 
using the relation reported in \citet{Valenti2005},
\begin{equation}\label{eq:macroT}
\zeta_{\rm{RT}} = \left( 3.98 - \frac{T_{\rm{eff}} - 5770 \rm{K}}{650 \rm{K}} \right) \ .
\end{equation}
Using these equations, we derived $v_{\rm{rt}}$, and finally $v \sin i$ by using the relation $v_{\rm{rt}}\sim 0.94 v \sin i$ \citep{Gray2005}. 
The resultant $v \sin i$ values of the target stars and the comparison stars 
are listed in Tables \ref{table:activity} and \ref{table:comp-stpara}, respectively. 
\\ \\
\ \ \ \ \ \ \
\authorcite{Hirano2012} (\yearcite{Hirano2012}, \yearcite{Hirano2014}) 
estimated the systematic uncertainty of $v \sin i$ by changing $\zeta_{\mathrm{RT}}$ by
$\pm$15\% from Equation (\ref{eq:macroT}) for cool stars ($T_{\rm{eff}} \leq $6100K), 
on the basis of observed distribution of $\zeta_{\rm{RT}}$ (See also Figure 3 in \cite{Valenti2005}). 
We used this type of error values arising from $\zeta_{\rm{RT}}$ for estimating errors of $v\sin i$, as in \citet{YNotsu2015a}.
In addition, we also took error values of wavelength resolution 
$(R\sim 59000\pm 4000)$ mentioned above into consideration.
As final error values of $v \sin i$ listed in Tables \ref{table:activity} and \ref{table:comp-stpara}, 
we used root sum squares of these two types of error values (error value from $\zeta_{\rm{RT}}$ and that from $R$).
\\ \\
\ \ \ \ \ \ \
As we mentioned in \citet{YNotsu2015a}, 
\citet{Doyle2014} derived the following new equation between macroturblence velocity, $T_{\rm{eff}}$, and $\log g$:
\begin{eqnarray}\label{eq:Doyle}
\zeta_{\rm{RT}} &=& 3.21 + 2.33 \times 10^{-3} (T_{\rm{eff}} - 5777 ) \nonumber \\
&&+ 2.00 \times  10^{-6} (T_{\rm{eff}} - 5777)^{2} - 2.00 (\log g - 4.44 ) \ .
\end{eqnarray}
For comparison, we then derived the new $v \sin i$ value of our targets by using this new equation (Equation (\ref{eq:Doyle}))
in stead of Equation (\ref{eq:macroT}).
The resultant values are listed in Supplementary Table 3.
The error value of this new $v \sin i$ is estimated from the errors of Equation (\ref{eq:Doyle}) ($\Delta\zeta_{\rm{RT}}\sim$0.73 km s$^{-1}$) 
reported in \citet{Doyle2014} and that of wavelength resolution $(R\sim 59000\pm 4000)$ mentioned above.
In Figure \ref{fig:vsini-vmac}, we compare this new $v \sin i$ value estimated by using Equation (\ref{eq:Doyle}) with the original $v \sin i$ value by Equation (\ref{eq:macroT}).
As shown in this figure, the difference between these two $v \sin i$ values is not so large ($<$1 km s$^{-1}$) for most of the target stars.
We have already used the values with Equation (\ref{eq:macroT}) in our previous researches (\cite{SNotsu2013} \& \yearcite{YNotsu2015a}).
Because of these things, in the discussions of this paper, 
we only use the original $v \sin i$ value estimated in the above paragraphs using Equation (\ref{eq:macroT}).
This value with Equation (\ref{eq:macroT}) is plotted in the horizontal axis of Figure \ref{fig:vsini-vmac}  
and listed in Tables \ref{table:activity} and \ref{table:comp-stpara}.

\subsection{Chromospheric activity indicators}\label{subsec-apen:ana-CaHa}
\begin{figure}[htbp]
 \begin{center}
  \FigureFile(65mm,65mm){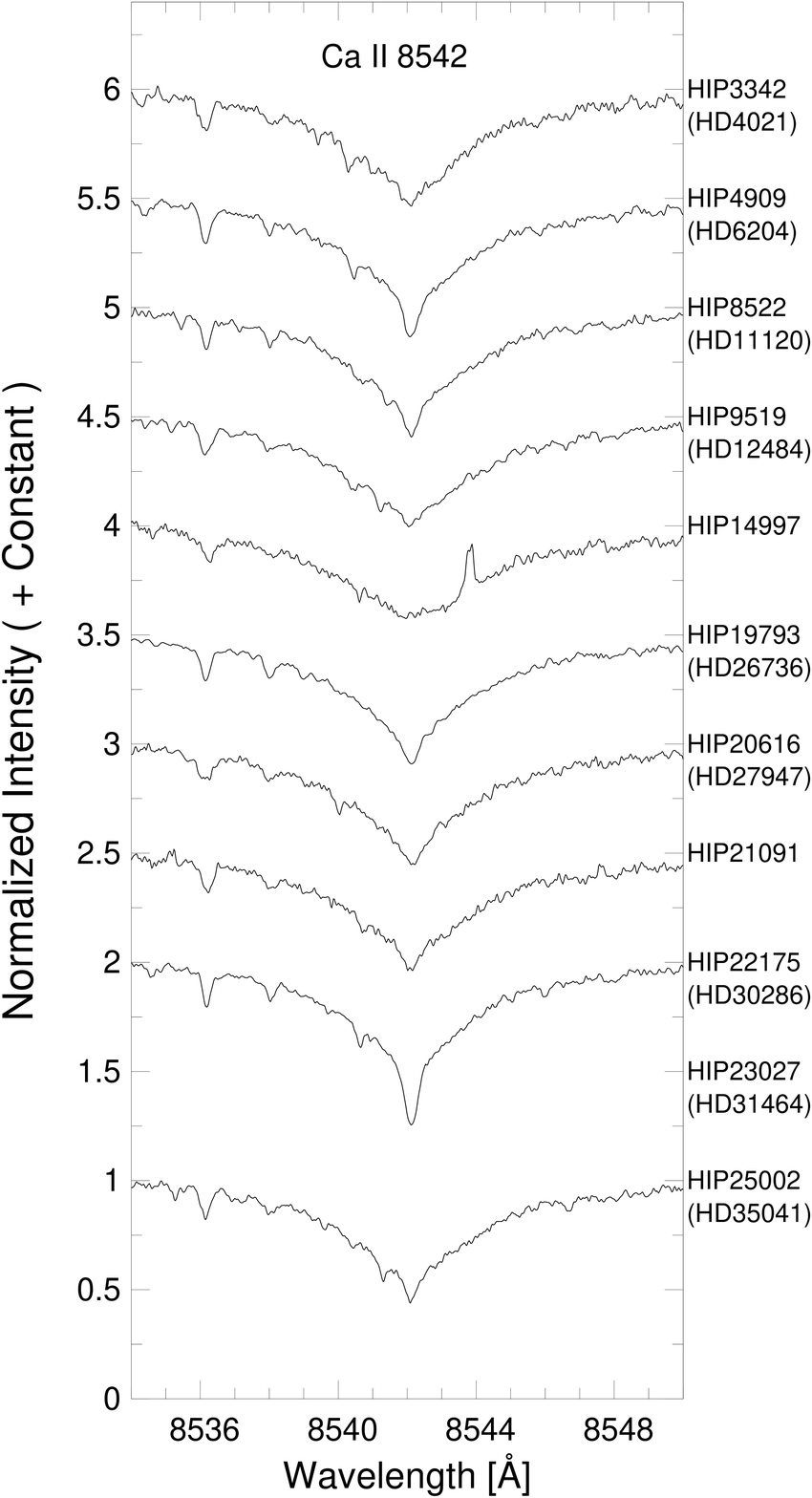}
  \FigureFile(65mm,65mm){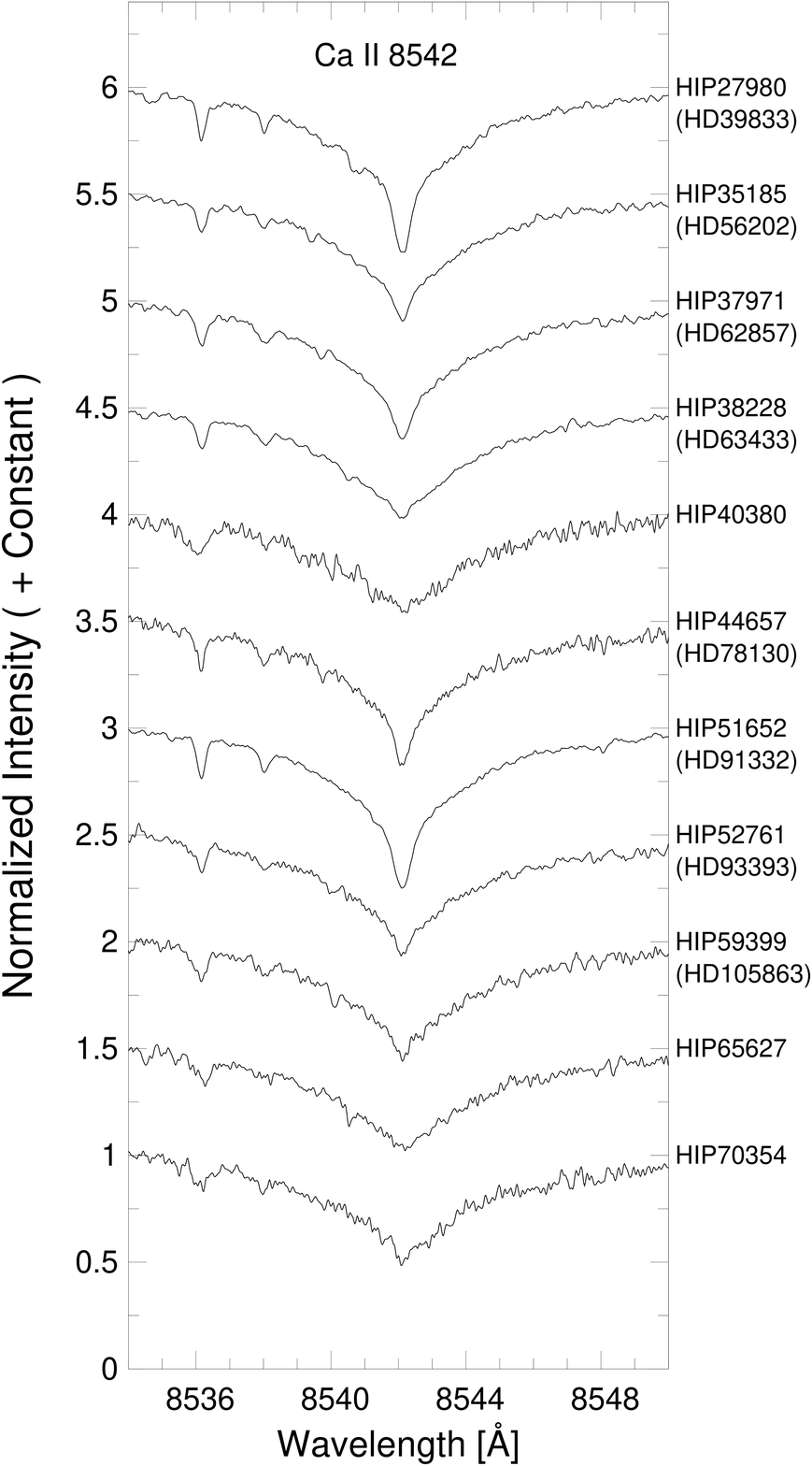}
  \FigureFile(65mm,65mm){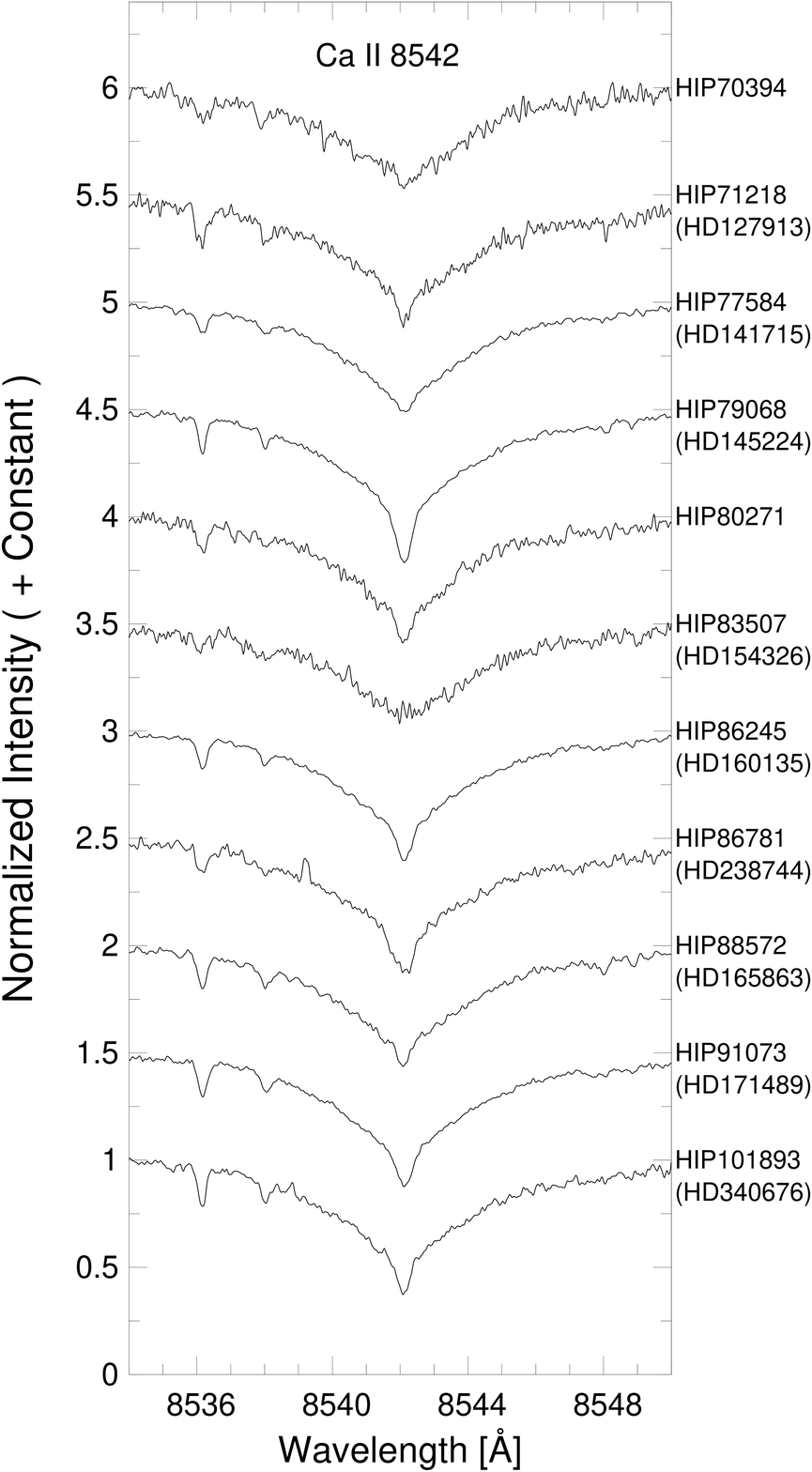}
  \FigureFile(65mm,65mm){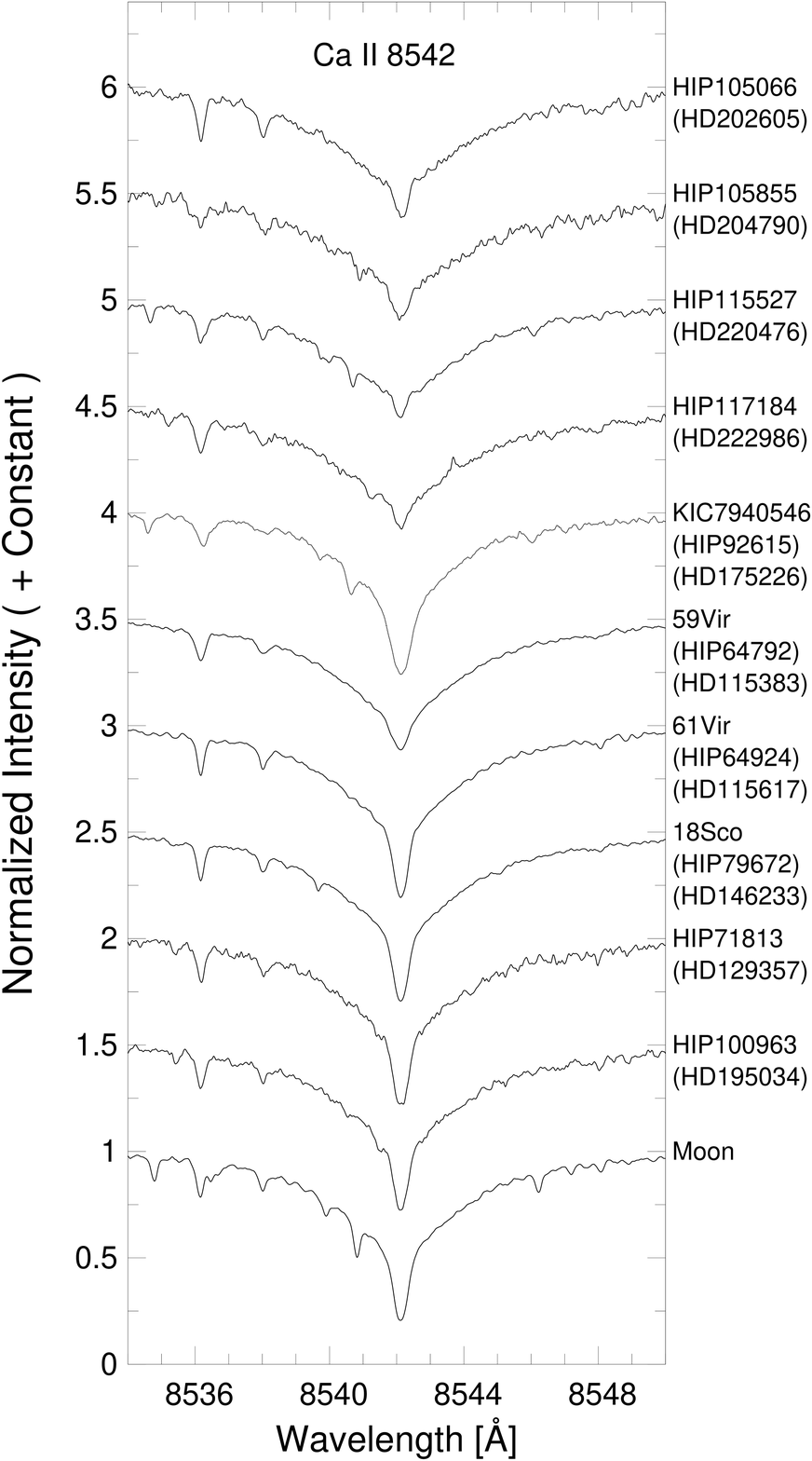}
 \end{center}
\caption{Spectra around Ca II 8542 line of the 37 target stars that show no evidence of binarity, 6 comparison stars, and Moon. 
The wavelength scale is adjusted to the laboratory frame. 
Co-added spectra are used here in case the star was observed multiple times. 
We have no Ca II 8542 data of HIP23027 since the wavelength range of observation on 2014 December 30 did not include the range longer than 7400\AA.
}\label{fig:specsg-Ca8542}
\end{figure}

\begin{figure}[htbp]
 \begin{center}
  \FigureFile(70mm,70mm){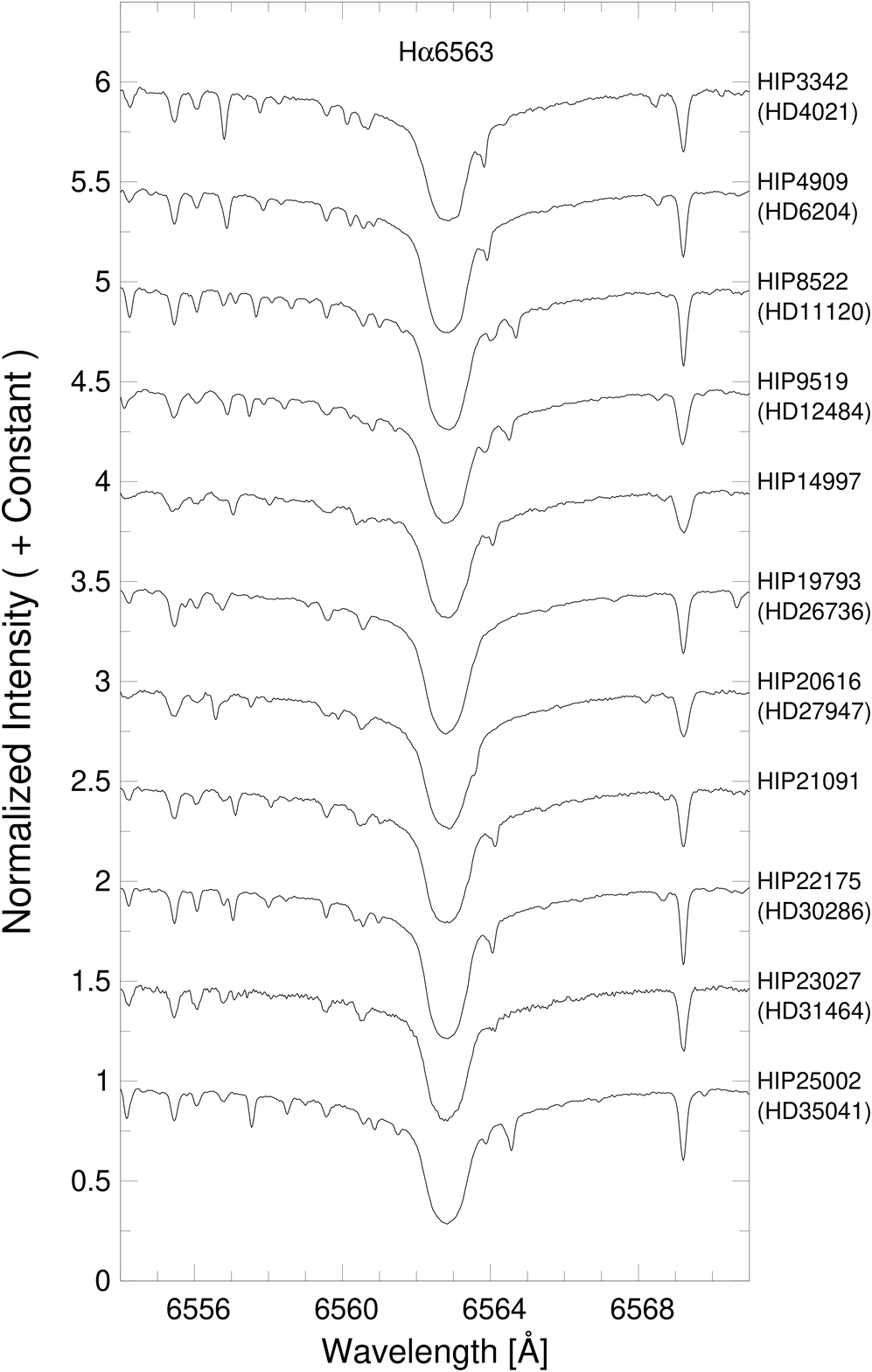}
  \FigureFile(70mm,70mm){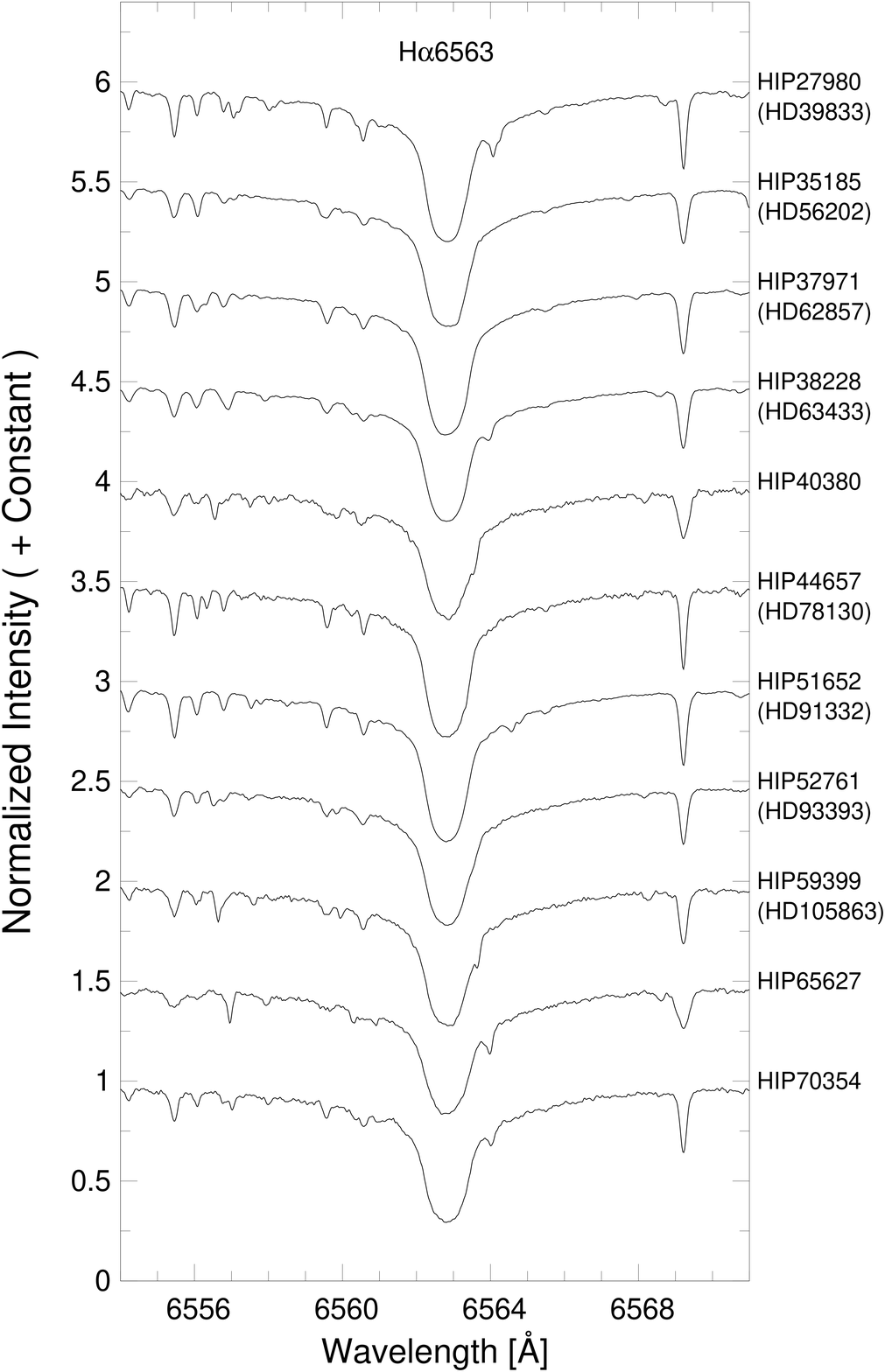}
  \FigureFile(70mm,70mm){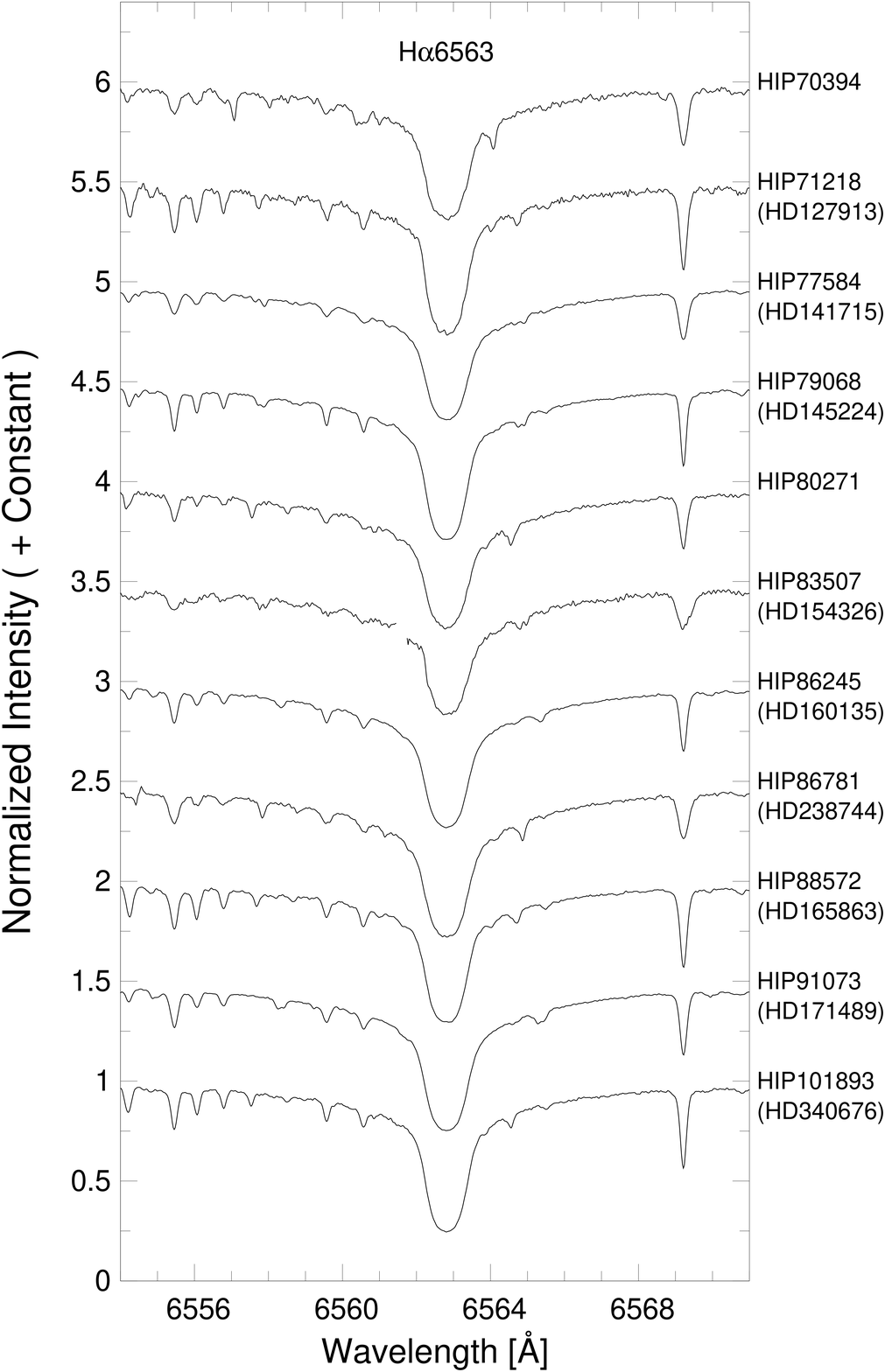}
  \FigureFile(70mm,70mm){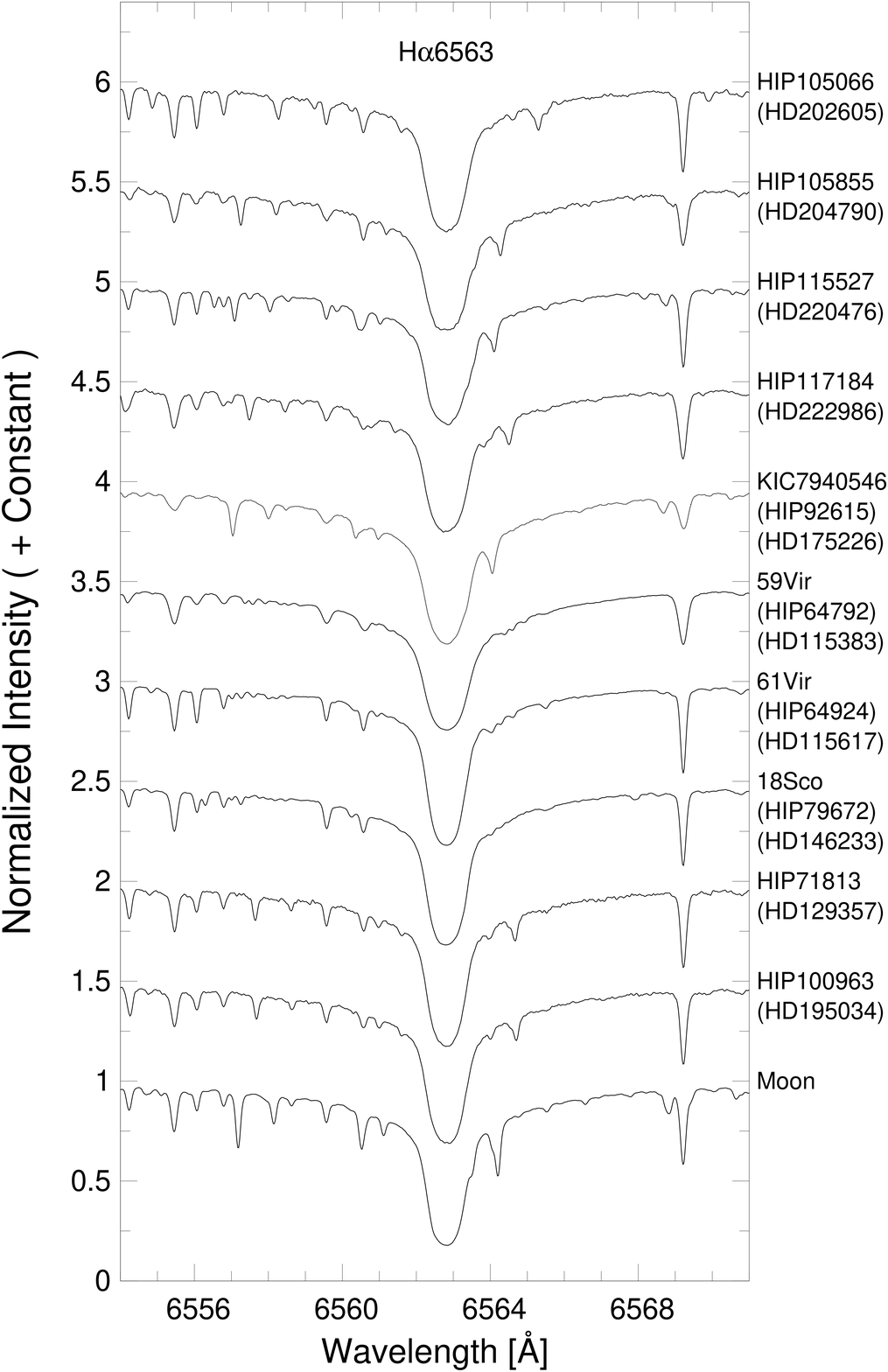}
 \end{center}
\caption{Spectra around H$\alpha$ 6563 line of the 37 target stars that show no evidence of binarity, 6 comparison stars, and Moon. 
The wavelength scale is adjusted to the laboratory frame. 
Co-added spectra are used here in case the star was observed multiple times.
}\label{fig:specsg-Ha}
\end{figure}

\ \ \ \ \ \ \
The observed spectra of the single target stars and the comparison stars 
around Ca II 8542 and H$\alpha$ 6563 are shown in Figures \ref{fig:specsg-Ca8542} and \ref{fig:specsg-Ha}, respectively. 
We have no Ca II 8542 data of HIP23027 since the wavelength range of observation on 2014 December 30 did not include the range longer than 7400\AA.
In order to investigate the chromospheric activity of the target stars, 
we measured $r_{0}$(8542) and $r_{0}$(H$\alpha$) index, 
which are the residual core flux normalized by the continuum level 
at the line cores of the Ca II 8542 and H$\alpha$, respectively.
As we have already introduced the previous researches in detail in Section 3.3 of \citet{SNotsu2013}, 
these indexes are known to be good indicators of stellar chromospheric activity (e.g., \cite{Linsky1979}; \cite{Takeda2010}).
As the chromospheric activity is enhanced, the intensity of these indicators becomes large 
since a greater amount of emission from the chromosphere fills in the core of the lines. 
The values of $r_{0}$(8542) and $r_{0}$(H$\alpha$) indexes of the target single stars and comparison stars are listed 
in Tables \ref{table:activity} and \ref{table:comp-stpara}, respectively.
For reference, we also measured the emission flux of Ca II 8542 and H$\alpha$ lines in Appendix \ref{subsec-apen:ExcessFlux}. 
\\

\noindent
\ \ \ \ \ \ \
\citet{YNotsu2015b} estimated a rough empirical relation between $r_{0}$(8542) index and average stellar magnetic field ($\langle fB\rangle)$
on the basis of spectroheliographic observation of a solar active region.
We then roughly estimate  $(\langle fB\rangle)$ values of the target stars, by using that relation (Equation (1) of \cite{YNotsu2015b}) and $r_{0}$ (8542) values. 
The estimated $\langle fB\rangle$ values are also listed in Tables \ref{table:activity} and \ref{table:comp-stpara}, respectively.
The error value of each $\langle fB\rangle$ value in these tables is estimated with the same way as \citet{YNotsu2015b}.
In addition, if we apply that relation to solar $r_{0}$(8542) value ($r_{0}$(8542)=0.20), 
$\langle fB\rangle$ is estimated to be $0.2\pm6$ [Gauss]. 
This value is consistent with the values of solar mean magnetic field (a few Gauss) within its error range, though the error range is large.

\subsection{Discussion of chromospheric activities using ``Flux Method"}\label{subsec-apen:ExcessFlux}
\begin{figure}[htbp]
 \begin{center}
   \FigureFile(82mm,82mm){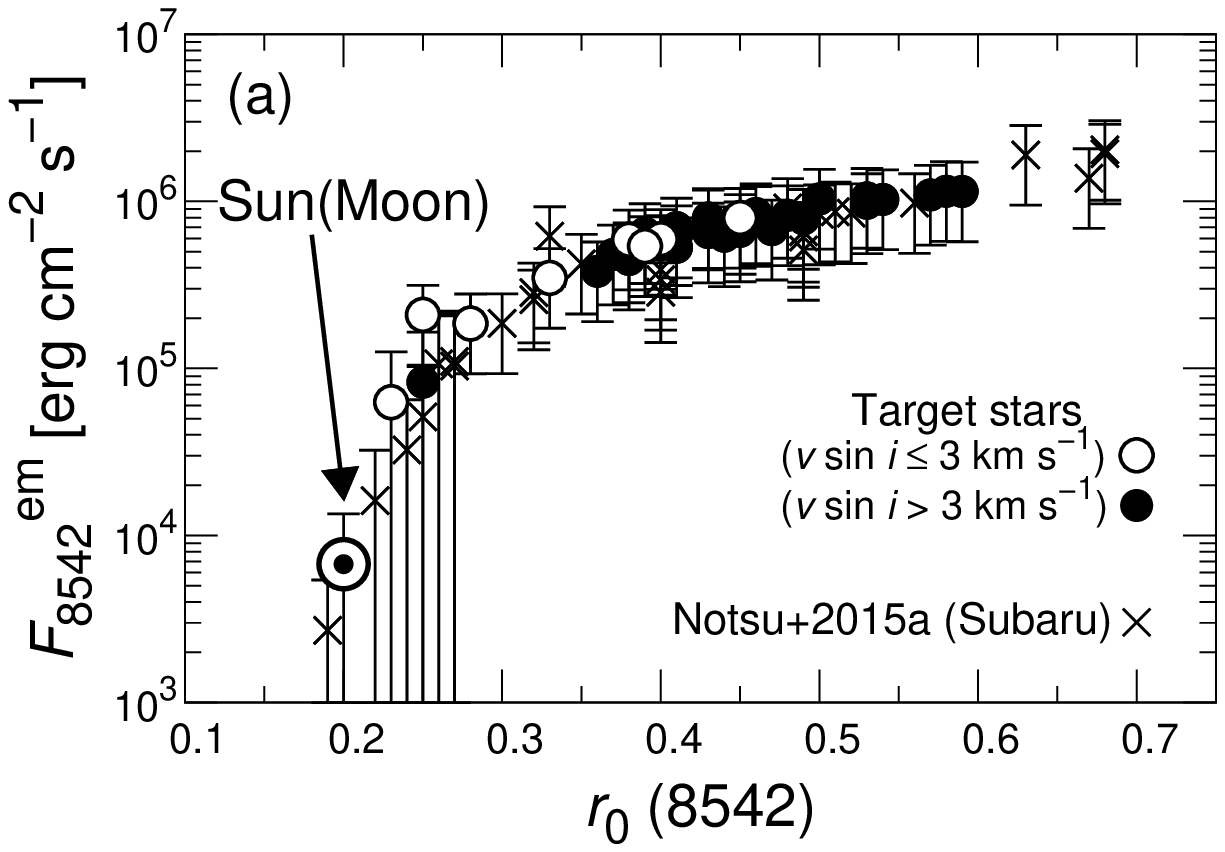}
   \FigureFile(82mm,82mm){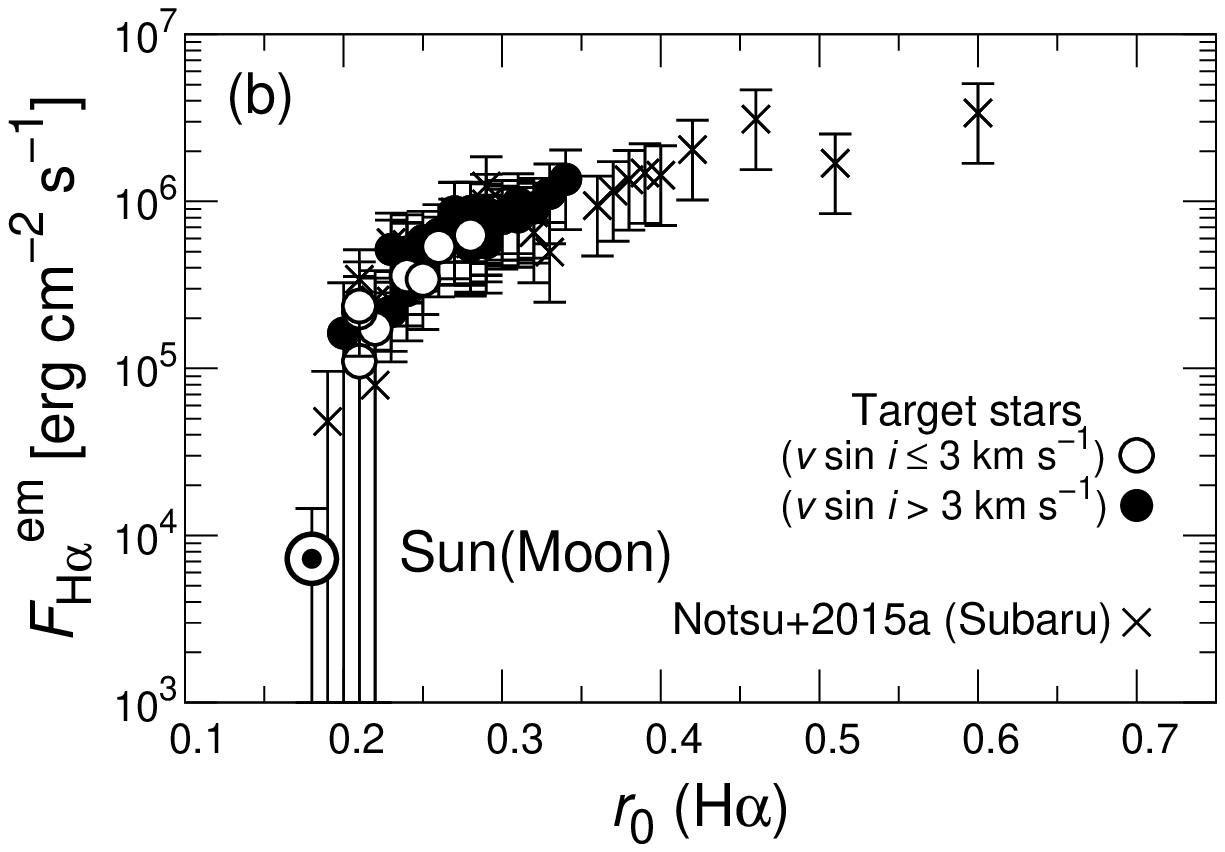}
 \end{center}
\caption{
(a) $F^{\rm{em}}_{8542}$ as a function of $r_{0}$(8542).  \\ 
(b)$F^{\rm{em}}_{H\alpha}$ as a function of $r_{0}$(H$\alpha$). \\
The data points of the target stars in this observation and the stars observed by \citet{YNotsu2015a} with Subaru/HDS 
(superflare stars) are plotted for reference, using circles and cross marks, respectively, in both (a) and (b).
As for circles, target stars with $v\sin i\leq$3 km s$^{-1}$ and with $v\sin i>$3 km s$^{-1}$are 
shown by using open circles and filled circles, respectively.
The Moon value in this paper is also plotted as a solar value for reference. 
}\label{fig:r0-flux}
\end{figure}

\ \ \ \ \ \ \
We measured the $r_{0}$ index values of Ca II 8542 and H$\alpha$ 6563 in Appendix \ref{subsec-apen:ana-CaHa}, 
and discussed chromospheric activity of superflare stars by using the $r_{0}$(8542) index in Section \ref{sec:discussion}. 
This index is an indicator of chromospheric activity as described in the above sections, 
but it also depends on $v \sin i$ (e.g., Figure 5(a) of \cite{Takeda2010}).
A large value of $v \sin i$ can indeed increase the residual flux, mimicking
the effect of filling the line core with chromospheric emission.
Because of this, we here roughly estimated the emission flux of the Ca II 8542 and H$\alpha$ lines 
in order to remove this influence of $v \sin i$.
The methods of estimating are basically the same 
as we did in \authorcite{SNotsu2013} (\yearcite{SNotsu2013} and \yearcite{YNotsu2015b}), 
and we summarize the method in the following. 
\\ \\
\ \ \ \ \ \ \
We used the spectral subtraction technique 
(e.g., \cite{Frasca1994}; \cite{Frasca2011}; \cite{Martinez-Arnaiz2011}; \cite{YNotsu2015b}). 
With this subtraction process, we can subtract the underlying photospheric contribution from the spectrum of the star, 
and can investigate the spectral emission originating from the chromosphere in detail \citep{Martinez-Arnaiz2011}.
As also done in \citet{YNotsu2015b}, we used the spectrum of 61Vir obtained in this observation 
as an inactive template to be subtracted from the spectrum of the target stars. 
We measured the excess equivalent width ($W^{\rm{em}}_{\lambda}$) of the Ca II 8542 and H$\alpha$ lines 
by using the residual spectrum around the line core resulting from this subtraction process.
We then derived emission fluxes ($F^{\rm{em}}_{\lambda}$) of Ca II 8542 and H$\alpha$ lines 
from $W^{\rm{em}}_{\lambda}$ of these lines by the following relation,
$F^{\rm{em}}_{\lambda}=W^{\rm{em}}_{\lambda}F^{\rm{cont}}_{\lambda}$ \citep{Martinez-Arnaiz2011},
where $F^{\rm{cont}}_{\lambda}$ is the continuum flux around the wavelength of each line. 
We calculated $F^{\rm{cont}}_{\lambda}$ by using the empirical 
relationships between $F^{\rm{cont}}_{\lambda}$ and color index ($B-V$) shown in \citet{Hall1996}.
We here did not use $B-V$ values in Hipparcos Catalogue ($(B-V)_{\rm{HIP}}$ in Tables \ref{table:basic-data}\&\ref{table:comp-stpara}) 
in order to avoid effects of interstellar extinction.
Instead, we used the following empirical relations of low-mass main-sequence stars (F0V$\sim$K5V) 
among $B-V$, $T_{\rm{eff}}$, and [Fe/H] derived by \citet{Alonso1996}.
\\ \\ 
\ \ \ \ \ \ \
The estimated values of the emission flux of Ca II 8542 ($ F^{\rm{em}}_{8542}$) and H$\alpha$ ($F^{\rm{em}}_{\rm{H}\alpha}$) are listed 
in Supplementary Table 6 of this paper.
The excess equivalent width values ($W^{\rm{em}}_{8542}$ and $W^{\rm{em}}_{\rm{H}\alpha}$) are also listed in this table.
In Figure \ref{fig:r0-flux}, the $F^{\rm{em}}_{8542}$ and $F^{\rm{em}}_{\rm{H}\alpha}$ are plotted as a function of $r_{0}$(8542) and $r_{0}$(H$\alpha$), respectively.
The error values of $F^{\rm{em}}_{8542}$ and $F^{\rm{em}}_{\rm{H}\alpha}$ in Figure \ref{fig:r0-flux} are 
plotted in the following method. 
Roughly assuming the upper limit of subtraction error, 
we here consider the typical error value of the emission flux ($F^{\rm{em}}$) is roughly $\pm50$\%, 
if excess equivalent width ($W^{\rm{em}}$) is larger than 30m\AA~(sufficiently high compared to 61Vir).
On the other hand, if the excess equivalent width ($W^{\rm{em}}$) is less than 30m\AA~(not so high compared to 61Vir), 
we consider that the typical error value of the emission flux ($F^{\rm{em}}$) is about $\pm100$\%~.
We can see a positive correlation between emission flux and $r_{0}$ index in this figure,  
and all of the target stars with high $r_{0}$(8542) and $r_{0}$(H$\alpha$) values show high $F^{\rm{em}}_{8542}$ and $F^{\rm{em}}_{\rm{H}\alpha}$ values.
This positive correlation is consistent with that of superflare stars we measured with Subaru/HDS in \citet{YNotsu2015b}.
This figure shows us that by using values of the emission flux,
we can reach the same basic conclusions we did with $r_{0}$(8542) index in Section \ref{sec:discussion}.

\subsection{Li abundances}\label{subsec-apen:ana-Li}
\begin{figure}[htbp]
 \begin{center}
  \FigureFile(70mm,70mm){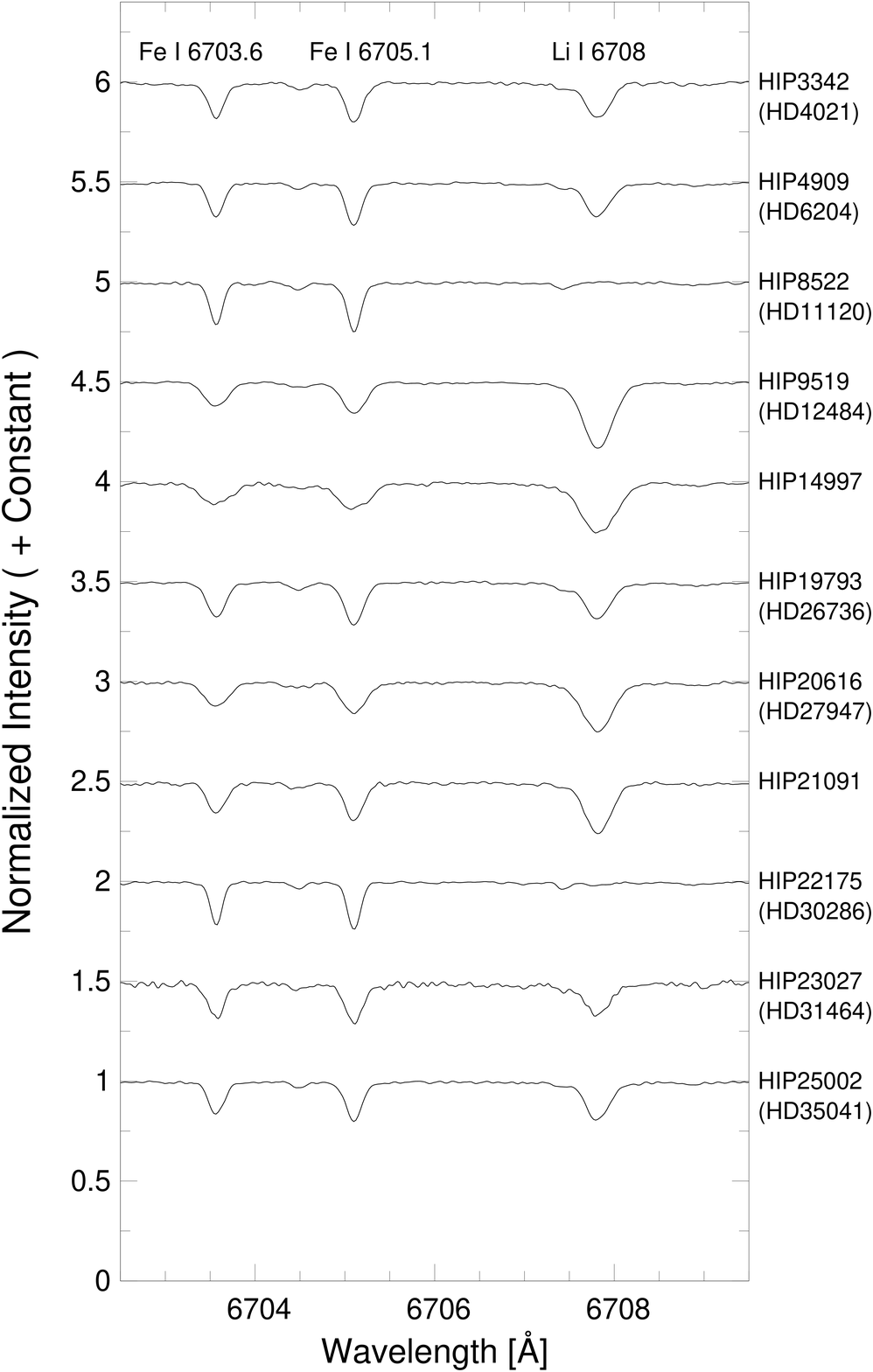}
  \FigureFile(70mm,70mm){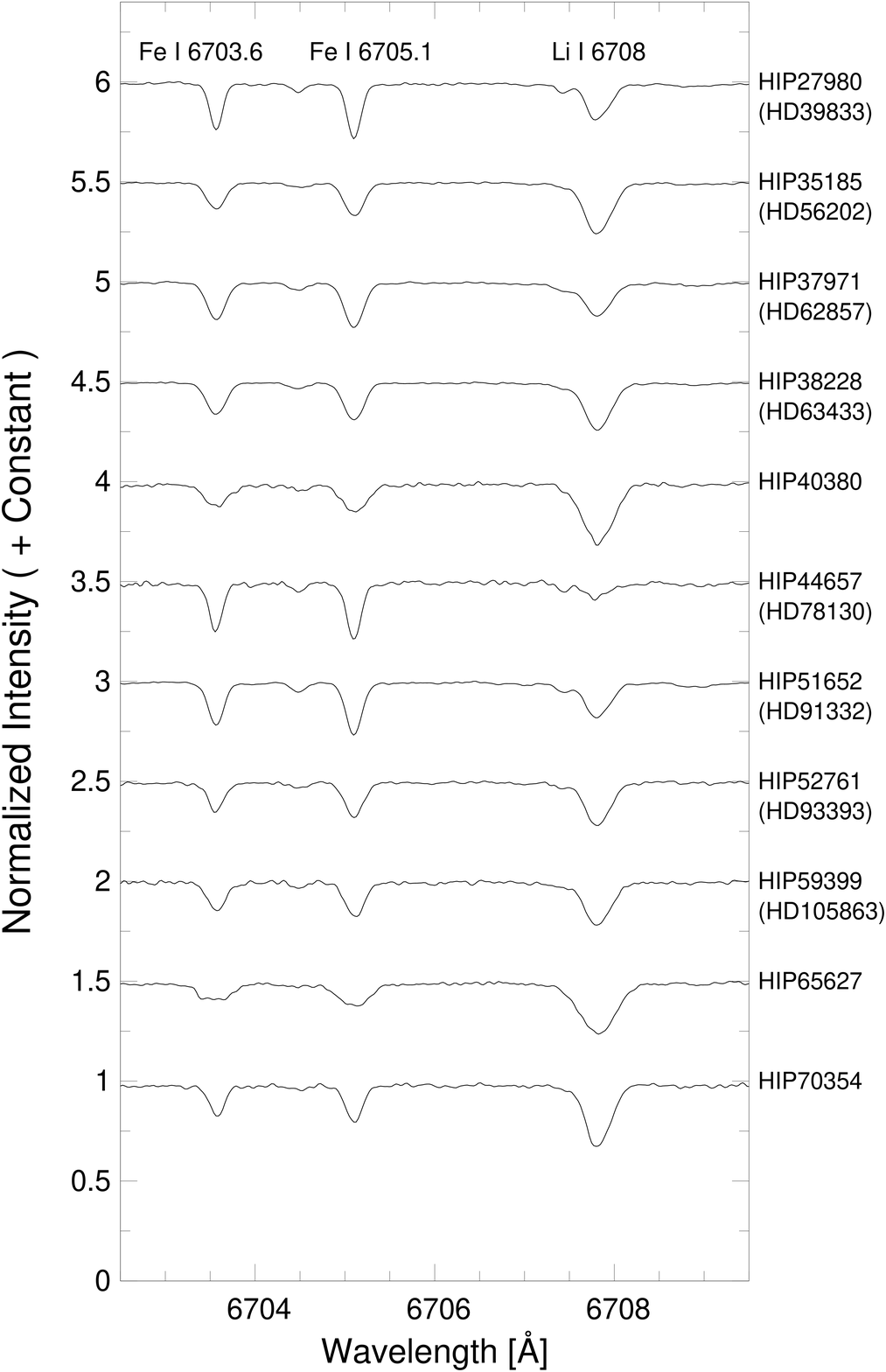}
  \FigureFile(70mm,70mm){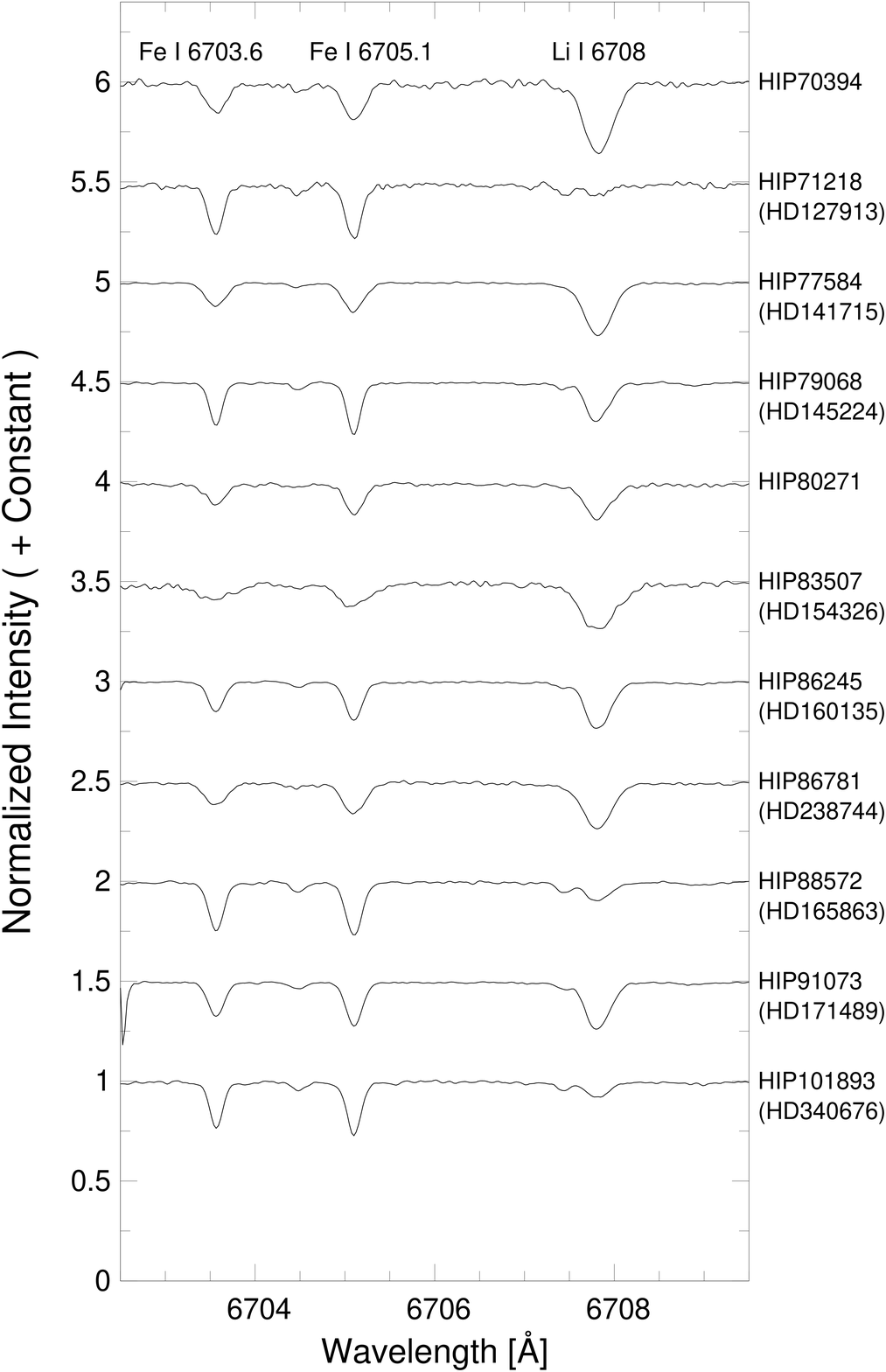}
  \FigureFile(70mm,70mm){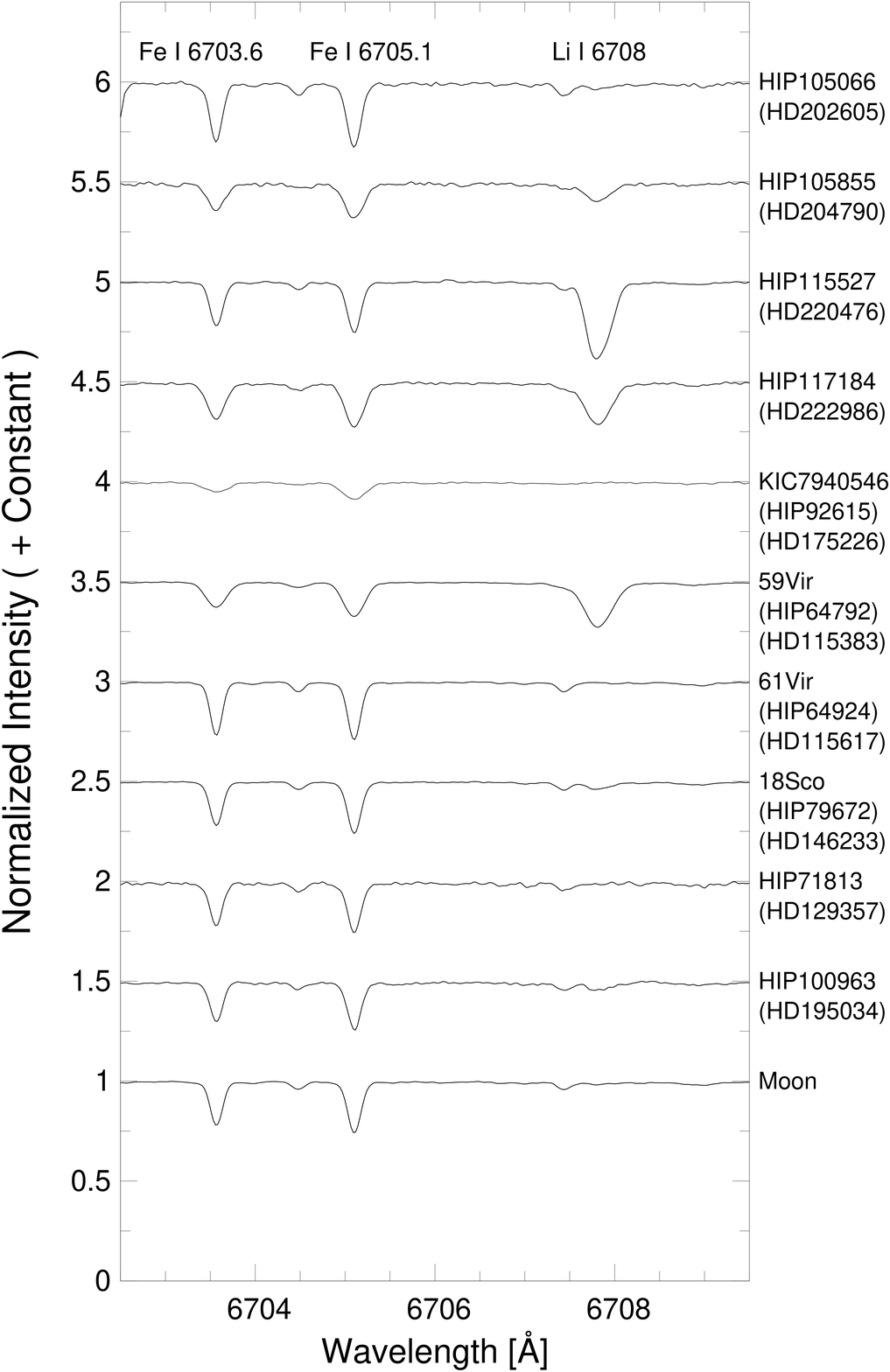}
 \end{center}
\caption{Spectra around Li I 6708 line of the 37 target stars that show no evidence of binarity, 6 comparison stars, and Moon. 
The wavelength scale is adjusted to the laboratory frame. 
Co-added spectra are used here in case that the star was observed multiple times.
}\label{fig:specsg-Li6708}
\end{figure}

\ \ \ \ \ \ \
The observed spectra of the single target stars and the comparison stars 
around Li I 6708 are shown in Figure \ref{fig:specsg-Li6708}. 
We measured Li abundances [$A$(Li)] of these stars from these spectra using the atmospheric parameters determined in Appendix \ref{subsec-apen:ana-atmos-para}.
We here used the automatic profile fitting method that is basically the same as in our previous studies (\cite{SNotsu2013}; \cite{Honda2015}). 
The method is originally based on the one described in \citet{TakedaKawanomoto2005}.
We summarize the method in the following.
\\

\noindent
\ \ \ \ \ \ \
In this process of calculating $A$(Li), we used the MPFIT program contained in SPTOOL software package, which are also used above.
We assumed local thermodynamic equilibrium (LTE) and derived Li abundances 
by using the synthesis spectrum with interpolated model atmospheres taken from \citet{Kurucz1993}.
We also assumed $^{6}\rm{Li}/^{7}\rm{Li}=0$ throughout this study.
The line data around the Li I 6708\AA~region adapted here are the same as those used in Section 2.2 of \citet{TakedaKawanomoto2005},
which are originally based on the lists of \citet{Smith1998} and \citet{Lambert1993}.  
The resultant Li abundances ($A$(Li)) of the single target stars and comparison stars 
are listed in Tables \ref{table:activity} and \ref{table:comp-stpara}, respectively.
\\ \\ 
\ \ \ \ \ \ \
For the stars where the Li feature is absent (e.g., below the detectable limit),
we estimated the upper limit of $A$(Li) by applying the method of \citet{TakedaKawanomoto2005}.
They used the average FWHM (full width at half maximum) of weak Fe lines and signal-to-noise ratios (S/N)
for deriving the upper limit of equivalent width of Li I 6708\AA~absorption line ($EW$(Li)).
Using this value and the WIDTH program in SPTOOL software package, we then estimated the corresponding upper limit of $A$(Li).
These upper limit values are also listed in Tables \ref{table:activity} and \ref{table:comp-stpara}.
\\ \\
\ \ \ \ \ \ \
Our previous paper, \citet{Honda2015} discussed the typical errors of Li abundances by considering errors arising from multiple causes 
(errors linked to atmospheric parameters, uncertainties arising from profile fitting errors, and non-LTE effects).
We here assume errors of $A$(Li) on the basis of \citet{Honda2015}, as in the following.
As for the errors linked to atmospheric parameters, we used the typical error of 0.08 dex, as used in \citet{Honda2015}.
This is because the method of estimating the atmospheric parameters in this study is also basically the same as that in \citet{Honda2015}.
As also for the uncertainties arising from profile fitting errors, we also use the results in \citet{Honda2015}.
In most cases, the errors of $A$(Li) are smaller than 0.05 dex, though some objects having low $A$(Li) values ($A$(Li)$\lesssim$1.0) show a error larger than 0.1 dex.
It is necessary to keep in mind that the low A(Li) stars have larger errors than high A(Li) stars.
\\ \\
\ \ \ \ \ \ \
In the case of typical atmospheric parameters and Li abundance 
[$T_{\rm{eff}}$ = 5800 K, $\log g$ = 4.5, [Fe/H]=0, and $A$(Li) = 2.4] of our target stars, 
the effect of non-LTE is about 0.03 dex, found by using the grid of corrections (\cite{Carlsson1994}).
However, it must be noted that the correction of $-0.12$ dex is the most effective case for $A$(Li)=3.0, 
which is the highest $A$(Li) value among our target stars.
\\ \\
\ \ \ \ \ \ \
Considering these things, we then assume the typical error of $A$(Li) is about 0.15 dex in the targets of this observation.

\subsection{Stellar luminosity and radius}\label{subsec-apen:ana-luminosity-radius}
\begin{figure}[htbp]
 \begin{center}
  \FigureFile(82mm,82mm){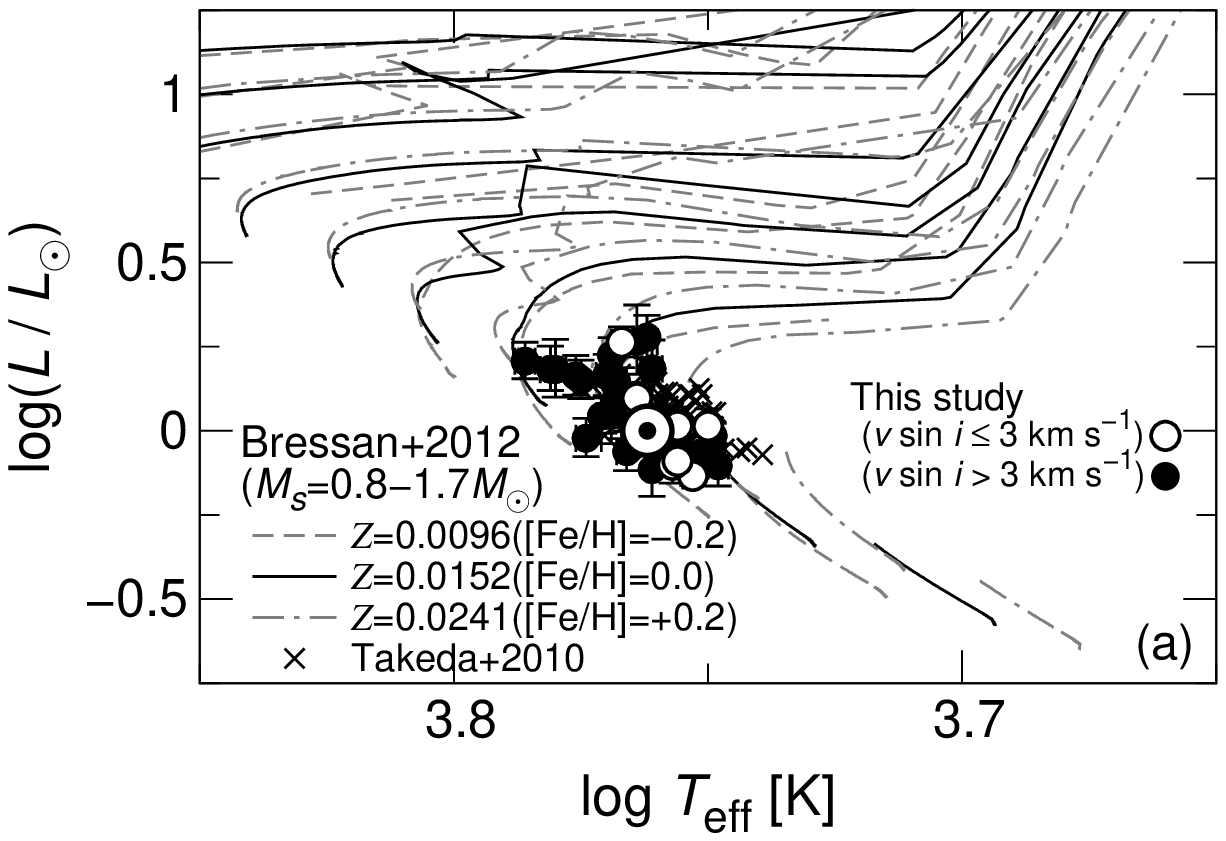}
  \FigureFile(82mm,82mm){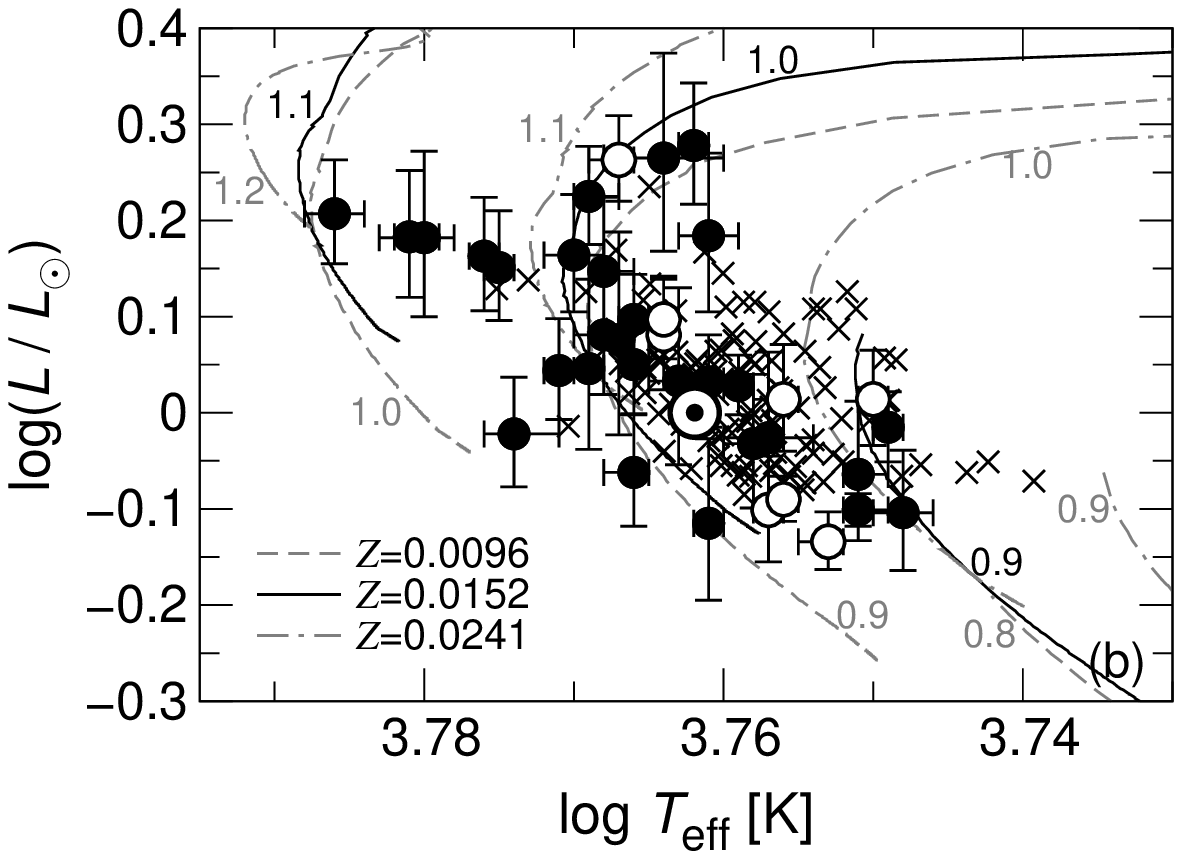}          
  \FigureFile(82mm,82mm){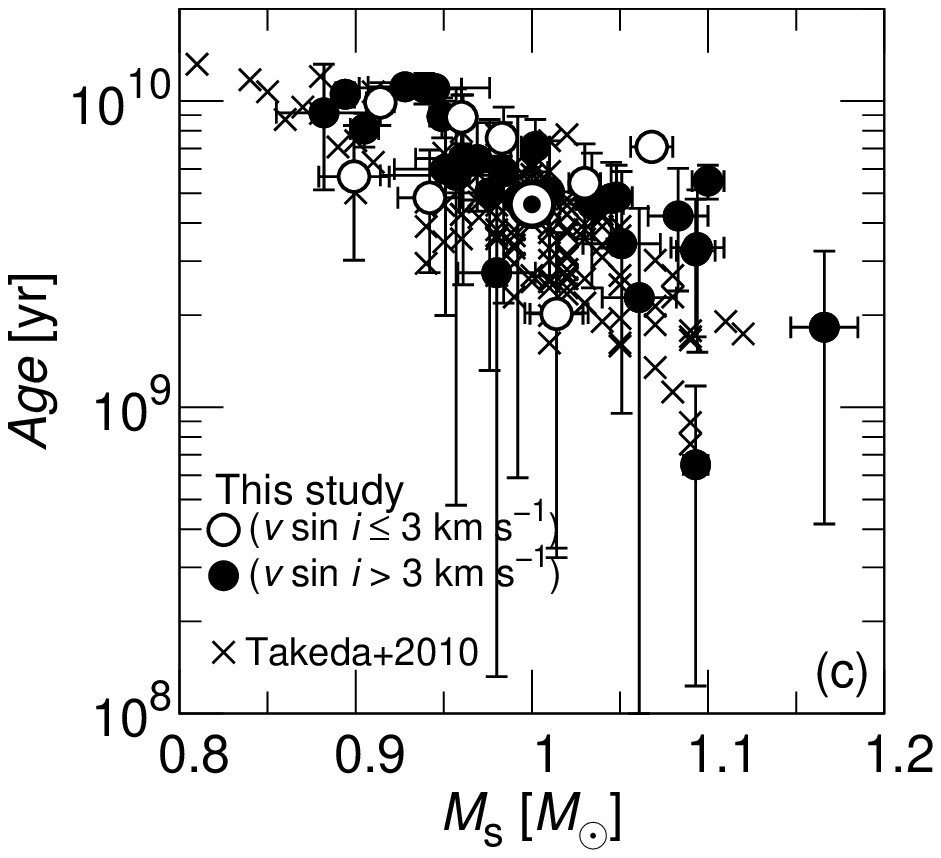}
  \FigureFile(82mm,82mm){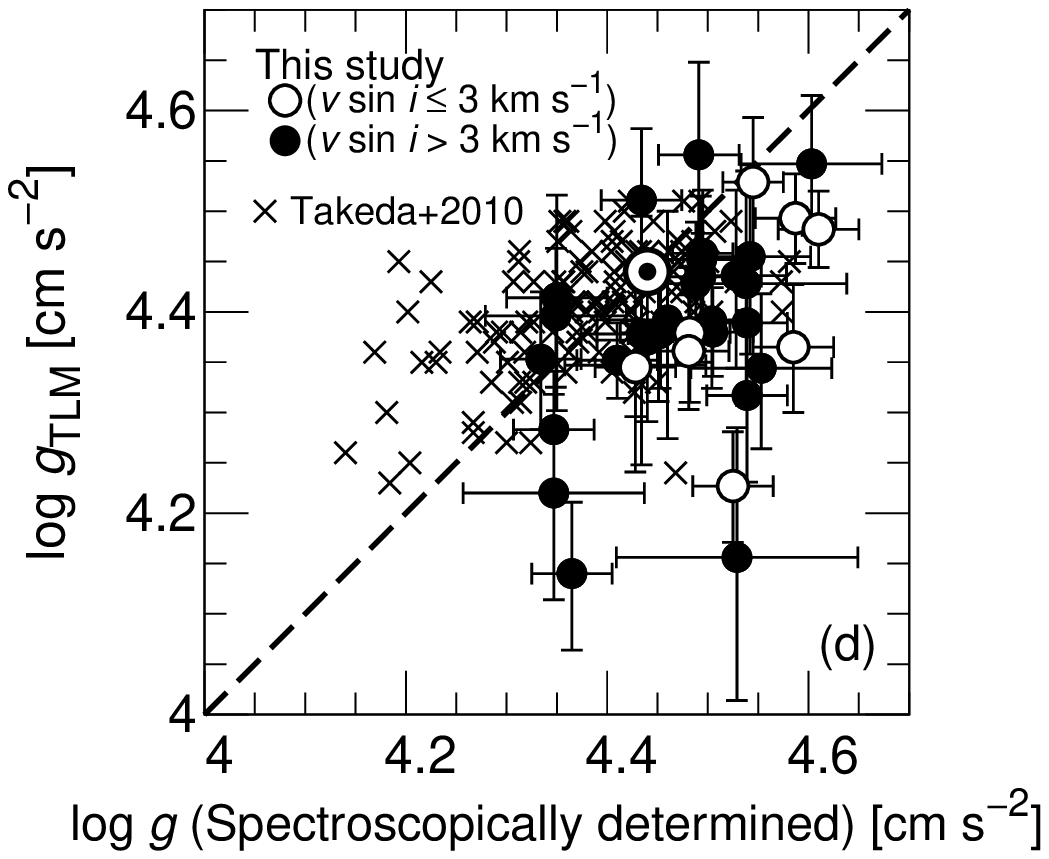}
\end{center}
\caption{(a) The 37 target single stars (open and filled circles) plotted on the theoretical HR diagram 
(absolute stellar luminosity $\log (L/L_{\odot})$ vs. temperature $T_{\rm{eff}}$) 
with the theoretical evolutionary tracks (lines) deduced from PARSEC isochrones of \citet{Bressan2012}.
Target stars with $v\sin i\leq$3 km s$^{-1}$ and with $v\sin i>$3 km s$^{-1}$are 
shown by using open circles and filled circles, respectively.
The theoretical tracks are drawn for 10 mass values ($M_{\rm{s}}$) ranging from 0.8 to 1.7 $M_{\odot}$ 
with a step of 0.1 $M_{\odot}$ for three different metallicities: $Z=0.0096$ ([Fe/H] $\approx -0.2$, dashed line),
$Z=0.0152$ ([Fe/H] $\approx 0.0$, solid line), $Z=0.0241$ ([Fe/H] $\approx +0.2$, dash dotted line).
The values of ordinary solar-analog stars investigated by \citet{Takeda2010} (cross marks) and 
that of the Sun (the circled dot point) are also plotted for reference.
\\
(b) Enlarged figure of (a).
\\
(c) Relation between the stellar mass ($M_{\rm{s}}$) and age ($Age$) derived from the evolutionary tracks. 
The classification of the data points are the same as in (a) and (b).
\\
(d) Comparison of the spectroscopically determined surface gravities ($\log g$) 
with those derived from stellar mass ($M_{\rm{s}}$) and radius ($R_{\rm{s}}$) by using Equation (\ref{eq:loggTLM}).
The classification of the data points are the same as in (a), (b) and (c).
\\
}\label{fig:HR}
\end{figure}

\ \ \ \ \ \ \ 
Using the atmospheric parameters ($T_{\rm{eff}}$, $\log g$, and [Fe/H]) estimated in Appendix \ref{subsec-apen:ana-atmos-para} 
and stellar distances from Hipparcos Catalogue ($d_{\rm{HIP}}$ in Table \ref{table:basic-data}),
we then determine the stellar positions of the target stars 
on the HR diagram in Appendix \ref{subsec-apen:ana-luminosity-radius},
and compare them with the theoretical evolutionary tracks in Appendix \ref{subsec-apen:ana-age-mass}.
The method here is basically on the basis of \authorcite{Takeda2002b} (\yearcite{Takeda2002b} \& \yearcite{Takeda2007}), 
though the estimation method of the interstellar extinction and the used theoretical evolutionary tracks are different.
\\ \\
\ \ \ \ \ \ \ 
At the first step, we estimated the interstellar extinction ($A_{V}$) 
by comparing the photometric $V-K$ color index ($(V-K)_{\rm{pho}}$) with that estimated from the atmospheric parameters ($(V-K)_{\rm{spec}}$) \footnote{
We did not use $B-V$ color here since the relation among $V-K$, $T_{\rm{eff}}$, and [Fe/H] 
(e.g., The relation among $V-K$, $T_{\rm{eff}}$, and [Fe/H] has lower dispersion values 
compared with that among $B-V$, $T_{\rm{eff}}$, and [Fe/H] \citep{Alonso1996}. 
}.
$(V-K)_{\rm{pho}}$ is calculated with $V$ and $K$ magnitude listed in Table \ref{table:basic-data}.
The former ($V$ magnitude) is originally taken from the Hipparcos Catalogue \citep{ESA1997}, and the latter ($K$ magnitude) is from the 2MASS All-Sky Catalog of Point Sources \citep{Cutri2003}.
For deriving $(V-K)_{\rm{spec}}$ values, we used the empirical relations of low-mass main-sequence stars (F0V$\sim$K5V) 
among $V-K$, $T_{\rm{eff}}$, and [Fe/H] shown in \citet{Alonso1996}.
By calculating the difference between $(V-K)_{\rm{pho}}$ and $(V-K)_{\rm{spec}}$, 
we can estimate how $V-K$ color values are affected by the interstellar extinction,
since $(V-K)_{\rm{spec}}$ is free from the interstellar reddening.
Then we calculated the color excess $E(V-K)$ by using these two color values:
\begin{eqnarray}\label{eq:E(V-K)}
E(V-K)=(V-K)_{\rm{pho}}-(V-K)_{\rm{spec}} \ .
\end{eqnarray}
The effect of interstellar reddening on $K$ magnitude value is small compared with 
that on $V$ magnitude (e.g., \cite{Cardelli1989}). 
Because of this, we here assume that interstellar extinction of $V$ magnitude $A_{V}$
is roughly equal to the color excess $E(V-K)$:
\begin{eqnarray}\label{eq:Av=E(V-K)}
A_{V}\approx E(V-K)=(V-K)_{\rm{pho}}-(V-K)_{\rm{spec}} \ .
\end{eqnarray}
The resultant $A_{V}$ values of the target stars and the comparison stars are listed in Tables \ref{table:atmos} and \ref{table:comp-stpara}, 
respectively.
\\
\\
\ \ \ \ \ \ \
We then derived the revised absolute $V$ magnitude ($M_{V}^{\rm{rev}}$) from the absolute $V$ magnitude ($M_{V}$) 
estimated in Section \ref{sec:target-and-obs} and $A_{V}$ estimated here:
\begin{eqnarray}\label{eq:Mv-rev}
M_{V}^{\rm{rev}} = M_{V} - A_{V} \ .
\end{eqnarray}
Applying the bolometric correction (B.C.) evaluated from 
Table 4 of \citet{Alonso1995} (in terms of $T_{\rm{eff}}$, $\log g$, and [Fe/H] listed in Tables \ref{table:atmos} and \ref{table:comp-stpara})
to this $M_{V}^{\rm{rev}}$, 
we obtained the absolute bolometric magnitude ($M_{\rm{bol}}$) and thus the absolute stellar luminosity ($L$). 
The resultant values of stellar luminosity are listed as $\log (L/L_{\odot})$ in Tables \ref{table:atmos} and \ref{table:comp-stpara}.
The error values of $\log (L/L_{\odot})$ listed in Tables \ref{table:atmos} and \ref{table:comp-stpara} correspond to errors of $M_{V}$,
which are originally caused by the errors of stellar distance ($d_{\rm{HIP}}$) in Tables \ref{table:basic-data} and \ref{table:comp-stpara}. 
The position of each target star on the $\log (L/L_{\odot})$ vs. $\log T_{\rm{eff}}$ diagram (HR diagram) 
is shown in Figure \ref{fig:HR} (a) \& (b).
\\
\\
\ \ \ \ \ \ \
Using $L$ and $T_{\rm{eff}}$, we estimated stellar radius ($R_{\rm{s}}$) of each star:
\begin{eqnarray}\label{eq:Rs}
\frac{R_{\rm{s}}}{R_{\odot}}= \biggl(\frac{L}{L_{\odot}}\biggl)^{1/2}\bigg/\biggl(\frac{T_{\rm{eff}}}{T_{\rm{eff}, \odot}}\biggl)^{2} \ .
\end{eqnarray}
The resultant values of $R_{\rm{s}}$ are listed in Tables \ref{table:atmos} and \ref{table:comp-stpara}.
The error values of $R_{\rm{s}}$ in these tables are estimated by considering the error values of $L$ and $T_{\rm{eff}}$.

\subsection{Age and mass}\label{subsec-apen:ana-age-mass}
\ \ \ \ \ \ \
In the above sections, we established $T_{\rm{eff}}$, $L$, the metallicity ($Z\equiv 0.152\times 10^{\rm{[Fe/H]}}$, where $Z_{\odot}=0.152$; \cite{Bressan2012}).
We then derived the stellar mass ($M_{\rm{s}}$) and stellar age ($Age$) by comparing the position 
on the HR diagram ($\log L$ vs. $T_{\rm{eff}}$ diagram) 
with the theoretical evolutionary tracks for various mass values and the corresponding metallicity (cf. Figure \ref{fig:HR} (a)\&(b)).
We here used the theoretical evolutionary tracks deduced from PARSEC isochrones \citep{Bressan2012}.
In this process, we selected all the data points that had possible sets of $L$, $T_{\rm{eff}}$, and $Z$ from PARSEC isochrones, 
taking into account the error values of $L$ and $T_{\rm{eff}}$ ($\Delta L$ and $\Delta T_{\rm{eff}}$, respectively) 
shown in Tables \ref{table:atmos} and \ref{table:comp-stpara}.
There were two stars (HIP20616 and 61Vir) that have no suitable isochrones within their original error range of $L$ and $T_{\rm{eff}}$.
For these two stars, we then took into account 2$\Delta L$ and 2$\Delta T_{\rm{eff}}$.
From the selected data points, we now have possible sets of $M_{\rm{s}}$ and $Age$ for each star.
The resultant values of $M_{\rm{s}}$ and $Age$, which we list in Tables \ref{table:atmos} and \ref{table:comp-stpara}, 
were calculated by taking medians between the maximum and minimum values among all the possible  $M_{\rm{s}}$ and $Age$ values 
selected from the isochrone data, respectively.
The error values of $M_{\rm{s}}$ and $Age$ in Tables \ref{table:atmos} and \ref{table:comp-stpara} correspond to these maximum and minimum values. 
The values of $M_{\rm{s}}$ and $Age$ of the 37 single target stars are also plotted in Figure \ref{fig:HR} (c). 
From this figure, we can see that $M_{\rm{s}}$ and $Age$ of the 37 target single stars are roughly anti-correlated with each other,
which are also seen for the ordinary solar-analog stars reported in \citet{Takeda2010}.
\\ \\
\ \ \ \ \ \ \
By using the $M_{\rm{s}}$ and $R_{\rm{s}}$ estimated above, 
we can compute the theoretical surface gravity $\log g_{TLM}$ as done in \citet{Takeda2007}:
\begin{eqnarray}\label{eq:loggTLM}
\log\biggl(\frac{g}{g_{\odot}}\biggl)_{TLM} &=& \log\biggl(\frac{M_{\rm{s}}}{M_{\odot}}\biggl) - 2\log\biggl(\frac{R_{\rm{s}}}{R_{\odot}}\biggl) \nonumber \\ 
&=& \log\biggl(\frac{M_{\rm{s}}}{M_{\odot}}\biggl) - \log\biggl(\frac{L}{L_{\odot}}\biggl) + 4\log\biggl(\frac{T_{\rm{eff}}}{T_{\rm{eff}, \odot}}\biggl) \ .
\end{eqnarray}
The resultant $\log g_{TLM}$ values of the target stars and
the comparison stars are listed in Tables \ref{table:atmos} and \ref{table:comp-stpara}, respectively. 
The error values of $\log g_{TLM}$ in these tables are estimated by considering the error values of $M_{\rm{s}}$ and $R_{\rm{s}}$.
Figure \ref{fig:HR} (d) shows a comparison of such evaluated $\log g_{TLM}$ 
with our spectroscopic surface gravity $\log g$ derived from Fe I/II lines.
We see that some of the spectroscopically determined $\log g$ can be somewhat overestimated by 0.1$\sim$0.2 dex, compared with $\log g_{TLM}$.
We should remember this error in the discussions of this paper, 
though this error is so small that it does not cause the essential problems for the following discussions of this paper 
(e.g., discussing whether the target stars are main sequence stars or not).

\subsection{X-ray Luminosity}\label{subsec-apen:ana-Xray}
\ \ \ \ \ \ \ 
In this section, we estimated X-ray luminosity ($L_{X}$) of the 37 single target stars from the ROSAT X-ray count rate and the stellar distance.
We evaluated the X-ray luminosity without correction for the effect of interstellar absorption of X-rays ($L_{X, 0}$).
We used the relation used by \citet{Fleming1995} and {\citet{Frohlich2012}: 
\begin{eqnarray}\label{eq:Lx-obs}
L_{X, 0} = 4\pi d_{\rm{HIP}}(8.31+5.30HR1) \times 10^{-12} X_{\rm{count}}~~[\rm{erg~s}^{-1}] \ ,
\end{eqnarray}
where $d_{\rm{HIP}}$ is stellar distance listed in Table \ref{table:basic-data}, derived from Hipparcos parallax.
$X_{\rm{count}}$ and $HR1$ are the X-ray count rate and hardness ratio 
taken from ROSAT All-Sky Survey Catalogues (\cite{Voges1999} \& \yearcite{Voges2000}), respectively, 
and $X_{\rm{count}}$ value of each star is listed in Table \ref{table:activity}.
\\ \\
\ \ \ \ \ \ \
We then estimated the effect of interstellar extinction and corrected the values of X-ray luminosity, with the following ways.
\citet{Bohlin1978} showed an empirical relation between total neutral hydrogen density ($N_{\rm{H}}$) and color excess $E(B-V)$:
\begin{eqnarray}\label{eq:NH-E(B-V)}
\frac{N_{\rm{H}}}{E(B-V)} = 5.8 \times 10^{21}~~[\rm{atoms}~\rm{cm}^{-2}~\rm{mag}^{-1}] \ .
\end{eqnarray}
According to \citet{Cardelli1989}, there is an empirical relation between $E(B-V)$ and $A_{V}$: $A_{V}/E(B-V) \sim 3.1$.
Using this relation and Equation (\ref{eq:NH-E(B-V)}), 
we estimated $N_{\rm{H}}$ of the target stars from $A_{V}$ listed in Table \ref{table:basic-data},
and listed the resultant values of $N_{\rm{H}}$ in Table \ref{table:activity}.
According to \citet{Wilms2000}, 
the final X-ray luminosity $L_{X}$ 
with correction for the effect of interstellar absorption can be estimated from $L_{X, 0}$ and $N_{\rm{H}}$
by the following relation:
\begin{eqnarray}\label{eq:Lx-Lxobs}
L_{X} = \exp (-\sigma N_{\rm{H}}) \times L_{X, 0} \ ,
\end{eqnarray}
where $\sigma$ is the photoionization cross section of the interstellar mediam (ISM) per hydrogen atom.
In the most range of the energy band of ROSAT (0.1$\sim$2.4keV),
$\sigma$ is $2\sim 4 \times 10^{-22}$ cm$^{2}$,
on the basis of Figure 1 of \citet{Wilms2000}.
We here assume $\sigma = 3\times 10^{-22}$ cm$^{2}$, 
and finally derived $L_{X}$ from $L_{X, 0}$, $\sigma$, and $N_{\rm{H}}$ with Equation (\ref{eq:Lx-Lxobs}).
\\ \\
\ \ \ \ \ \ \
The resultant value of $L_{X}$ are listed in in Table \ref{table:activity}.
As a final error value of $L_{X}$ listed in this table, 
we used a root sum square of the error values arising from three different error sources: 
$X_{\rm{count}}$, $HR1$, and $d_{\rm{HIP}}$.
Figure \ref{fig:LxVK} shows a comparison of $L_{X}$ and $L_{X, 0}$. 
We can see that as for the X-ray luminosities in this paper, 
the effects of interstellar absorption is much smaller than the other type of observational errors 
(errors arising from the stellar distance ($d_{\rm{HIP}}$), ROSAT X-ray count ($X_{\rm{count}}$), and hardness ratio ($HR1$)).
\\

\begin{figure}[htbp]
 \begin{center}
   \FigureFile(88mm,88mm){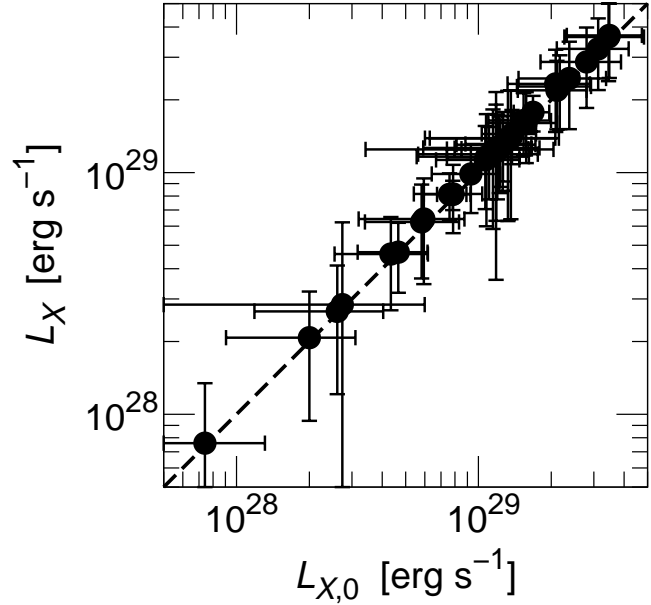} 
\end{center}
\caption{Comparizon of $L_{X}$ and $L_{X, 0}$. $L_{X}$ is the resultant X-ray luminosity with correction for the effect of interstellar absorption, 
while $L_{X, 0}$ is that without such correction.}\label{fig:LxVK}
\end{figure}

\subsection{Comparison of our estimated stellar parameters with the previous studies}\label{subsec-apen:para-comparison}
\ \ \ \ \ \ \
In Appendix \ref{subsec-apen:ana-atmos-para}, \ref{subsec-apen:ana-vsini}, and \ref{subsec-apen:ana-Li}, 
we estimated the stellar parameters such as $T_{\rm{eff}}$, $\log g$, [Fe/H], $v\sin i$, and $A$(Li) of the 37 target stars,
which are considered as single stars in Appendix \ref{subsec-apen:ana-binarity}, 
and the resultant values are in Tables \ref{table:atmos} and \ref{table:activity}.
We also estimated these values of 6 comparison stars and the Moon (Table \ref{table:comp-stpara}).
In the following of this subsection, we performed some analyses in order to check 
whether these spectroscopically derived values are good sources with which to discuss the actual properties of stars.
\\

\begin{figure}[htbp]
 \begin{center}
  \FigureFile(75mm,75mm){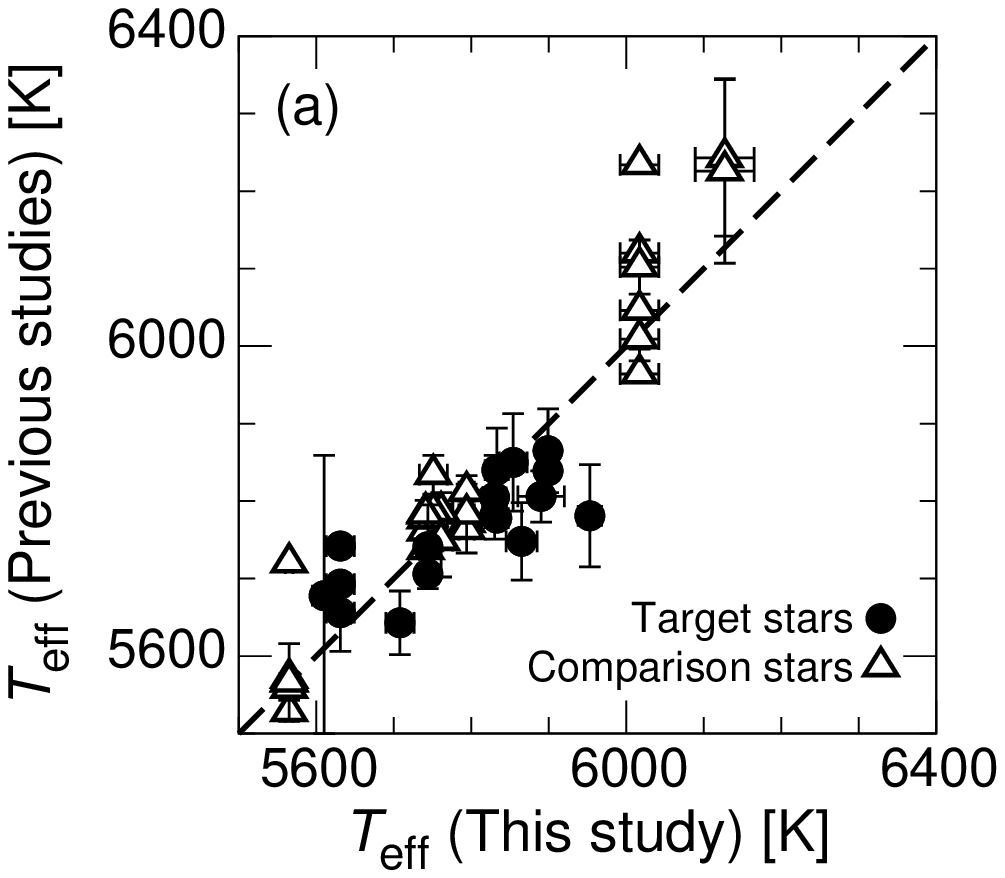}  
  \FigureFile(75mm,75mm){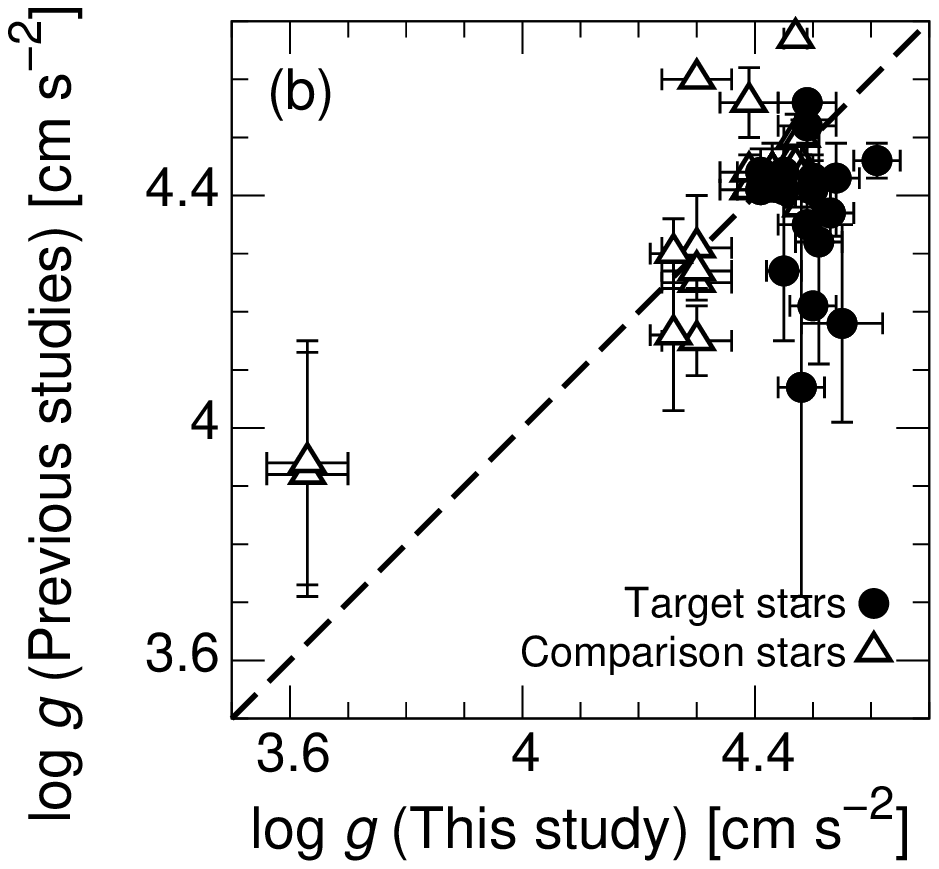}  
  \FigureFile(75mm,75mm){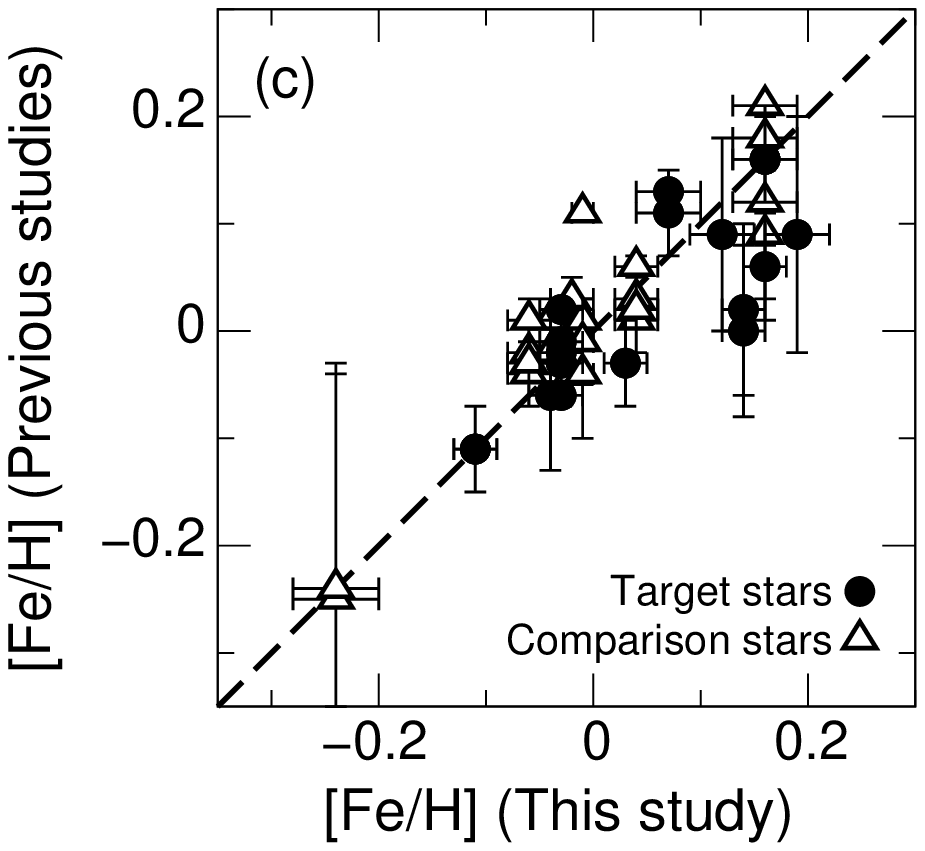}  
  \FigureFile(75mm,75mm){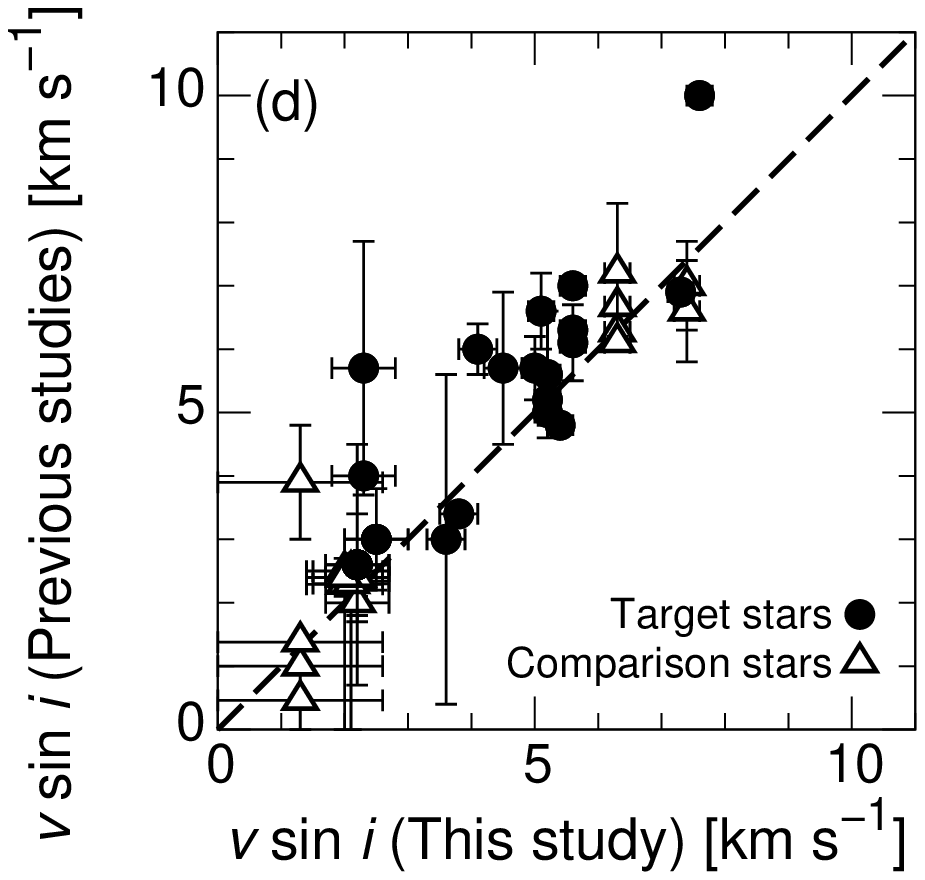}  
  \FigureFile(75mm,75mm){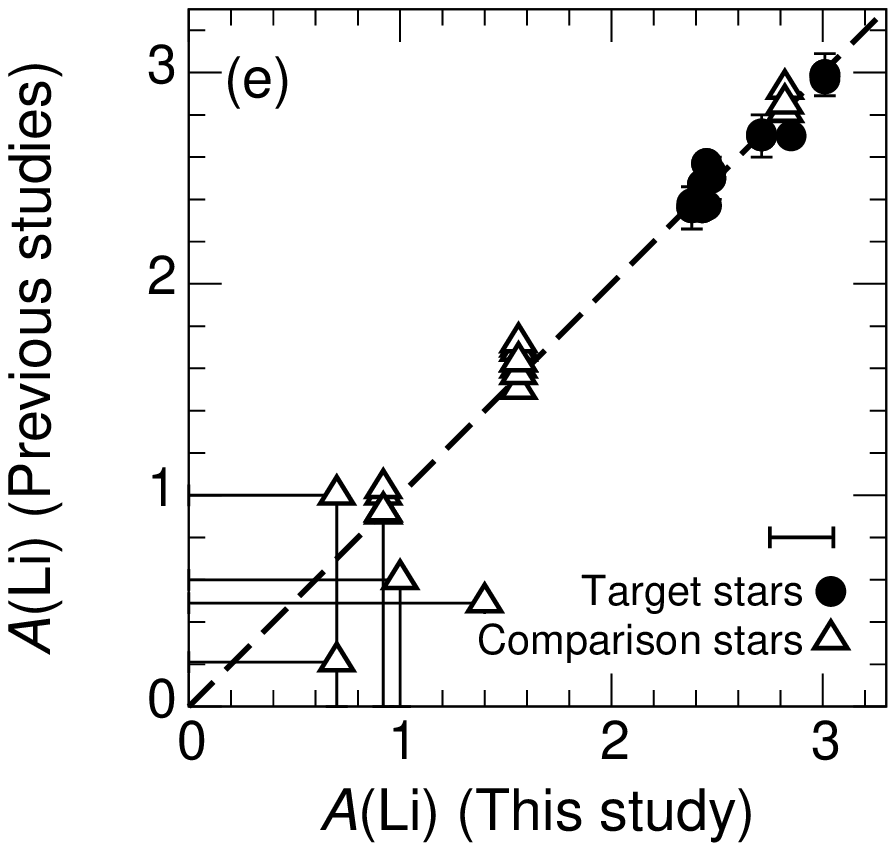}  
 \end{center}
\caption{
Comparison of the stellar parameters ($T_{\rm{eff}}$, $\log g$, [Fe/H], $v\sin i$, and $A$(Li)) in this paper 
with those reported by the previous researches. 
The values are listed in Supplementary Tables 4 \& 5. 
Filled circles correspond to the values of the target stars on the basis of Supplementary Table 4, 
while open triangles show those of comparison stars taken from Supplementary Table 5.
In (e), typical error value of $A$(Li) ($\sim$0.15 dex) mentioned in Appendix \ref{subsec-apen:ana-Li} is shown 
with the error bar at the lower right side.
}\label{fig:vsini-comp}
\end{figure}

\noindent
\ \ \ \ \ \ \
First, we compared these parameters that we estimated above with those reported by the previous studies.
In Figure \ref{fig:vsini-comp}, we compare our $T_{\rm{eff}}$, $\log g$, [Fe/H], $v\sin i$, and $A$(Li) 
of the target stars and comparison stars including the Moon 
with those reported by the previous studies.
The values of the target stars and comparison stars shown in Figure \ref{fig:vsini-comp} are listed in 
Supplementary Tables 4 \& 5 \footnote{
In Supplementary Tables 4 \& 5, we used the stellar parameters reported by the following studies: 
\citet{Ammler-vonEiff2012}, 
\citet{Anderson2010}, 
\citet{Asplund2009}, 
\citet{Cutispoto2002},
\citet{Guillout2009},
\citet{Herrero2012},
\citet{Honda2015},
\citet{Jenkins2011},
\citet{King2005},
\citet{Mermilliod2009},
\citet{Mishenina2012},
\citet{Molenda-Zakowic2013},
\citet{Nordstrom2004},
\citet{SNotsu2013},
\citet{YNotsu2015a},
\citet{Ramirez2012},
\citet{TakedaTajitsu2009},
\citet{Takeda2005},
\citet{Takeda2010},
\citet{Torres2006}, and
\citet{Valenti2005}.
}, respectively. 
Our values seem to be in good agreement with the previous values especially for
$T_{\rm{eff}}$, [Fe/H], $v\sin i$, and $A$(Li) values.
$\log g$ in Figure \ref{fig:vsini-comp} (b) have some dispersion, 
and similar order of dispersions can be seen in Figure \ref{fig:HR} (d),
where we compared our spectroscopic $\log g$ with $\log g_{TLM}$. 
We should remember this error in the discussions of this paper, 
though this dispersion is so small that it does not cause the essential problems for the following discussions of this paper 
(e.g., discussing whether the target stars are main sequence stars or not).
\\
\\
\ \ \ \ \ \ \
Next, we compared our $v\sin i$ values with the value of rotation period 
in GCVS (General Catalogue of Variable Stars) database \citep{Samus2015}. 
In this database, four target stars (HIP19793, HIP38228, HIP105066, and  HIP115527) 
are identified as BY Dra type variable stars, and among them, 
rotation period ($P_{\rm{rot}}$) of three stars (HIP19793, HIP38228, and HIP105066) 
are reported (See Table \ref{table:period-BYDra}).
Using this $P_{\rm{rot}}$ and stellar radius $R_{\rm{s}}$ estimated in Appendix \ref{subsec-apen:ana-luminosity-radius},
we can estimate the rotational velocity ($v_{\rm{lc}}$):
\begin{equation}\label{eq:vlc}
v_{\rm{lc}}=\frac{2\pi R_{\rm{s}}}{P_{\rm{rot}}} \ ,
\end{equation}
as in Equation (2) of \citet{YNotsu2015b}. 
Comparing $v\sin i$ and $v_{\rm{lc}}$ in Table \ref{table:period-BYDra}, 
the relation ``$v \sin i\lesssim v_{\rm{lc}}$" is satisfied for at least these three stars.
This suggests our $v\sin i$ values are consistent with the photometric data, 
though only three stars are investigated.
\\
\\
\ \ \ \ \ \ \
On the basis of the discussions above, 
we can confirm that our spectroscopically derived values are good sources to discuss the stellar properties.
We only use the values that we derived spectroscopically in the above when we discuss stellar properties in Section \ref{sec:discussion}. 
However, we need to remember that $v\sin i$ values estimated in Appendix \ref{subsec-apen:ana-vsini} 
can be strongly affected by the way of estimating macroturbulence ($\zeta_{\rm{RT}}$ in Appendix \ref{subsec-apen:ana-vsini}), 
as we have also mentioned in Section 4.3 of \citet{YNotsu2015a}. 
We assumed Equation (\ref{eq:macroT}) in this paper, but this equation is an only rough approximation.
The effect of errors in macroturbulence is especially large for slowly-rotating stars ($v\sin i\lesssim2-3$ km s$^{-1}$),
as we can see in Figure \ref{fig:vsini-vmac}.
In this figure, the difference between the two $v \sin i$ value, which originally comes from 
the difference in the estimation method of macroturbulence velocity 
is a bit larger especially in the range of $v\sin i\lesssim2-3$ km s$^{-1}$.
More accurate estimation of macroturbulence is difficult (e.g., \cite{Takeda1995b}), 
and we think that this is beyond the scope of this paper.
Because of this, we need to remember that the uncertainty of $v\sin i$ is large 
especially for slowly-rotating stars ($v\sin i\lesssim2-3$ km s$^{-1}$).

\section{Tables}\label{sec-apen:tables}

\begin{center}
\begin{table}[htbp]
\rotatebox{-90}{
\begin{minipage}{\textheight}
  \caption{Basic data of the target stars.}\label{table:basic-data}{%
  \begin{tabular}{lcccccccccccc}
      \hline
Name(HIP) \footnotemark[a] & Name(HD) \footnotemark[a] & ROSAT-ID \footnotemark[b] 
& $\theta_{\rm{ROSAT-HIP}}$ \footnotemark[c] & Pos. Err \footnotemark[d] 
& $V$ \footnotemark[a] 
& $ (B-V)_{\rm{HIP}}$ \footnotemark[a] & $K$\footnotemark[e] 
& $\pi$ \footnotemark[a] & $d_{\rm{HIP}}$ \footnotemark[f] & $M_{V}$ \footnotemark[g] & binary \footnotemark[h] & Remarks \footnotemark[i]  \\ 
 & & & [arcsec] & [arcsec] & [mag] & [mag] & [mag] 
& [mas] & [pc] & [mag] & & \\   
\hline
\footnotesize{HIP3342} & \footnotesize{HD4021} & \footnotesize{1RXS-J004236.9-164214} & \footnotesize{12.5} & \footnotesize{11} & \footnotesize{8.87} & \footnotesize{$0.702\pm 0.021$} & \footnotesize{$7.17\pm 0.02$} & \footnotesize{$18.19\pm 1.45$}  & \footnotesize{$55.0_{-4.1}^{+4.8}$} & \footnotesize{$5.17_{-0.18}^{+0.17}$} & \footnotesize{no} & \scriptsize{(1)} \\ 
\footnotesize{HIP4909} & \footnotesize{HD6204} & \footnotesize{1RXS-J010257.2-095145} & \footnotesize{5.6} & \footnotesize{18} & \footnotesize{8.52} & \footnotesize{$0.649\pm 0.024$} & \footnotesize{$6.96\pm 0.02$} & \footnotesize{$18.29\pm 1.35$}  & \footnotesize{$54.7_{-3.8}^{+4.4}$} & \footnotesize{$4.83_{-0.17}^{+0.15}$} & \footnotesize{no} & \scriptsize{-} \\ 
\footnotesize{HIP8522} & \footnotesize{HD11120} & \footnotesize{1RXS-J014955.2+254443} & \footnotesize{4.0} & \footnotesize{25} & \footnotesize{8.56} & \footnotesize{$0.648\pm 0.023$} & \footnotesize{$6.94\pm 0.02$} & \footnotesize{$19.04\pm 1.21$}  & \footnotesize{$52.5_{-3.1}^{+3.6}$} & \footnotesize{$4.96_{-0.14}^{+0.13}$} & \footnotesize{no} & \scriptsize{(2)} \\ 
\footnotesize{HIP9519} & \footnotesize{HD12484} & \footnotesize{1RXS-J020227.1+024902} & \footnotesize{7.8} & \footnotesize{11} & \footnotesize{8.18} & \footnotesize{$0.649\pm 0.020$} & \footnotesize{$6.71\pm 0.03$} & \footnotesize{$21.24\pm 1.28$}  & \footnotesize{$47.1_{-2.7}^{+3.0}$} & \footnotesize{$4.82_{-0.13}^{+0.13}$} & \footnotesize{no} & \scriptsize{-} \\ 
\footnotesize{HIP14997} & \footnotesize{-} & \footnotesize{1RXS-J031318.0+113308} & \footnotesize{12.1} & \footnotesize{30} & \footnotesize{9.18} & \footnotesize{$0.712\pm 0.034$} & \footnotesize{$7.62\pm 0.03$} & \footnotesize{$16.30\pm 1.57$}  & \footnotesize{$61.3_{-5.4}^{+6.5}$} & \footnotesize{$5.24_{-0.22}^{+0.20}$} & \footnotesize{no} & \scriptsize{-} \\ 
\footnotesize{HIP19793} & \footnotesize{HD26736} & \footnotesize{1RXS-J041432.2+233448} & \footnotesize{18.3} & \footnotesize{12} & \footnotesize{8.05} & \footnotesize{$0.657\pm 0.007$} & \footnotesize{$6.52\pm 0.02$} & \footnotesize{$21.69\pm 1.14$}  & \footnotesize{$46.1_{-2.3}^{+2.6}$} & \footnotesize{$4.73_{-0.12}^{+0.11}$} & \footnotesize{no} & \scriptsize{(3)} \\ 
\footnotesize{HIP20616} & \footnotesize{HD27947} & \footnotesize{1RXS-J042500.6+115834} & \footnotesize{7.7} & \footnotesize{12} & \footnotesize{8.41} & \footnotesize{$0.639\pm 0.015$} & \footnotesize{$6.93\pm 0.02$} & \footnotesize{$21.00\pm 1.37$}  & \footnotesize{$47.6_{-2.9}^{+3.3}$} & \footnotesize{$5.02_{-0.15}^{+0.14}$} & \footnotesize{no} & \scriptsize{(2),(4)} \\ 
\footnotesize{HIP21091} & \footnotesize{-} & \footnotesize{1RXS-J043111.0+111428} & \footnotesize{11.4} & \footnotesize{12} & \footnotesize{8.70} & \footnotesize{$0.665\pm 0.021$} & \footnotesize{$7.24\pm 0.02$} & \footnotesize{$14.85\pm 1.27$}  & \footnotesize{$67.3_{-5.3}^{+6.3}$} & \footnotesize{$4.56_{-0.19}^{+0.18}$} & \footnotesize{no} & \scriptsize{(2)} \\ 
\footnotesize{HIP22175} & \footnotesize{HD30286} & \footnotesize{1RXS-J044616.2+031624} & \footnotesize{17.5} & \footnotesize{20} & \footnotesize{7.81} & \footnotesize{$0.679\pm 0.015$} & \footnotesize{$6.21\pm 0.02$} & \footnotesize{$31.11\pm 1.07$}  & \footnotesize{$32.1_{-1.1}^{+1.1}$} & \footnotesize{$5.27_{-0.08}^{+0.07}$} & \footnotesize{no} & \scriptsize{(2)} \\ 
\footnotesize{HIP23027} & \footnotesize{HD31464} & \footnotesize{1RXS-J045705.3+244504} & \footnotesize{15.8} & \footnotesize{11} & \footnotesize{8.60} & \footnotesize{$0.714\pm 0.026$} & \footnotesize{$6.88\pm 0.02$} & \footnotesize{$22.09\pm 1.58$}  & \footnotesize{$45.3_{-3.0}^{+3.5}$} & \footnotesize{$5.32_{-0.16}^{+0.15}$} & \footnotesize{no} & \scriptsize{(2)} \\ 
\footnotesize{HIP25002} & \footnotesize{HD35041} & \footnotesize{1RXS-J052113.0-140854} & \footnotesize{14.4} & \footnotesize{18} & \footnotesize{7.68} & \footnotesize{$0.636\pm 0.015$} & \footnotesize{$6.17\pm 0.03$} & \footnotesize{$26.89\pm 0.96$}  & \footnotesize{$37.2_{-1.3}^{+1.4}$} & \footnotesize{$4.83_{-0.08}^{+0.08}$} & \footnotesize{no} & \scriptsize{-} \\ 
\footnotesize{HIP27980} & \footnotesize{HD39833} & \footnotesize{1RXS-J055501.9-003016} & \footnotesize{12.2} & \footnotesize{16} & \footnotesize{7.65} & \footnotesize{$0.629\pm 0.015$} & \footnotesize{$6.15\pm 0.02$} & \footnotesize{$21.41\pm 1.10$}  & \footnotesize{$46.7_{-2.3}^{+2.5}$} & \footnotesize{$4.30_{-0.11}^{+0.11}$} & \footnotesize{no} & \scriptsize{(2)} \\ 
\footnotesize{HIP35185} & \footnotesize{HD56202} & \footnotesize{1RXS-J071619.2+050443} & \footnotesize{13.6} & \footnotesize{11} & \footnotesize{8.42} & \footnotesize{$0.641\pm 0.019$} & \footnotesize{$6.93\pm 0.03$} & \footnotesize{$18.99\pm 1.18$}  & \footnotesize{$52.7_{-3.1}^{+3.5}$} & \footnotesize{$4.81_{-0.14}^{+0.13}$} & \footnotesize{no} & \scriptsize{-} \\ 
\footnotesize{HIP37971} & \footnotesize{HD62857} & \footnotesize{1RXS-J074658.6+260142} & \footnotesize{12.9} & \footnotesize{14} & \footnotesize{8.51} & \footnotesize{$0.687\pm 0.005$} & \footnotesize{$6.94\pm 0.02$} & \footnotesize{$18.86\pm 1.04$}  & \footnotesize{$53.0_{-2.8}^{+3.1}$} & \footnotesize{$4.89_{-0.12}^{+0.12}$} & \footnotesize{no} & \scriptsize{(2)} \\ 
\footnotesize{HIP38228} & \footnotesize{HD63433} & \footnotesize{1RXS-J074955.3+272152} & \footnotesize{5.5} & \footnotesize{8} & \footnotesize{6.90} & \footnotesize{$0.682\pm 0.000$} & \footnotesize{$5.26\pm 0.02$} & \footnotesize{$45.84\pm 0.89$}  & \footnotesize{$21.8_{-0.4}^{+0.4}$} & \footnotesize{$5.21_{-0.04}^{+0.04}$} & \footnotesize{no} & \scriptsize{(2),(3)} \\ 
\footnotesize{HIP40380} & \footnotesize{-} & \footnotesize{1RXS-J081437.0+625609} & \footnotesize{18.4} & \footnotesize{19} & \footnotesize{9.34} & \footnotesize{$0.638\pm 0.026$} & \footnotesize{$7.81\pm 0.02$} & \footnotesize{$12.72\pm 1.34$}  & \footnotesize{$78.6_{-7.5}^{+9.3}$} & \footnotesize{$4.86_{-0.24}^{+0.22}$} & \footnotesize{no} & \scriptsize{-} \\ 
\footnotesize{HIP44657} & \footnotesize{HD78130} & \footnotesize{1RXS-J090558.3-213145} & \footnotesize{15.9} & \footnotesize{33} & \footnotesize{8.66} & \footnotesize{$0.674\pm 0.014$} & \footnotesize{$7.13\pm 0.02$} & \footnotesize{$16.54\pm 1.12$}  & \footnotesize{$60.5_{-3.8}^{+4.4}$} & \footnotesize{$4.75_{-0.15}^{+0.14}$} & \footnotesize{no} & \scriptsize{(2)} \\ 
\footnotesize{HIP51652} & \footnotesize{HD91332} & \footnotesize{1RXS-J103310.0+304519} & \footnotesize{4.7} & \footnotesize{14} & \footnotesize{8.25} & \footnotesize{$0.646\pm 0.013$} & \footnotesize{$6.77\pm 0.02$} & \footnotesize{$16.86\pm 0.99$}  & \footnotesize{$59.3_{-3.3}^{+3.7}$} & \footnotesize{$4.38_{-0.13}^{+0.12}$} & \footnotesize{no} & \scriptsize{(2)} \\ 
\footnotesize{HIP52761} & \footnotesize{HD93393} & \footnotesize{1RXS-J104719.4+204718} & \footnotesize{7.2} & \footnotesize{17} & \footnotesize{9.16} & \footnotesize{$0.616\pm 0.027$} & \footnotesize{$7.66\pm 0.02$} & \footnotesize{$13.58\pm 1.38$}  & \footnotesize{$73.6_{-6.8}^{+8.3}$} & \footnotesize{$4.82_{-0.23}^{+0.21}$} & \footnotesize{no} & \scriptsize{(2)} \\ 
\footnotesize{HIP59399} & \footnotesize{HD105863} & \footnotesize{1RXS-J121106.8+255933} & \footnotesize{11.5} & \footnotesize{18} & \footnotesize{9.54} & \footnotesize{$0.624\pm 0.002$} & \footnotesize{$8.07\pm 0.03$} & \footnotesize{$10.92\pm 1.32$}  & \footnotesize{$91.6_{-9.9}^{+12.6}$} & \footnotesize{$4.73_{-0.28}^{+0.25}$} & \footnotesize{no} & \scriptsize{(2),(4)} \\ 
\hline
    \end{tabular}
}    
\begin{tabnote}
\end{tabnote}
\end{minipage}
}
\end{table}
\end{center}

\begin{center}
\addtocounter{table}{-1}
\begin{table}[htbp]
\rotatebox{-90}{
\begin{minipage}{\textheight}
  \caption{(Continued.)}{%
  \begin{tabular}{lcccccccccccc}
      \hline
Name(HIP) \footnotemark[a] & Name(HD) \footnotemark[a] & ROSAT-ID \footnotemark[b] 
& $\theta_{\rm{ROSAT-HIP}}$ \footnotemark[c] & Pos. Err \footnotemark[d] 
& $V$ \footnotemark[a] 
& $ (B-V)_{\rm{HIP}}$ \footnotemark[a] & $K$\footnotemark[e] 
& $\pi$ \footnotemark[a] & $d_{\rm{HIP}}$ \footnotemark[f] & $M_{V}$ \footnotemark[g] & binary \footnotemark[h] & Remarks \footnotemark[i]  \\ 
 & & & [arcsec] & [arcsec] & [mag] & [mag] & [mag] 
& [mas] & [pc] & [mag] & & \\  
     \hline
\footnotesize{HIP65627} & \footnotesize{-} & \footnotesize{1RXS-J132718.3+464911} & \footnotesize{2.6} & \footnotesize{12} & \footnotesize{9.13} & \footnotesize{$0.640\pm 0.015$} & \footnotesize{$7.54\pm 0.02$} & \footnotesize{$12.06\pm 1.14$}  & \footnotesize{$82.9_{-7.2}^{+8.7}$} & \footnotesize{$4.54_{-0.22}^{+0.20}$} & \footnotesize{no} & \scriptsize{(2)} \\ 
\footnotesize{HIP70354} & \footnotesize{-} & \footnotesize{1RXS-J142341.3+652337} & \footnotesize{7.3} & \footnotesize{11} & \footnotesize{9.33} & \footnotesize{$0.640\pm 0.026$} & \footnotesize{$7.81\pm 0.02$} & \footnotesize{$14.42\pm 0.97$}  & \footnotesize{$69.3_{-4.4}^{+5.0}$} & \footnotesize{$5.12_{-0.15}^{+0.14}$} & \footnotesize{no} & \scriptsize{-} \\ 
\footnotesize{HIP70394} & \footnotesize{-} & \footnotesize{1RXS-J142407.7+290932} & \footnotesize{10.3} & \footnotesize{13} & \footnotesize{9.56} & \footnotesize{$0.635\pm 0.035$} & \footnotesize{$7.95\pm 0.02$} & \footnotesize{$12.55\pm 1.22$}  & \footnotesize{$79.7_{-7.1}^{+8.6}$} & \footnotesize{$5.05_{-0.22}^{+0.20}$} & \footnotesize{no} & \scriptsize{(2)} \\ 
\footnotesize{HIP71218} & \footnotesize{HD127913} & \footnotesize{1RXS-J143350.9+023430} & \footnotesize{9.0} & \footnotesize{37} & \footnotesize{9.14} & \footnotesize{$0.724\pm 0.037$} & \footnotesize{$7.43\pm 0.02$} & \footnotesize{$16.21\pm 1.35$}  & \footnotesize{$61.7_{-4.7}^{+5.6}$} & \footnotesize{$5.19_{-0.19}^{+0.17}$} & \footnotesize{no} & \scriptsize{-} \\ 
\footnotesize{HIP77584} & \footnotesize{HD141715} & \footnotesize{1RXS-J155026.6+014853} & \footnotesize{16.9} & \footnotesize{13} & \footnotesize{8.28} & \footnotesize{$0.623\pm 0.016$} & \footnotesize{$6.74\pm 0.02$} & \footnotesize{$15.69\pm 1.13$}  & \footnotesize{$63.7_{-4.3}^{+4.9}$} & \footnotesize{$4.26_{-0.16}^{+0.15}$} & \footnotesize{no} & \scriptsize{-} \\ 
\footnotesize{HIP79068} & \footnotesize{HD145224} & \footnotesize{1RXS-J160827.3+334356} & \footnotesize{5.3} & \footnotesize{17} & \footnotesize{8.37} & \footnotesize{$0.644\pm 0.013$} & \footnotesize{$6.87\pm 0.03$} & \footnotesize{$18.31\pm 0.87$}  & \footnotesize{$54.6_{-2.5}^{+2.7}$} & \footnotesize{$4.68_{-0.11}^{+0.10}$} & \footnotesize{no} & \scriptsize{(2)} \\ 
\footnotesize{HIP80271} & \footnotesize{-} & \footnotesize{1RXS-J162309.8+353512} & \footnotesize{9.9} & \footnotesize{12} & \footnotesize{9.33} & \footnotesize{$0.604\pm 0.025$} & \footnotesize{$7.97\pm 0.02$} & \footnotesize{$10.54\pm 1.04$}  & \footnotesize{$94.9_{-8.5}^{+10.4}$} & \footnotesize{$4.44_{-0.23}^{+0.20}$} & \footnotesize{no} & \scriptsize{(2)} \\ 
\footnotesize{HIP83507} & \footnotesize{HD154326} & \footnotesize{1RXS-J170359.2+205243} & \footnotesize{7.9} & \footnotesize{13} & \footnotesize{9.04} & \footnotesize{$0.653\pm 0.022$} & \footnotesize{$7.49\pm 0.02$} & \footnotesize{$11.41\pm 1.35$}  & \footnotesize{$87.6_{-9.3}^{+11.8}$} & \footnotesize{$4.33_{-0.27}^{+0.24}$} & \footnotesize{asymmetry?} & \scriptsize{-} \\ 
\footnotesize{HIP86245} & \footnotesize{HD160135} & \footnotesize{1RXS-J173727.2+222100} & \footnotesize{12.4} & \footnotesize{12} & \footnotesize{8.38} & \footnotesize{$0.623\pm 0.011$} & \footnotesize{$6.90\pm 0.02$} & \footnotesize{$17.27\pm 1.18$}  & \footnotesize{$57.9_{-3.7}^{+4.2}$} & \footnotesize{$4.57_{-0.15}^{+0.14}$} & \footnotesize{no} & \scriptsize{(2)} \\ 
\footnotesize{HIP86781} & \footnotesize{HD238744} & \footnotesize{1RXS-J174357.1+550931} & \footnotesize{10.2} & \footnotesize{12} & \footnotesize{9.06} & \footnotesize{$0.644\pm 0.024$} & \footnotesize{$7.71\pm 0.02$} & \footnotesize{$11.78\pm 0.73$}  & \footnotesize{$84.9_{-5.0}^{+5.6}$} & \footnotesize{$4.42_{-0.14}^{+0.13}$} & \footnotesize{no} & \scriptsize{(2)} \\ 
\footnotesize{HIP88572} & \footnotesize{HD165863} & \footnotesize{1RXS-J180506.2+532157} & \footnotesize{4.8} & \footnotesize{11} & \footnotesize{8.77} & \footnotesize{$0.725\pm 0.020$} & \footnotesize{$7.06\pm 0.02$} & \footnotesize{$18.46\pm 0.77$}  & \footnotesize{$54.2_{-2.2}^{+2.4}$} & \footnotesize{$5.10_{-0.09}^{+0.09}$} & \footnotesize{no} & \scriptsize{-} \\ 
\footnotesize{HIP91073} & \footnotesize{HD171489} & \footnotesize{1RXS-J183435.3+123221} & \footnotesize{5.4} & \footnotesize{16} & \footnotesize{8.28} & \footnotesize{$0.646\pm 0.012$} & \footnotesize{$6.85\pm 0.02$} & \footnotesize{$17.85\pm 1.17$}  & \footnotesize{$56.0_{-3.4}^{+3.9}$} & \footnotesize{$4.54_{-0.15}^{+0.14}$} & \footnotesize{no} & \scriptsize{-} \\ 
\footnotesize{HIP101893} & \footnotesize{HD340676} & \footnotesize{1RXS-J203855.8+264209} & \footnotesize{10.0} & \footnotesize{14} & \footnotesize{8.85} & \footnotesize{$0.680\pm 0.015$} & \footnotesize{$7.23\pm 0.02$} & \footnotesize{$19.06\pm 1.21$}  & \footnotesize{$52.5_{-3.1}^{+3.6}$} & \footnotesize{$5.25_{-0.14}^{+0.13}$} & \footnotesize{no} & \scriptsize{(2)} \\ 
\footnotesize{HIP105066} & \footnotesize{HD202605} & \footnotesize{1RXS-J211702.5-010439} & \footnotesize{5.5} & \footnotesize{18} & \footnotesize{8.08} & \footnotesize{$0.737\pm 0.015$} & \footnotesize{$6.43\pm 0.02$} & \footnotesize{$23.33\pm 1.32$}  & \footnotesize{$42.9_{-2.3}^{+2.6}$} & \footnotesize{$4.92_{-0.13}^{+0.12}$} & \footnotesize{no} & \scriptsize{(3)} \\ 
\footnotesize{HIP105855} & \footnotesize{HD204790} & \footnotesize{1RXS-J212623.1+734006} & \footnotesize{12.6} & \footnotesize{21} & \footnotesize{9.21} & \footnotesize{$0.595\pm 0.003$} & \footnotesize{$7.77\pm 0.02$} & \footnotesize{$11.51\pm 0.87$}  & \footnotesize{$86.9_{-6.1}^{+7.1}$} & \footnotesize{$4.52_{-0.17}^{+0.16}$} & \footnotesize{no} & \scriptsize{(2)} \\ 
\footnotesize{HIP115527} & \footnotesize{HD220476} & \footnotesize{1RXS-J232405.7-073305} & \footnotesize{9.8} & \footnotesize{9} & \footnotesize{7.62} & \footnotesize{$0.682\pm 0.002$} & \footnotesize{$5.99\pm 0.02$} & \footnotesize{$33.10\pm 0.91$}  & \footnotesize{$30.2_{-0.8}^{+0.9}$} & \footnotesize{$5.22_{-0.06}^{+0.06}$} & \footnotesize{no} & \scriptsize{(3)} \\ 
\footnotesize{HIP117184} & \footnotesize{HD222986} & \footnotesize{1RXS-J234535.4+402626} & \footnotesize{11.0} & \footnotesize{12} & \footnotesize{8.81} & \footnotesize{$0.684\pm 0.020$} & \footnotesize{$7.33\pm 0.02$} & \footnotesize{$13.88\pm 0.98$}  & \footnotesize{$72.0_{-4.8}^{+5.5}$} & \footnotesize{$4.52_{-0.16}^{+0.15}$} & \footnotesize{no} & \scriptsize{-} \\ 
\footnotesize{HIP10972} & \footnotesize{HD14554} & \footnotesize{1RXS-J022121.1+094420} & \footnotesize{12.6} & \footnotesize{18} & \footnotesize{8.77} & \footnotesize{$0.701\pm 0.023$} & \footnotesize{$7.16\pm 0.02$} & \footnotesize{$14.81\pm 1.24$}  & \footnotesize{$67.5_{-5.2}^{+6.2}$} & \footnotesize{$4.62_{-0.19}^{+0.17}$} & \footnotesize{yes(RV)} & \scriptsize{(2)} \\ 
\footnotesize{HIP58071} & \footnotesize{HD103418} & \footnotesize{1RXS-J115433.6+234854} & \footnotesize{17.7} & \footnotesize{21} & \footnotesize{9.27} & \footnotesize{$0.665\pm 0.029$} & \footnotesize{$7.38\pm 0.02$} & \footnotesize{$14.30\pm 1.32$}  & \footnotesize{$69.9_{-5.9}^{+7.1}$} & \footnotesize{$5.05_{-0.21}^{+0.19}$} & \footnotesize{yes(RV)} & \scriptsize{(2),(5)} \\ 
\footnotesize{HIP83099} & \footnotesize{HD153190} & \footnotesize{1RXS-J165852.9-234502} & \footnotesize{17.4} & \footnotesize{17} & \footnotesize{8.51} & \footnotesize{$0.661\pm 0.020$} & \footnotesize{$6.85\pm 0.02$} & \footnotesize{$18.24\pm 1.38$}  & \footnotesize{$54.8_{-3.9}^{+4.5}$} & \footnotesize{$4.82_{-0.17}^{+0.16}$} & \footnotesize{yes(RV)} & \scriptsize{(1)} \\ 
     \hline
    \end{tabular}
}    
\begin{tabnote}
\end{tabnote}
\end{minipage}
}
\end{table}
\end{center}

\begin{center}
\addtocounter{table}{-1}
\begin{table}[htbp]
\rotatebox{-90}{
\begin{minipage}{\textheight}
  \caption{(Conitnued.)}{%
  \begin{tabular}{lcccccccccccc}
      \hline
Name(HIP) \footnotemark[a] & Name(HD) \footnotemark[a] & ROSAT-ID \footnotemark[b] 
& $\theta_{\rm{ROSAT-HIP}}$ \footnotemark[c] & Pos. Err \footnotemark[d] 
& $V$ \footnotemark[a] 
& $ (B-V)_{\rm{HIP}}$ \footnotemark[a] & $K$\footnotemark[e] 
& $\pi$ \footnotemark[a] & $d_{\rm{HIP}}$ \footnotemark[f] & $M_{V}$ \footnotemark[g] & binary \footnotemark[h] & Remarks \footnotemark[i]  \\ 
 & & & [arcsec] & [arcsec] & [mag] & [mag] & [mag] 
& [mas] & [pc] & [mag] & & \\  
     \hline 
\footnotesize{HIP112364} & \footnotesize{-} & \footnotesize{1RXS-J224532.3+390534} & \footnotesize{11.0} & \footnotesize{15} & \footnotesize{8.56} & \footnotesize{$0.634\pm 0.018$} & \footnotesize{$7.08\pm 0.02$} & \footnotesize{$12.96\pm 1.21$}  & \footnotesize{$77.2_{-6.6}^{+7.9}$} & \footnotesize{$4.12_{-0.21}^{+0.19}$} & \footnotesize{yes(RV)} & \scriptsize{(2)} \\ 
\footnotesize{HIP100259} & \footnotesize{HD193554} & \footnotesize{1RXS-J202003.8+233818} & \footnotesize{2.3} & \footnotesize{10} & \footnotesize{8.27} & \footnotesize{$0.631\pm 0.014$} & \footnotesize{$6.66\pm 0.02$} & \footnotesize{$19.43\pm 1.18$}  & \footnotesize{$51.5_{-2.9}^{+3.3}$} & \footnotesize{$4.71_{-0.14}^{+0.13}$} & \footnotesize{yes(RVarch)} & \scriptsize{-} \\ 
\footnotesize{HIP40562} & \footnotesize{HD69027} & \footnotesize{1RXS-J081654.1+431618} & \footnotesize{4.9} & \footnotesize{17} & \footnotesize{8.91} & \footnotesize{$0.746\pm 0.024$} & \footnotesize{$7.20\pm 0.02$} & \footnotesize{$12.61\pm 1.47$}  & \footnotesize{$79.3_{-8.3}^{+10.5}$} & \footnotesize{$4.41_{-0.27}^{+0.24}$} & \footnotesize{yes(SB2)} & \scriptsize{-} \\ 
\footnotesize{HIP43897} & \footnotesize{HD76446} & \footnotesize{1RXS-J085632.5+122607} & \footnotesize{9.1} & \footnotesize{12} & \footnotesize{8.39} & \footnotesize{$0.672\pm 0.018$} & \footnotesize{$6.80\pm 0.02$} & \footnotesize{$13.60\pm 1.09$}  & \footnotesize{$73.5_{-5.5}^{+6.4}$} & \footnotesize{$4.06_{-0.18}^{+0.17}$} & \footnotesize{yes(SB2)} & \scriptsize{-} \\ 
\footnotesize{HIP53980} & \footnotesize{-} & \footnotesize{1RXS-J110233.5+630730} & \footnotesize{15.0} & \footnotesize{10} & \footnotesize{9.23} & \footnotesize{$0.727\pm 0.025$} & \footnotesize{$7.65\pm 0.03$} & \footnotesize{$10.97\pm 1.25$}  & \footnotesize{$91.2_{-9.3}^{+11.7}$} & \footnotesize{$4.43_{-0.26}^{+0.23}$} & \footnotesize{yes(SB2)} & \scriptsize{(1)} \\ 
\footnotesize{HIP67458} & \footnotesize{HD120368} & \footnotesize{1RXS-J134927.7-262052} & \footnotesize{9.8} & \footnotesize{15} & \footnotesize{7.92} & \footnotesize{$0.709\pm 0.015$} & \footnotesize{$6.16\pm 0.02$} & \footnotesize{$23.06\pm 0.96$}  & \footnotesize{$43.4_{-1.7}^{+1.9}$} & \footnotesize{$4.73_{-0.09}^{+0.09}$} & \footnotesize{yes(SB2)} & \scriptsize{(2)} \\ 
\footnotesize{HIP77528} & \footnotesize{HD141970} & \footnotesize{1RXS-J154949.7+472828} & \footnotesize{5.9} & \footnotesize{7} & \footnotesize{8.59} & \footnotesize{$0.743\pm 0.016$} & \footnotesize{$6.71\pm 10.00$} & \footnotesize{$14.81\pm 0.81$}  & \footnotesize{$67.5_{-3.5}^{+3.9}$} & \footnotesize{$4.44_{-0.12}^{+0.12}$} & \footnotesize{yes(SB2)} & \scriptsize{(6)} \\ 
\footnotesize{HIP79491} & \footnotesize{HD146756} & \footnotesize{1RXS-J161317.1+663655} & \footnotesize{5.4} & \footnotesize{9} & \footnotesize{8.19} & \footnotesize{$0.664\pm 0.015$} & \footnotesize{$6.62\pm 0.02$} & \footnotesize{$16.38\pm 0.66$}  & \footnotesize{$61.1_{-2.4}^{+2.6}$} & \footnotesize{$4.26_{-0.09}^{+0.09}$} & \footnotesize{yes(SB2)} & \scriptsize{(2)} \\ 
\footnotesize{HIP97321} & \footnotesize{HD187160} & \footnotesize{1RXS-J194641.5+442110} & \footnotesize{16.0} & \footnotesize{15} & \footnotesize{7.07} & \footnotesize{$0.591\pm 0.008$} & \footnotesize{$5.62\pm 0.02$} & \footnotesize{$23.36\pm 0.58$}  & \footnotesize{$42.8_{-1.0}^{+1.1}$} & \footnotesize{$3.91_{-0.05}^{+0.05}$} & \footnotesize{yes(SB2)} & \scriptsize{(2),(7)} \\ 
     \hline
    \end{tabular}
}    
\begin{tabnote}
\footnotemark[a] These data (HIP number, HD number, $V$ magnitude, $B-V$ color, and parallax in milliarcsecond ($\pi$))
were taken from the Hipparcos Catalogue \citep{ESA1997}. 
\\
\footnotemark[b] ROSAT-ID is based on ROSAT All-Sky Survey Bright Source Catalogue \citep{Voges1999} 
and the ROSAT All-Sky Survey Faint Source Catalogue \citep{Voges2000}.  
\\
\footnotemark[c] 
Angular separation between the coordinate of the star in Hipparcos Catalogue and that of the corresponding ROSAT X-ray source.
\\
\footnotemark[d] 
Total positional error of each ROSAT All-Sky Survey Catalogue source shown in \authorcite{Voges1999} (\yearcite{Voges1999}, \yearcite{Voges2000})
\\
\footnotemark[e] $K$ magnitude was taken from the 2MASS All-Sky Catalog of Point Sources \citep{Cutri2003}.
\\
\footnotemark[f] Stellar distance derived from Hipparcos parallax ($\pi$). Error value of $d_{\rm{HIP}}$ corresponds to that of parallax ($\pi$).
\\
\footnotemark[g] $M_{V}$(absolute magnitude) was derived from $V$ and $d_{\rm{HIP}}$. Error value of $M_{V}$ corresponds to that of stellar distance ($d_{\rm{HIP}}$).
\\
\footnotemark[h] 
For the details of this column, see Appendix \ref{subsec-apen:ana-binarity}. 
``no" means that the star show no evidence of binary system. 
``asymmetry?" means that the star show a bit asymmetry profiles, but we treat that star as a single star in this paper.
``yes (SB2)" corresponds to stars that have double-lined profile, 
``yes (RV)" means that the star show radial velocity changes within our observations, 
and ``yes (RVarch)" means that the stars show long-term radial velocity changes compared with the previous studies (archive data). 
\\
\footnotemark[i] Object classification in SIMBAD Astronomical Database (http://simbad.u-strasbg.fr/simbad/), considering an error range of 30 arcsec. 
\\ \ \ \ \ \ \ (1) pre-main sequence Star Candidate (2) X-ray source (3) Variable of BY Dra type (4) Star in Cluster (5) Spectroscopic binary
\\ \ \ \ \ \ \ (6) AGN (7) Rotationally variable Star 
\end{tabnote}
\end{minipage}
}
\end{table}
\end{center}

  \begin{center} 
\begin{table}[htbp]
  \caption{Target stars that are also identified as the Swift X-ray source.}\label{table:Swift-catalog}
    \begin{tabular}{lccccc}
      \hline
       Starname (HIP) & Swift ID \footnotemark[a] & $\theta_{\rm{Swift-HIP}}$ \footnotemark[b] & Pos. Err \footnotemark[c] & $X_{\rm{count}}$ \footnotemark[d] & Remarks \\
        & & [arcsec] & [arcsec] & [s$^{-1}$] & \\
      \hline 
HIP25002 & 1SXPS-J052112.7-140907 & 3.0 & 4.9 & $0.012^{+0.005}_{-0.004}  $ & \\
HIP112364 & 1SXPS-J224532.9+390538 & 1.8 & 5.9 & $0.002^{+0.001}_{-0.001}  $ & \\
HIP40562 & 1SXPS-J081654.6+431612 & 2.3 & 3.9 & $0.003^{+0.001}_{-0.001} $ & \\ 
HIP77528 & 1SXPS-J154949.6+472822 & 0.1 & 3.9 & $0.088^{+0.012}_{-0.012} $ & \\
KIC7940546 (HIP92615) & 1SXPS J185216.6+434231 & 4.6 & 5.5 & $0.004^{+0.002}_{-0.002} $ & comparison star \\
      \hline     
    \end{tabular}
\begin{tabnote}
\footnotemark[a] Swift ID is based on 1SXPS Swift X-ray Telescope Point Source Catalogue \citep{Evans2014}.  
\\
\footnotemark[b] 
Angular separation between the coordinate of the star in Hipparcos Catalogue and that of the corresponding Swift X-ray source.
\\
\footnotemark[c] 
90\% radial position error of each Swift source shown in \citet{Evans2014}. 
\\
\footnotemark[d] Total (0.3-10keV)-band count-rate of Swift data.
\end{tabnote}    
\end{table} 
  \end{center}

\begin{center}
\begin{table}[htbp]
\rotatebox{-90}{
\begin{minipage}{\textheight}
\caption{Atmospheric parameters of the target stars estimated from spectroscopic data.}\label{table:atmos}{%
  \begin{tabular}{lcccccccccccc}
\hline
Name(HIP) & Name(HD) & $ T_{\rm{eff}}$ & $\log g$ & $v_{\rm{t}}$ & [Fe/H] & $A_{V}$\footnotemark[a] & $B.C.$\footnotemark[b] & $\log (L/L_{\odot})$ \footnotemark[c] & $R_{\rm{s}}$ \footnotemark[c] & $\log Age$ \footnotemark[c] & $M_{s}$ \footnotemark[c] & $\log g_{\rm{TLM}}$ \footnotemark[c]  \\
 & & [K] & [cm s$^{-2}$] & [km s$^{-1}$] & & [mag] & [mag] & & [$R_{\odot}$] & [yr] & [$M_{\odot}$] &[cm s$^{-2}$]  \\ 
\hline
\footnotesize{HIP3342} & \footnotesize{HD4021} & \footnotesize{$5733\pm 20$} & \footnotesize{$4.53\pm 0.05$} & \footnotesize{$1.11\pm 0.09$} & \footnotesize{$-0.02\pm 0.02$} & \footnotesize{0.19} & \footnotesize{-0.14} & \footnotesize{$-0.03_{-0.07}^{+0.07}$} & \footnotesize{$0.98_{-0.08}^{+0.09}$} & \footnotesize{$9.78_{-0.48}^{+0.22}$} & \footnotesize{$0.95\pm 0.03$} & \footnotesize{$4.44_{-0.09}^{+0.09}$} \\ 
\footnotesize{HIP4909} & \footnotesize{HD6204} & \footnotesize{$5865\pm 20$} & \footnotesize{$4.54\pm 0.04$} & \footnotesize{$1.05\pm 0.08$} & \footnotesize{$0.03\pm 0.02$} & \footnotesize{0.14} & \footnotesize{-0.14} & \footnotesize{$0.08_{-0.06}^{+0.07}$} & \footnotesize{$1.07_{-0.08}^{+0.09}$} & \footnotesize{$9.70_{-0.27}^{+0.17}$} & \footnotesize{$1.01\pm 0.02$} & \footnotesize{$4.39_{-0.08}^{+0.08}$} \\ 
\footnotesize{HIP8522} & \footnotesize{HD11120} & \footnotesize{$5707\pm 15$} & \footnotesize{$4.59\pm 0.04$} & \footnotesize{$1.01\pm 0.08$} & \footnotesize{$-0.04\pm 0.02$} & \footnotesize{0.10} & \footnotesize{-0.14} & \footnotesize{$0.01_{-0.05}^{+0.06}$} & \footnotesize{$1.04_{-0.07}^{+0.08}$} & \footnotesize{$10.00_{-0.08}^{+0.06}$} & \footnotesize{$0.91\pm 0.01$} & \footnotesize{$4.37_{-0.07}^{+0.06}$} \\ 
\footnotesize{HIP9519} & \footnotesize{HD12484} & \footnotesize{$5899\pm 13$} & \footnotesize{$4.50\pm 0.03$} & \footnotesize{$1.39\pm 0.13$} & \footnotesize{$0.07\pm 0.03$} & \footnotesize{0.07} & \footnotesize{-0.10} & \footnotesize{$0.04_{-0.05}^{+0.05}$} & \footnotesize{$1.01_{-0.06}^{+0.07}$} & \footnotesize{$9.36_{-1.36}^{+0.29}$} & \footnotesize{$1.06\pm 0.02$} & \footnotesize{$4.46_{-0.07}^{+0.06}$} \\ 
\footnotesize{HIP14997} & \footnotesize{-} & \footnotesize{$5764\pm 15$} & \footnotesize{$4.49\pm 0.04$} & \footnotesize{$1.49\pm 0.13$} & \footnotesize{$0.05\pm 0.03$} & \footnotesize{0.07} & \footnotesize{-0.13} & \footnotesize{$-0.11_{-0.08}^{+0.09}$} & \footnotesize{$0.88_{-0.08}^{+0.10}$} & \footnotesize{$9.31_{-0.77}^{+0.26}$} & \footnotesize{$1.01\pm 0.02$} & \footnotesize{$4.56_{-0.10}^{+0.09}$} \\ 
\footnotesize{HIP19793} & \footnotesize{HD26736} & \footnotesize{$5833\pm 15$} & \footnotesize{$4.45\pm 0.04$} & \footnotesize{$1.22\pm 0.12$} & \footnotesize{$0.16\pm 0.03$} & \footnotesize{0.09} & \footnotesize{-0.13} & \footnotesize{$0.10_{-0.04}^{+0.05}$} & \footnotesize{$1.10_{-0.06}^{+0.07}$} & \footnotesize{$9.69_{-0.14}^{+0.10}$} & \footnotesize{$1.05\pm 0.01$} & \footnotesize{$4.38_{-0.06}^{+0.05}$} \\ 
\footnotesize{HIP20616} & \footnotesize{HD27947} & \footnotesize{$5938\pm 33$} & \footnotesize{$4.60\pm 0.07$} & \footnotesize{$1.23\pm 0.16$} & \footnotesize{$0.15\pm 0.04$} & \footnotesize{0.10} & \footnotesize{-0.11} & \footnotesize{$-0.02_{-0.05}^{+0.06}$} & \footnotesize{$0.92_{-0.07}^{+0.08}$} & \footnotesize{$8.81_{-0.72}^{+0.26}$} & \footnotesize{$1.09\pm 0.01$} & \footnotesize{$4.55_{-0.07}^{+0.07}$} \\ 
\footnotesize{HIP21091} & \footnotesize{-} & \footnotesize{$5857\pm 18$} & \footnotesize{$4.54\pm 0.04$} & \footnotesize{$1.20\pm 0.10$} & \footnotesize{$0.03\pm 0.03$} & \footnotesize{0.03} & \footnotesize{-0.14} & \footnotesize{$0.15_{-0.07}^{+0.08}$} & \footnotesize{$1.15_{-0.10}^{+0.12}$} & \footnotesize{$9.85_{-0.11}^{+0.09}$} & \footnotesize{$1.00\pm 0.01$} & \footnotesize{$4.32_{-0.09}^{+0.08}$} \\ 
\footnotesize{HIP22175} & \footnotesize{HD30286} & \footnotesize{$5667\pm 15$} & \footnotesize{$4.59\pm 0.04$} & \footnotesize{$0.95\pm 0.08$} & \footnotesize{$-0.12\pm 0.02$} & \footnotesize{0.05} & \footnotesize{-0.14} & \footnotesize{$-0.13_{-0.03}^{+0.03}$} & \footnotesize{$0.89_{-0.03}^{+0.04}$} & \footnotesize{$9.75_{-0.27}^{+0.17}$} & \footnotesize{$0.90\pm 0.02$} & \footnotesize{$4.49_{-0.04}^{+0.04}$} \\ 
\footnotesize{HIP23027} & \footnotesize{HD31464} & \footnotesize{$5601\pm 23$} & \footnotesize{$4.49\pm 0.06$} & \footnotesize{$1.08\pm 0.12$} & \footnotesize{$-0.07\pm 0.04$} & \footnotesize{0.13} & \footnotesize{-0.18} & \footnotesize{$-0.10_{-0.06}^{+0.06}$} & \footnotesize{$0.94_{-0.07}^{+0.08}$} & \footnotesize{$9.96_{-0.25}^{+0.16}$} & \footnotesize{$0.88\pm 0.03$} & \footnotesize{$4.43_{-0.09}^{+0.08}$} \\ 
\footnotesize{HIP25002} & \footnotesize{HD35041} & \footnotesize{$5744\pm 13$} & \footnotesize{$4.41\pm 0.04$} & \footnotesize{$1.11\pm 0.09$} & \footnotesize{$-0.11\pm 0.02$} & \footnotesize{0.01} & \footnotesize{-0.13} & \footnotesize{$0.03_{-0.03}^{+0.03}$} & \footnotesize{$1.05_{-0.04}^{+0.04}$} & \footnotesize{$10.02_{-0.04}^{+0.04}$} & \footnotesize{$0.89\pm 0.01$} & \footnotesize{$4.35_{-0.04}^{+0.04}$} \\ 
\footnotesize{HIP27980} & \footnotesize{HD39833} & \footnotesize{$5854\pm 18$} & \footnotesize{$4.53\pm 0.04$} & \footnotesize{$0.99\pm 0.09$} & \footnotesize{$0.16\pm 0.02$} & \footnotesize{0.07} & \footnotesize{-0.14} & \footnotesize{$0.26_{-0.04}^{+0.05}$} & \footnotesize{$1.32_{-0.07}^{+0.08}$} & \footnotesize{$9.85_{-0.02}^{+0.02}$} & \footnotesize{$1.07\pm 0.01$} & \footnotesize{$4.23_{-0.06}^{+0.05}$} \\ 
\footnotesize{HIP35185} & \footnotesize{HD56202} & \footnotesize{$5830\pm 15$} & \footnotesize{$4.50\pm 0.04$} & \footnotesize{$1.19\pm 0.09$} & \footnotesize{$-0.03\pm 0.02$} & \footnotesize{0.04} & \footnotesize{-0.14} & \footnotesize{$0.05_{-0.05}^{+0.06}$} & \footnotesize{$1.04_{-0.07}^{+0.07}$} & \footnotesize{$9.81_{-0.17}^{+0.12}$} & \footnotesize{$0.97\pm 0.02$} & \footnotesize{$4.39_{-0.07}^{+0.06}$} \\ 
\footnotesize{HIP37971} & \footnotesize{HD62857} & \footnotesize{$5774\pm 20$} & \footnotesize{$4.49\pm 0.05$} & \footnotesize{$1.15\pm 0.11$} & \footnotesize{$0.20\pm 0.03$} & \footnotesize{0.08} & \footnotesize{-0.13} & \footnotesize{$0.03_{-0.05}^{+0.05}$} & \footnotesize{$1.04_{-0.06}^{+0.07}$} & \footnotesize{$9.53_{-0.55}^{+0.24}$} & \footnotesize{$1.05\pm 0.02$} & \footnotesize{$4.43_{-0.06}^{+0.06}$} \\ 
\footnotesize{HIP38228} & \footnotesize{HD63433} & \footnotesize{$5631\pm 18$} & \footnotesize{$4.49\pm 0.05$} & \footnotesize{$1.26\pm 0.09$} & \footnotesize{$-0.03\pm 0.02$} & \footnotesize{0.06} & \footnotesize{-0.13} & \footnotesize{$-0.10_{-0.02}^{+0.02}$} & \footnotesize{$0.94_{-0.02}^{+0.02}$} & \footnotesize{$9.91_{-0.06}^{+0.05}$} & \footnotesize{$0.90\pm 0.01$} & \footnotesize{$4.45_{-0.03}^{+0.03}$} \\ 
\footnotesize{HIP40380} & \footnotesize{-} & \footnotesize{$5789\pm 30$} & \footnotesize{$4.46\pm 0.07$} & \footnotesize{$1.48\pm 0.19$} & \footnotesize{$-0.04\pm 0.04$} & \footnotesize{0.06} & \footnotesize{-0.13} & \footnotesize{$0.03_{-0.09}^{+0.10}$} & \footnotesize{$1.04_{-0.11}^{+0.13}$} & \footnotesize{$9.81_{-0.41}^{+0.21}$} & \footnotesize{$0.96\pm 0.03$} & \footnotesize{$4.39_{-0.12}^{+0.11}$} \\ 
\footnotesize{HIP44657} & \footnotesize{HD78130} & \footnotesize{$5803\pm 15$} & \footnotesize{$4.48\pm 0.04$} & \footnotesize{$1.10\pm 0.10$} & \footnotesize{$0.14\pm 0.03$} & \footnotesize{0.07} & \footnotesize{-0.13} & \footnotesize{$0.08_{-0.06}^{+0.06}$} & \footnotesize{$1.09_{-0.07}^{+0.09}$} & \footnotesize{$9.73_{-0.18}^{+0.13}$} & \footnotesize{$1.03\pm 0.01$} & \footnotesize{$4.38_{-0.07}^{+0.07}$} \\ 
\footnotesize{HIP51652} & \footnotesize{HD91332} & \footnotesize{$5870\pm 15$} & \footnotesize{$4.35\pm 0.04$} & \footnotesize{$1.12\pm 0.09$} & \footnotesize{$0.24\pm 0.02$} & \footnotesize{0.06} & \footnotesize{-0.13} & \footnotesize{$0.22_{-0.05}^{+0.05}$} & \footnotesize{$1.26_{-0.08}^{+0.09}$} & \footnotesize{$9.74_{-0.06}^{+0.05}$} & \footnotesize{$1.10\pm 0.01$} & \footnotesize{$4.28_{-0.06}^{+0.06}$} \\ 
\footnotesize{HIP52761} & \footnotesize{HD93393} & \footnotesize{$5881\pm 20$} & \footnotesize{$4.35\pm 0.05$} & \footnotesize{$1.29\pm 0.11$} & \footnotesize{$-0.09\pm 0.03$} & \footnotesize{0.08} & \footnotesize{-0.10} & \footnotesize{$0.05_{-0.08}^{+0.09}$} & \footnotesize{$1.02_{-0.10}^{+0.12}$} & \footnotesize{$9.70_{-0.58}^{+0.24}$} & \footnotesize{$0.98\pm 0.03$} & \footnotesize{$4.41_{-0.11}^{+0.10}$} \\ 
\footnotesize{HIP59399} & \footnotesize{HD105863} & \footnotesize{$5851\pm 23$} & \footnotesize{$4.43\pm 0.06$} & \footnotesize{$1.28\pm 0.12$} & \footnotesize{$-0.03\pm 0.03$} & \footnotesize{0.03} & \footnotesize{-0.13} & \footnotesize{$0.08_{-0.10}^{+0.11}$} & \footnotesize{$1.07_{-0.12}^{+0.16}$} & \footnotesize{$9.77_{-0.43}^{+0.21}$} & \footnotesize{$0.98\pm 0.02$} & \footnotesize{$4.38_{-0.13}^{+0.12}$} \\ 
\footnotesize{HIP65627} & \footnotesize{-} & \footnotesize{$5768\pm 33$} & \footnotesize{$4.35\pm 0.09$} & \footnotesize{$1.52\pm 0.16$} & \footnotesize{$-0.08\pm 0.04$} & \footnotesize{0.11} & \footnotesize{-0.13} & \footnotesize{$0.18_{-0.08}^{+0.09}$} & \footnotesize{$1.24_{-0.12}^{+0.15}$} & \footnotesize{$10.05_{-0.04}^{+0.03}$} & \footnotesize{$0.93\pm 0.02$} & \footnotesize{$4.22_{-0.11}^{+0.10}$} \\ 
\footnotesize{HIP70354} & \footnotesize{-} & \footnotesize{$5840\pm 15$} & \footnotesize{$4.43\pm 0.04$} & \footnotesize{$1.31\pm 0.09$} & \footnotesize{$-0.08\pm 0.02$} & \footnotesize{0.08} & \footnotesize{-0.13} & \footnotesize{$-0.06_{-0.06}^{+0.06}$} & \footnotesize{$0.91_{-0.06}^{+0.07}$} & \footnotesize{$9.44_{-1.32}^{+0.29}$} & \footnotesize{$0.98\pm 0.02$} & \footnotesize{$4.51_{-0.07}^{+0.07}$} \\ 
\hline
\end{tabular}
}    
\begin{tabnote}
\end{tabnote}
\end{minipage}
}
\end{table}
\end{center}

\begin{center}
\addtocounter{table}{-1}
\begin{table}[htbp]
\rotatebox{-90}{
\begin{minipage}{\textheight}
\caption{(Conitnued.)}{%
  \begin{tabular}{lcccccccccccc}
\hline
Name(HIP) & Name(HD) & $ T_{\rm{eff}}$ & $\log g$ & $v_{\rm{t}}$ & [Fe/H] & $A_{V}$\footnotemark[a] & $B.C.$\footnotemark[b] & $\log (L/L_{\odot})$ \footnotemark[c] & $R_{\rm{s}}$ \footnotemark[c] & $\log Age$ \footnotemark[c] & $M_{s}$ \footnotemark[c] & $\log g_{\rm{TLM}}$ \footnotemark[c]  \\ 
 & & [K] & [cm s$^{-2}$] & [km s$^{-1}$] & & [mag] & [mag] & & [$R_{\odot}$] & [yr] & [$M_{\odot}$] & [cm s$^{-2}$]  \\ 
\hline
\footnotesize{HIP70394} & \footnotesize{-} & \footnotesize{$5718\pm 40$} & \footnotesize{$4.54\pm 0.10$} & \footnotesize{$1.29\pm 0.15$} & \footnotesize{$-0.01\pm 0.04$} & \footnotesize{0.09} & \footnotesize{-0.14} & \footnotesize{$-0.03_{-0.08}^{+0.09}$} & \footnotesize{$0.99_{-0.10}^{+0.12}$} & \footnotesize{$9.76_{-1.08}^{+0.28}$} & \footnotesize{$0.96\pm 0.04$} & \footnotesize{$4.43_{-0.12}^{+0.11}$} \\ 
\footnotesize{HIP71218} & \footnotesize{HD127913} & \footnotesize{$5632\pm 23$} & \footnotesize{$4.54\pm 0.06$} & \footnotesize{$1.10\pm 0.16$} & \footnotesize{$0.17\pm 0.04$} & \footnotesize{0.13} & \footnotesize{-0.14} & \footnotesize{$-0.06_{-0.07}^{+0.08}$} & \footnotesize{$0.98_{-0.08}^{+0.10}$} & \footnotesize{$9.68_{-0.91}^{+0.27}$} & \footnotesize{$0.99\pm 0.03$} & \footnotesize{$4.46_{-0.10}^{+0.09}$} \\ 
\footnotesize{HIP77584} & \footnotesize{HD141715} & \footnotesize{$5787\pm 13$} & \footnotesize{$4.37\pm 0.04$} & \footnotesize{$1.20\pm 0.11$} & \footnotesize{$-0.12\pm 0.02$} & \footnotesize{0.06} & \footnotesize{-0.13} & \footnotesize{$0.28_{-0.06}^{+0.06}$} & \footnotesize{$1.37_{-0.10}^{+0.11}$} & \footnotesize{$10.04_{-0.02}^{+0.02}$} & \footnotesize{$0.95\pm 0.02$} & \footnotesize{$4.14_{-0.08}^{+0.07}$} \\ 
\footnotesize{HIP79068} & \footnotesize{HD145224} & \footnotesize{$5809\pm 15$} & \footnotesize{$4.43\pm 0.04$} & \footnotesize{$1.02\pm 0.08$} & \footnotesize{$0.04\pm 0.02$} & \footnotesize{0.04} & \footnotesize{-0.13} & \footnotesize{$0.10_{-0.04}^{+0.04}$} & \footnotesize{$1.11_{-0.06}^{+0.06}$} & \footnotesize{$9.88_{-0.06}^{+0.05}$} & \footnotesize{$0.98\pm 0.00$} & \footnotesize{$4.34_{-0.05}^{+0.05}$} \\ 
\footnotesize{HIP80271} & \footnotesize{-} & \footnotesize{$6032\pm 30$} & \footnotesize{$4.43\pm 0.07$} & \footnotesize{$1.21\pm 0.16$} & \footnotesize{$-0.06\pm 0.04$} & \footnotesize{0.04} & \footnotesize{-0.10} & \footnotesize{$0.18_{-0.08}^{+0.09}$} & \footnotesize{$1.13_{-0.11}^{+0.14}$} & \footnotesize{$9.66_{-0.27}^{+0.17}$} & \footnotesize{$1.03\pm 0.02$} & \footnotesize{$4.35_{-0.11}^{+0.10}$} \\ 
\footnotesize{HIP83507} & \footnotesize{HD154326} & \footnotesize{$5806\pm 50$} & \footnotesize{$4.53\pm 0.12$} & \footnotesize{$1.74\pm 0.28$} & \footnotesize{$-0.13\pm 0.06$} & \footnotesize{0.09} & \footnotesize{-0.14} & \footnotesize{$0.26_{-0.10}^{+0.11}$} & \footnotesize{$1.34_{-0.16}^{+0.21}$} & \footnotesize{$10.04_{-0.05}^{+0.05}$} & \footnotesize{$0.94\pm 0.04$} & \footnotesize{$4.16_{-0.14}^{+0.13}$} \\ 
\footnotesize{HIP86245} & \footnotesize{HD160135} & \footnotesize{$5973\pm 18$} & \footnotesize{$4.33\pm 0.04$} & \footnotesize{$1.16\pm 0.09$} & \footnotesize{$0.03\pm 0.02$} & \footnotesize{0.12} & \footnotesize{-0.10} & \footnotesize{$0.16_{-0.06}^{+0.06}$} & \footnotesize{$1.13_{-0.08}^{+0.09}$} & \footnotesize{$9.68_{-0.16}^{+0.12}$} & \footnotesize{$1.04\pm 0.01$} & \footnotesize{$4.35_{-0.07}^{+0.07}$} \\ 
\footnotesize{HIP86781} & \footnotesize{HD238744} & \footnotesize{$6110\pm 30$} & \footnotesize{$4.35\pm 0.07$} & \footnotesize{$1.35\pm 0.13$} & \footnotesize{$0.16\pm 0.04$} & \footnotesize{0.08} & \footnotesize{-0.10} & \footnotesize{$0.21_{-0.05}^{+0.06}$} & \footnotesize{$1.14_{-0.08}^{+0.09}$} & \footnotesize{$9.26_{-0.64}^{+0.25}$} & \footnotesize{$1.17\pm 0.02$} & \footnotesize{$4.40_{-0.07}^{+0.07}$} \\ 
\footnotesize{HIP88572} & \footnotesize{HD165863} & \footnotesize{$5610\pm 15$} & \footnotesize{$4.51\pm 0.04$} & \footnotesize{$1.15\pm 0.10$} & \footnotesize{$0.12\pm 0.03$} & \footnotesize{0.12} & \footnotesize{-0.19} & \footnotesize{$-0.02_{-0.04}^{+0.04}$} & \footnotesize{$1.04_{-0.05}^{+0.05}$} & \footnotesize{$9.95_{-0.07}^{+0.06}$} & \footnotesize{$0.95\pm 0.01$} & \footnotesize{$4.38_{-0.04}^{+0.04}$} \\ 
\footnotesize{HIP91073} & \footnotesize{HD171489} & \footnotesize{$5953\pm 15$} & \footnotesize{$4.45\pm 0.03$} & \footnotesize{$1.12\pm 0.09$} & \footnotesize{$0.14\pm 0.02$} & \footnotesize{0.06} & \footnotesize{-0.10} & \footnotesize{$0.15_{-0.06}^{+0.06}$} & \footnotesize{$1.12_{-0.07}^{+0.08}$} & \footnotesize{$9.51_{-0.28}^{+0.17}$} & \footnotesize{$1.09\pm 0.01$} & \footnotesize{$4.38_{-0.07}^{+0.06}$} \\ 
\footnotesize{HIP101893} & \footnotesize{HD340676} & \footnotesize{$5719\pm 13$} & \footnotesize{$4.55\pm 0.03$} & \footnotesize{$1.03\pm 0.08$} & \footnotesize{$0.08\pm 0.02$} & \footnotesize{0.10} & \footnotesize{-0.14} & \footnotesize{$-0.10_{-0.05}^{+0.06}$} & \footnotesize{$0.91_{-0.06}^{+0.07}$} & \footnotesize{$9.31_{-0.80}^{+0.26}$} & \footnotesize{$1.01\pm 0.01$} & \footnotesize{$4.53_{-0.07}^{+0.06}$} \\ 
\footnotesize{HIP105066} & \footnotesize{HD202605} & \footnotesize{$5627\pm 15$} & \footnotesize{$4.48\pm 0.04$} & \footnotesize{$1.19\pm 0.09$} & \footnotesize{$0.14\pm 0.02$} & \footnotesize{0.07} & \footnotesize{-0.13} & \footnotesize{$0.01_{-0.05}^{+0.05}$} & \footnotesize{$1.07_{-0.06}^{+0.07}$} & \footnotesize{$9.95_{-0.09}^{+0.07}$} & \footnotesize{$0.96\pm 0.01$} & \footnotesize{$4.36_{-0.06}^{+0.06}$} \\ 
\footnotesize{HIP105855} & \footnotesize{HD204790} & \footnotesize{$6038\pm 20$} & \footnotesize{$4.44\pm 0.05$} & \footnotesize{$1.22\pm 0.09$} & \footnotesize{$0.06\pm 0.03$} & \footnotesize{0.12} & \footnotesize{-0.10} & \footnotesize{$0.18_{-0.06}^{+0.07}$} & \footnotesize{$1.13_{-0.09}^{+0.10}$} & \footnotesize{$9.52_{-0.34}^{+0.19}$} & \footnotesize{$1.09\pm 0.02$} & \footnotesize{$4.37_{-0.08}^{+0.08}$} \\ 
\footnotesize{HIP115527} & \footnotesize{HD220476} & \footnotesize{$5708\pm 18$} & \footnotesize{$4.61\pm 0.04$} & \footnotesize{$1.15\pm 0.09$} & \footnotesize{$-0.04\pm 0.02$} & \footnotesize{0.10} & \footnotesize{-0.14} & \footnotesize{$-0.09_{-0.02}^{+0.02}$} & \footnotesize{$0.92_{-0.03}^{+0.03}$} & \footnotesize{$9.68_{-0.24}^{+0.16}$} & \footnotesize{$0.94\pm 0.02$} & \footnotesize{$4.48_{-0.04}^{+0.04}$} \\ 
\footnotesize{HIP117184} & \footnotesize{HD222986} & \footnotesize{$5890\pm 30$} & \footnotesize{$4.55\pm 0.07$} & \footnotesize{$1.21\pm 0.13$} & \footnotesize{$0.19\pm 0.03$} & \footnotesize{0.07} & \footnotesize{-0.11} & \footnotesize{$0.16_{-0.06}^{+0.06}$} & \footnotesize{$1.16_{-0.09}^{+0.10}$} & \footnotesize{$9.62_{-0.24}^{+0.16}$} & \footnotesize{$1.08\pm 0.02$} & \footnotesize{$4.34_{-0.08}^{+0.07}$} \\ 
\hline
\end{tabular}
}    
\begin{tabnote}
\footnotemark[a] Interstellar extension estimated in Appendix \ref{subsec-apen:ana-luminosity-radius}.
\\
\footnotemark[b] Bolometric Correction ($V$ band) on the basis of Table 4 of \citet{Alonso1995}.
\\
\footnotemark[c] The Stellar luminosity ($\log (L/L_{\odot})$), the stellar radius ($R_{\rm{s}}$), the stellar age ($\log Age$), 
the stellar mass ($M_{s}$), and the theoretical surface gravity ($\log g_{\rm{TLM}}$)
estimated in Appendix \ref{subsec-apen:ana-luminosity-radius} and \ref{subsec-apen:ana-age-mass}.
\end{tabnote}
\end{minipage}
}
\end{table}
\end{center}

\begin{center}
\small
\begin{longtable}{lccccccccc}
\caption{Rotation velocity, activity indicators, and Lithium abundance of the target stars}\label{table:activity}
\hline
Name(HIP) & $v\sin i$ & $A$(Li) & $r_{0}$(8542) & $\langle fB\rangle$ \footnotemark[a] & $r_{0}$(H$\alpha$) &ROSAT-ID \footnotemark[b] & 
$ X_{\rm{count}}$ \footnotemark[b] & $\log N_{\rm{H}}$ \footnotemark[c] & $\log L_{X} $ \footnotemark[d]  \\ 
 & [km s$^{-1}$] & & & [Gauss] & & & [s$^{-1}$] & [cm$^{-2}$] & [erg s$^{-1}$]  \\ 
\hline
\endhead
\hline
\endfoot
\multicolumn{10}{l}{\hbox to 0pt{\parbox{165mm}{\footnotesize
\footnotemark[a] Mean intensity of the stellar magnetic field estimated from $r_{0}(8542)$ index on the basis of Equation (1) of \citet{YNotsu2015b}.
}}}\\
\multicolumn{10}{l}{\hbox to 0pt{\parbox{165mm}{\footnotesize
\footnotemark[b] ROSAT-ID and X-ray count rate ($ X_{\rm{count}}$) is based on ROSAT All-Sky Survey Bright Source Catalogue \citep{Voges1999} 
and the ROSAT All-Sky Survey Faint Source Catalogue \citep{Voges2000}.  
}}}\\
\multicolumn{10}{l}{\hbox to 0pt{\parbox{165mm}{\footnotesize
\footnotemark[c] The total neutral hydrogen number density ($N_{\rm{H}}$) estimated from the interstellar extension ($A_{V}$) (See Appendix \ref{subsec-apen:ana-Xray} for the details).
}}}\\
\multicolumn{10}{l}{\hbox to 0pt{\parbox{165mm}{\footnotesize
\footnotemark[d] X-ray Luminosity (0.1$\sim$2.4 keV) on the basis of ROSAT data (See Appendix \ref{subsec-apen:ana-Xray} for the details).
}}}\\
\multicolumn{10}{l}{\hbox to 0pt{\parbox{165mm}{\footnotesize
\footnotemark[e] The wavelengh range of Dec. 29th\&30th observation does not include Ca II 8542.
}}} 
\endlastfoot
 \footnotesize{HIP3342} & \footnotesize{$ 4.5\pm 0.3$} & \footnotesize{2.34} &\footnotesize{0.47} & \footnotesize{$46\pm 41$} &\footnotesize{0.30} & \footnotesize{1RXS-J004236.9-164214} & \footnotesize{$0.061\pm 0.018$} & \footnotesize{20.6} & \footnotesize{$29.37_{-0.20}^{+0.14}$}  \\ 
 \footnotesize{HIP4909} & \footnotesize{$ 3.6\pm 0.3$} & \footnotesize{2.43} &\footnotesize{0.37} & \footnotesize{$10\pm 24$} &\footnotesize{0.24} & \footnotesize{1RXS-J010257.2-095145} & \footnotesize{$0.032\pm 0.011$} & \footnotesize{20.4} & \footnotesize{$28.81_{-0.27}^{+0.17}$}  \\ 
 \footnotesize{HIP8522} & \footnotesize{$ 2.6\pm 0.4$} & \footnotesize{$ < $1.3} &\footnotesize{0.40} & \footnotesize{$16\pm 24$} &\footnotesize{0.26} & \footnotesize{1RXS-J014955.2+254443} & \footnotesize{$0.054\pm 0.016$} & \footnotesize{20.3} & \footnotesize{$29.05_{-0.20}^{+0.14}$}  \\ 
 \footnotesize{HIP9519} & \footnotesize{$ 7.3\pm 0.2$} & \footnotesize{3.01} &\footnotesize{0.50} & \footnotesize{$68\pm 41$} &\footnotesize{0.29} & \footnotesize{1RXS-J020227.1+024902} & \footnotesize{$0.076\pm 0.017$} & \footnotesize{20.1} & \footnotesize{$29.12_{-0.16}^{+0.12}$}  \\ 
 \footnotesize{HIP14997} & \footnotesize{$ 10.1\pm 0.1$} & \footnotesize{2.79} &\footnotesize{0.58} & \footnotesize{$176\pm 123$} &\footnotesize{0.31} & \footnotesize{1RXS-J031318.0+113308} & \footnotesize{$0.034\pm 0.014$} & \footnotesize{20.1} & \footnotesize{$29.08_{-0.31}^{+0.18}$}  \\ 
 \footnotesize{HIP19793} & \footnotesize{$ 5.2\pm 0.2$} & \footnotesize{2.47} &\footnotesize{0.41} & \footnotesize{$19\pm 41$} &\footnotesize{0.24} & \footnotesize{1RXS-J041432.2+233448} & \footnotesize{$0.085\pm 0.016$} & \footnotesize{20.2} & \footnotesize{$29.16_{-0.13}^{+0.10}$}  \\ 
 \footnotesize{HIP20616} & \footnotesize{$ 7.6\pm 0.2$} & \footnotesize{2.85} &\footnotesize{0.45} & \footnotesize{$35\pm 41$} &\footnotesize{0.27} & \footnotesize{1RXS-J042500.6+115834} & \footnotesize{$0.060\pm 0.014$} & \footnotesize{20.3} & \footnotesize{$28.99_{-0.16}^{+0.12}$}  \\ 
 \footnotesize{HIP21091} & \footnotesize{$ 5.1\pm 0.2$} & \footnotesize{2.71} &\footnotesize{0.46} & \footnotesize{$40\pm 41$} &\footnotesize{0.29} & \footnotesize{1RXS-J043111.0+111428} & \footnotesize{$0.047\pm 0.011$} & \footnotesize{19.8} & \footnotesize{$29.20_{-0.17}^{+0.13}$}  \\ 
 \footnotesize{HIP22175} & \footnotesize{$ 1.6\pm 0.8$} & \footnotesize{1.02} &\footnotesize{0.25} & \footnotesize{$0.8\pm 6$} &\footnotesize{0.21} & \footnotesize{1RXS-J044616.2+031624} & \footnotesize{$0.020\pm 0.009$} & \footnotesize{19.9} & \footnotesize{$27.88_{-0.64}^{+0.25}$}  \\ 
 \footnotesize{HIP23027} & \footnotesize{$ 4.5\pm 0.3$} & \footnotesize{2.15} & - \footnotemark[e] & - \footnotemark[e] &\footnotesize{0.29} & \footnotesize{1RXS-J045705.3+244504} & \footnotesize{$0.030\pm 0.010$} & \footnotesize{20.4} & \footnotesize{$28.80_{-0.23}^{+0.15}$}  \\ 
 \footnotesize{HIP25002} & \footnotesize{$ 3.8\pm 0.3$} & \footnotesize{2.38} &\footnotesize{0.43} & \footnotesize{$26\pm 41$} &\footnotesize{0.28} & \footnotesize{1RXS-J052113.0-140854} & \footnotesize{$0.052\pm 0.014$} & \footnotesize{19.3} & \footnotesize{$28.67_{-0.17}^{+0.12}$}  \\ 
 \footnotesize{HIP27980} & \footnotesize{$ 2.5\pm 0.5$} & \footnotesize{2.43} &\footnotesize{0.23} & \footnotesize{$0.5\pm 6$} &\footnotesize{0.21} & \footnotesize{1RXS-J055501.9-003016} & \footnotesize{$0.014\pm 0.006$} & \footnotesize{20.1} & \footnotesize{$28.32_{-0.34}^{+0.19}$}  \\ 
 \footnotesize{HIP35185} & \footnotesize{$ 5.4\pm 0.2$} & \footnotesize{2.71} &\footnotesize{0.40} & \footnotesize{$16\pm 24$} &\footnotesize{0.28} & \footnotesize{1RXS-J071619.2+050443} & \footnotesize{$0.056\pm 0.013$} & \footnotesize{19.9} & \footnotesize{$29.10_{-0.16}^{+0.12}$}  \\ 
 \footnotesize{HIP37971} & \footnotesize{$ 5.2\pm 0.2$} & \footnotesize{2.38} &\footnotesize{0.36} & \footnotesize{$8\pm 24$} &\footnotesize{0.24} & \footnotesize{1RXS-J074658.6+260142} & \footnotesize{$0.051\pm 0.013$} & \footnotesize{20.2} & \footnotesize{$29.06_{-0.17}^{+0.13}$}  \\ 
 \footnotesize{HIP38228} & \footnotesize{$ 5.6\pm 0.2$} & \footnotesize{2.45} &\footnotesize{0.48} & \footnotesize{$52\pm 41$} &\footnotesize{0.29} & \footnotesize{1RXS-J074955.3+272152} & \footnotesize{$0.224\pm 0.024$} & \footnotesize{20.1} & \footnotesize{$28.91_{-0.06}^{+0.06}$}  \\ 
 \footnotesize{HIP40380} & \footnotesize{$ 7.7\pm 0.2$} & \footnotesize{2.87} &\footnotesize{0.57} & \footnotesize{$158\pm 123$} &\footnotesize{0.32} & \footnotesize{1RXS-J081437.0+625609} & \footnotesize{$0.045\pm 0.013$} & \footnotesize{20.0} & \footnotesize{$29.39_{-0.21}^{+0.15}$}  \\ 
 \footnotesize{HIP44657} & \footnotesize{$ 2.2\pm 0.5$} & \footnotesize{1.96} &\footnotesize{0.33} & \footnotesize{$5\pm 24$} &\footnotesize{0.22} & \footnotesize{1RXS-J090558.3-213145} & \footnotesize{$0.021\pm 0.008$} & \footnotesize{20.1} & \footnotesize{$28.45_{-28.45}^{+0.34}$}  \\ 
 \footnotesize{HIP51652} & \footnotesize{$ 3.8\pm 0.3$} & \footnotesize{2.46} &\footnotesize{0.25} & \footnotesize{$0.8\pm 6$} &\footnotesize{0.20} & \footnotesize{1RXS-J103310.0+304519} & \footnotesize{$0.044\pm 0.012$} & \footnotesize{20.0} & \footnotesize{$29.10_{-0.19}^{+0.13}$}  \\ 
 \footnotesize{HIP52761} & \footnotesize{$ 4.5\pm 0.3$} & \footnotesize{2.62} &\footnotesize{0.45} & \footnotesize{$35\pm 41$} &\footnotesize{0.28} & \footnotesize{1RXS-J104719.4+204718} & \footnotesize{$0.022\pm 0.010$} & \footnotesize{20.2} & \footnotesize{$29.14_{-0.34}^{+0.20}$}  \\ 
 \footnotesize{HIP59399} & \footnotesize{$ 5.1\pm 0.2$} & \footnotesize{2.61} &\footnotesize{0.46} & \footnotesize{$40\pm 41$} &\footnotesize{0.28} & \footnotesize{1RXS-J121106.8+255933} & \footnotesize{$0.021\pm 0.009$} & \footnotesize{19.8} & \footnotesize{$29.14_{-0.34}^{+0.20}$}  \\ 
 \footnotesize{HIP65627} & \footnotesize{$ 10.2\pm 0.1$} & \footnotesize{2.85} &\footnotesize{0.53} & \footnotesize{$99\pm 123$} &\footnotesize{0.33} & \footnotesize{1RXS-J132718.3+464911} & \footnotesize{$0.041\pm 0.010$} & \footnotesize{20.3} & \footnotesize{$29.57_{-0.18}^{+0.13}$}  \\ 
 \footnotesize{HIP70354} & \footnotesize{$ 4.0\pm 0.3$} & \footnotesize{2.82} &\footnotesize{0.53} & \footnotesize{$99\pm 123$} &\footnotesize{0.30} & \footnotesize{1RXS-J142341.3+652337} & \footnotesize{$0.028\pm 0.008$} & \footnotesize{20.2} & \footnotesize{$29.36_{-0.18}^{+0.13}$}  \\ 
 \footnotesize{HIP70394} & \footnotesize{$ 6.4\pm 0.2$} & \footnotesize{2.88} &\footnotesize{0.54} & \footnotesize{$112\pm 123$} &\footnotesize{0.31} & \footnotesize{1RXS-J142407.7+290932} & \footnotesize{$0.012\pm 0.006$} & \footnotesize{20.2} & \footnotesize{$29.10_{-0.54}^{+0.24}$}  \\ 
 \footnotesize{HIP71218} & \footnotesize{$ 3.9\pm 0.3$} & \footnotesize{1.63} &\footnotesize{0.43} & \footnotesize{$26\pm 41$} &\footnotesize{0.23} & \footnotesize{1RXS-J143350.9+023430} & \footnotesize{$0.032\pm 0.013$} & \footnotesize{20.4} & \footnotesize{$29.07_{-0.29}^{+0.18}$}  \\ 
 \footnotesize{HIP77584} & \footnotesize{$ 6.2\pm 0.2$} & \footnotesize{2.73} &\footnotesize{0.49} & \footnotesize{$60\pm 41$} &\footnotesize{0.31} & \footnotesize{1RXS-J155026.6+014853} & \footnotesize{$0.052\pm 0.013$} & \footnotesize{20.1} & \footnotesize{$29.34_{-0.16}^{+0.12}$}  \\ 
 \footnotesize{HIP79068} & \footnotesize{$ 2.2\pm 0.5$} & \footnotesize{2.43} &\footnotesize{0.28} & \footnotesize{$2\pm 6$} &\footnotesize{0.21} & \footnotesize{1RXS-J160827.3+334356} & \footnotesize{$0.016\pm 0.007$} & \footnotesize{19.8} & \footnotesize{$28.43_{-0.34}^{+0.19}$}  \\ 
 \footnotesize{HIP80271} & \footnotesize{$ 4.7\pm 0.2$} & \footnotesize{2.64} &\footnotesize{0.41} & \footnotesize{$19\pm 41$} &\footnotesize{0.27} & \footnotesize{1RXS-J162309.8+353512} & \footnotesize{$0.031\pm 0.008$} & \footnotesize{19.9} & \footnotesize{$29.46_{-0.19}^{+0.14}$}  \\ 
 \footnotesize{HIP83507} & \footnotesize{$ 10.6\pm 0.1$} & \footnotesize{2.79} &\footnotesize{0.59} & \footnotesize{$197\pm 123$} &\footnotesize{0.34} & \footnotesize{1RXS-J170359.2+205243} & \footnotesize{$0.040\pm 0.010$} & \footnotesize{20.2} & \footnotesize{$29.56_{-0.18}^{+0.15}$}  \\ 
 \footnotesize{HIP86245} & \footnotesize{$ 4.0\pm 0.3$} & \footnotesize{2.74} &\footnotesize{0.39} & \footnotesize{$14\pm 24$} &\footnotesize{0.27} & \footnotesize{1RXS-J173727.2+222100} & \footnotesize{$0.049\pm 0.010$} & \footnotesize{20.3} & \footnotesize{$29.22_{-0.15}^{+0.11}$}  \\ 
 \footnotesize{HIP86781} & \footnotesize{$ 7.2\pm 0.2$} & \footnotesize{2.94} &\footnotesize{0.38} & \footnotesize{$12\pm 24$} &\footnotesize{0.23} & \footnotesize{1RXS-J174357.1+550931} & \footnotesize{$0.010\pm 0.003$} & \footnotesize{20.2} & \footnotesize{$29.10_{-0.21}^{+0.14}$}  \\ 
 \footnotesize{HIP88572} & \footnotesize{$ 3.6\pm 0.3$} & \footnotesize{1.89} &\footnotesize{0.44} & \footnotesize{$30\pm 41$} &\footnotesize{0.29} & \footnotesize{1RXS-J180506.2+532157} & \footnotesize{$0.037\pm 0.006$} & \footnotesize{20.4} & \footnotesize{$28.91_{-0.11}^{+0.09}$}  \\ 
 \footnotesize{HIP91073} & \footnotesize{$ 4.1\pm 0.3$} & \footnotesize{2.72} &\footnotesize{0.38} & \footnotesize{$12\pm 24$} &\footnotesize{0.25} & \footnotesize{1RXS-J183435.3+123221} & \footnotesize{$0.054\pm 0.015$} & \footnotesize{20.1} & \footnotesize{$29.12_{-0.19}^{+0.14}$}  \\ 
 \footnotesize{HIP101893} & \footnotesize{$ 2.7\pm 0.4$} & \footnotesize{1.88} &\footnotesize{0.38} & \footnotesize{$12\pm 24$} &\footnotesize{0.24} & \footnotesize{1RXS-J203855.8+264209} & \footnotesize{$0.025\pm 0.009$} & \footnotesize{20.3} & \footnotesize{$28.66_{-0.23}^{+0.15}$}  \\ 
 \footnotesize{HIP105066} & \footnotesize{$ 2.2\pm 0.5$} & \footnotesize{1.25} &\footnotesize{0.39} & \footnotesize{$14\pm 24$} &\footnotesize{0.25} & \footnotesize{1RXS-J211702.5-010439} & \footnotesize{$0.055\pm 0.013$} & \footnotesize{20.1} & \footnotesize{$28.91_{-0.16}^{+0.12}$}  \\ 
 \footnotesize{HIP105855} & \footnotesize{$ 5.1\pm 0.2$} & \footnotesize{2.27} &\footnotesize{0.41} & \footnotesize{$19\pm 41$} &\footnotesize{0.26} & \footnotesize{1RXS-J212623.1+734006} & \footnotesize{$0.017\pm 0.007$} & \footnotesize{20.3} & \footnotesize{$29.10_{-0.30}^{+0.18}$}  \\ 
 \footnotesize{HIP115527} & \footnotesize{$ 2.3\pm 0.5$} & \footnotesize{2.79} &\footnotesize{0.45} & \footnotesize{$35\pm 41$} &\footnotesize{0.28} & \footnotesize{1RXS-J232405.7-073305} & \footnotesize{$0.195\pm 0.026$} & \footnotesize{20.3} & \footnotesize{$29.25_{-0.08}^{+0.07}$}  \\ 
 \footnotesize{HIP117184} & \footnotesize{$ 5.0\pm 0.2$} & \footnotesize{2.62} &\footnotesize{0.43} & \footnotesize{$26\pm 41$} &\footnotesize{0.25} & \footnotesize{1RXS-J234535.4+402626} & \footnotesize{$0.061\pm 0.015$} & \footnotesize{20.1} & \footnotesize{$29.51_{-0.17}^{+0.13}$} 
\\ 
\hline
\end{longtable}
\end{center}

\begin{center}
\begin{table}[htbp]
\rotatebox{-90}{
\begin{minipage}{\textheight}
  \caption{Stellar parameters of the comparison stars.}\label{table:comp-stpara}{%
  \begin{tabular}{lcccccccccccc}
      \hline
      \hline  
Starname & Name(HIP) \footnotemark[a] & Name(HD) \footnotemark[a] & $V$ \footnotemark[a] & 
$(B-V)_{\rm{HIP}}$ \footnotemark[a] & $K$ \footnotemark[b] & $\pi$ \footnotemark[a] & $d_{\rm{HIP}}$ \footnotemark[c] & $M_{V}$ \footnotemark[d] &
$ T_{\rm{eff}}$ & $\log g$ & $v_{\rm{t}}$ & [Fe/H] \\
 & & & [mag] & [mag] & [mag] & [mas] & [pc] & [mag] & [K] & [cm s$^{-2}$] & [km s$^{-1}$] & \\
     \hline
\footnotesize{KIC7940546\footnotemark[f]} & \footnotesize{HIP92615} & \footnotesize{HD175226} & \footnotesize{7.44} & \footnotesize{$0.495\pm 0.010$} & \footnotesize{$6.17\pm 0.02$} & \footnotesize{$11.50\pm 0.60$} & \footnotesize{$87.0_{-4.3}^{+4.8}$} & \footnotesize{$2.74_{-0.12}^{+0.11}$} & \footnotesize{$6127\pm 38$} & \footnotesize{$3.63\pm 0.07$} & \footnotesize{$1.40\pm 0.16$} & \footnotesize{$-0.24\pm 0.04$} \\ 
\footnotesize{59Vir} & \footnotesize{HIP64792} & \footnotesize{HD115383} & \footnotesize{5.19} & \footnotesize{$0.585\pm 0.007$} & \footnotesize{$4.03\pm 0.24$} & \footnotesize{$55.71\pm 0.85$} & \footnotesize{$18.0_{-0.3}^{+0.3}$} & \footnotesize{$3.92_{-0.03}^{+0.03}$} & \footnotesize{$6017\pm 25$} & \footnotesize{$4.30\pm 0.06$} & \footnotesize{$1.21\pm 0.11$} & \footnotesize{$0.16\pm 0.03$} \\ 
\footnotesize{61Vir} & \footnotesize{HIP64924} & \footnotesize{HD115617} & \footnotesize{4.74} & \footnotesize{$0.709\pm 0.007$} & \footnotesize{$2.96\pm 0.24$} & \footnotesize{$117.30\pm 0.71$} & \footnotesize{$8.5_{-0.1}^{+0.1}$} & \footnotesize{$5.09_{-0.01}^{+0.01}$} & \footnotesize{$5565\pm 5$} & \footnotesize{$4.47\pm 0.02$} & \footnotesize{$0.88\pm 0.06$} & \footnotesize{$-0.01\pm 0.01$} \\ 
\footnotesize{18Sco} & \footnotesize{HIP79672} & \footnotesize{HD146233} & \footnotesize{5.49} & \footnotesize{$0.652\pm 0.009$} & \footnotesize{$4.19\pm 0.29$} & \footnotesize{$71.30\pm 0.89$} & \footnotesize{$14.0_{-0.2}^{+0.2}$} & \footnotesize{$4.76_{-0.03}^{+0.03}$} & \footnotesize{$5794\pm 10$} & \footnotesize{$4.48\pm 0.03$} & \footnotesize{$0.98\pm 0.07$} & \footnotesize{$0.04\pm 0.02$} \\ 
\footnotesize{HIP71813} & \footnotesize{HIP71813} & \footnotesize{HD129357} & \footnotesize{7.81} & \footnotesize{$0.642\pm 0.011$} & \footnotesize{$6.19\pm 0.02$} & \footnotesize{$21.22\pm 1.02$} & \footnotesize{$47.1_{-2.2}^{+2.4}$} & \footnotesize{$4.44_{-0.11}^{+0.10}$} & \footnotesize{$5761\pm 15$} & \footnotesize{$4.26\pm 0.04$} & \footnotesize{$1.05\pm 0.08$} & \footnotesize{$-0.04\pm 0.02$} \\ 
\footnotesize{HIP100963} & \footnotesize{HIP100963} & \footnotesize{HD195034} & \footnotesize{7.09} & \footnotesize{$0.642\pm 0.009$} & \footnotesize{$5.58\pm 0.02$} & \footnotesize{$35.41\pm 0.80$} & \footnotesize{$28.2_{-0.6}^{+0.7}$} & \footnotesize{$4.84_{-0.05}^{+0.05}$} & \footnotesize{$5751\pm 18$} & \footnotesize{$4.39\pm 0.05$} & \footnotesize{$1.01\pm 0.11$} & \footnotesize{$-0.06\pm 0.02$} \\ 
\footnotesize{Moon} & \footnotesize{-} & \footnotesize{-} & \footnotesize{-} & \footnotesize{-} & \footnotesize{-} & \footnotesize{-} & \footnotesize{-} & \footnotesize{-} & \footnotesize{$5741\pm 8$} & \footnotesize{$4.43\pm 0.02$} & \footnotesize{$0.98\pm 0.06$} & \footnotesize{$-0.02\pm 0.02$} \\ 
\hline 
\hline 
Starname  & $A_{V}$ & $BC$ & $\log (L/L_{\odot})$ & $R_{\rm{s}}$ & $\log Age$ & $ M_{s} $ & $\log g_{\rm{TLM}}$ &
$v\sin i$ & $r_{0}$(8542) & $\langle fB\rangle$ \footnotemark[e] & $r_{0}$(H$\alpha$) & $A$(Li) \\     
 & [mag] & [mag] & & [$R_{\odot}$] & [yr] & [$M_{\odot}$] & [cm s$^{-2}$] & [km s$^{-1}$] & & [Gauss] & & \\ 
     \hline
\footnotesize{KIC7940546\footnotemark[f]} & \footnotesize{0.00} & \footnotesize{-0.08} & \footnotesize{$0.84_{-0.04}^{+0.05}$} & \footnotesize{$2.33_{-0.14}^{+0.16}$} & \footnotesize{$9.58_{-0.03}^{+0.03}$} & \footnotesize{$1.26\pm 0.02$} & \footnotesize{$3.80_{-0.07}^{+0.06}$} & \footnotesize{$7.4\pm 0.2$} & \footnotesize{0.23} & \footnotesize{$0.5\pm 6$} & \footnotesize{0.19} & \footnotesize{$<1.6$} \\ 
\footnotesize{59Vir} & \footnotesize{-0.18} & \footnotesize{-0.10} & \footnotesize{$0.37_{-0.01}^{+0.01}$} & \footnotesize{$1.42_{-0.03}^{+0.03}$} & \footnotesize{$9.70_{-0.03}^{+0.02}$} & \footnotesize{$1.15\pm 0.01$} & \footnotesize{$4.20_{-0.02}^{+0.02}$} & \footnotesize{$6.3\pm 0.2$} & \footnotesize{0.38} & \footnotesize{$12\pm 24$} & \footnotesize{0.26} & \footnotesize{2.82} \\ 
\footnotesize{61Vir} & \footnotesize{0.18} & \footnotesize{-0.18} & \footnotesize{$-0.06_{-0.01}^{+0.01}$} & \footnotesize{$1.01_{-0.01}^{+0.01}$} & \footnotesize{$10.11_{-0.02}^{+0.02}$} & \footnotesize{$0.87\pm 0.01$} & \footnotesize{$4.37_{-0.01}^{+0.01}$} & \footnotesize{$1.3\pm 1.3$} & \footnotesize{0.19} & \footnotesize{$0.1\pm 6$} & \footnotesize{0.18} & \footnotesize{$<0.7$} \\ 
\footnotesize{18Sco} & \footnotesize{-0.17} & \footnotesize{-0.13} & \footnotesize{$0.05_{-0.01}^{+0.01}$} & \footnotesize{$1.06_{-0.02}^{+0.02}$} & \footnotesize{$9.83_{-0.01}^{+0.01}$} & \footnotesize{$0.98\pm 0.01$} & \footnotesize{$4.38_{-0.01}^{+0.01}$} & \footnotesize{$2.2\pm 0.5$} & \footnotesize{0.20} & \footnotesize{$0.2\pm 6$} & \footnotesize{0.18} & \footnotesize{1.56} \\ 
\footnotesize{HIP71813} & \footnotesize{0.13} & \footnotesize{-0.13} & \footnotesize{$0.18_{-0.04}^{+0.04}$} & \footnotesize{$1.24_{-0.06}^{+0.07}$} & \footnotesize{$10.04_{-0.02}^{+0.01}$} & \footnotesize{$0.94\pm 0.01$} & \footnotesize{$4.23_{-0.05}^{+0.05}$} & \footnotesize{$2.1\pm 0.6$} & \footnotesize{0.22} & \footnotesize{$0.4\pm 6$} & \footnotesize{0.18} & \footnotesize{$<1.4$} \\ 
\footnotesize{HIP100963} & \footnotesize{0.01} & \footnotesize{-0.13} & \footnotesize{$0.02_{-0.02}^{+0.02}$} & \footnotesize{$1.03_{-0.03}^{+0.03}$} & \footnotesize{$9.97_{-0.03}^{+0.03}$} & \footnotesize{$0.92\pm 0.01$} & \footnotesize{$4.37_{-0.03}^{+0.03}$} & \footnotesize{$2.2\pm 0.5$} & \footnotesize{0.22} & \footnotesize{$0.4\pm 6$} & \footnotesize{0.19} & \footnotesize{1.56} \\ 
\footnotesize{Moon} & \footnotesize{-} & \footnotesize{-} & \footnotesize{0.00} & \footnotesize{$1.01_{-0.01}^{+0.01}$} & \footnotesize{$9.90_{-0.03}^{+0.02}$} & \footnotesize{$0.94\pm 0.01$} & \footnotesize{$4.40_{-0.01}^{+0.01}$} & \footnotesize{$2.0\pm 0.6$} & \footnotesize{0.20} & \footnotesize{$0.2\pm 6$} & \footnotesize{0.18} & \footnotesize{0.92} \\ 
      \hline
      \hline
    \end{tabular}
}
\begin{tabnote}
\footnotemark[a] These data (HIP number, HD number, spectral type (Sp.), $V$ magnitude, $B-V$ color, and parallax in milliarcsecond ($\pi$))
were taken from the Hipparcos Catalogue \citep{ESA1997}. 
\\
\footnotemark[b] $K$ magnitude was taken from the 2MASS All-Sky Catalog of Point Sources \citep{Cutri2003}.
\\
\footnotemark[c] Stellar distance derived from Hipparcos parallax ($\pi$). Error value of $d_{\rm{HIP}}$ corresponds to that of parallax ($\pi$).
\\
\footnotemark[d] $M_{V}$(absolute magnitude) was derived from $V$ and $d_{\rm{HIP}}$. Error value of $M_{V}$ corresponds to that of stellar distance ($d_{\rm{HIP}}$).
\\
\footnotemark[e] Mean intensity of the stellar magnetic field estimated from $r_{0}(8542)$ index on the basis of Equation (1) of \citet{YNotsu2015b}.
\\
\footnotemark[f] KIC7940546 (HIP92615) is identified as Swift X-ray source (see Table \ref{table:Swift-catalog}). 
\end{tabnote}
\end{minipage}
}
\end{table}
\end{center}

\clearpage

\begin{center} 
\begin{table}[htbp]
  \caption{Target stars identified as BY Dra type variable stars in GCVS (General Catalog of Variable Stars) database \footnotemark[a]}\label{table:period-BYDra}
    \begin{tabular}{llcccccc}
      \hline
Starname(HIP) & Starname(GCVS) & $v\sin i$ & $v_{\rm{lc}}$ & $R_{\rm{s}}$ & $P_{\rm{rot}}$ & Ref. of $P_{\rm{rot}}$ \footnotemark[b] & $r_{0}$(8542) \\
 & & [km s$^{-1}$] & [km s$^{-1}$] & [$R_{\odot}$] & [day] & & \\
      \hline 
HIP19793 & V1310 Tau & $5.2\pm 0.2$ & $6.6\pm 0.5$ & $1.10^{+0.07}_{-0.06}$ & $8.48\pm 0.35$ & (1) & 0.41 \\
HIP38228 & V0377 Gem & $5.6\pm 0.2$ & $7.3\pm 0.2$ & $0.94^{+0.02}_{-0.02}$ & $6.46\pm 0.01$ & (2) & 0.48 \\
HIP105066 & NS Aqr & $2.2\pm 0.5$ & $3.9$ & $1.07^{+0.07}_{-0.06}$ & $13.78$ & (3) & 0.39 \\ 
HIP115527 & NX Aqr & $2.3\pm 0.5$ & $-$ & $0.92^{+0.03}_{-0.03}$ & no value & (3) & 0.45 \\
    \hline     
    \end{tabular}
\begin{tabnote}
\footnotemark[a] \citet{Samus2015} 
\\
\footnotemark[b] (1) \citet{Paulson2004}; (2) \citet{Gaidos2000}; (3) \citet{Strassmeier2000}
\end{tabnote}    
\end{table} 
  \end{center}

\end{document}